\RequirePackage{fix-cm}
\documentclass[final]{svjour3}

\usepackage[square,numbers]{natbib}
\usepackage{booktabs}
\usepackage{morefloats}
\usepackage{tabularx}
\usepackage{longtable}
\usepackage{graphicx}
\usepackage[subrefformat=parens,labelformat=parens]{subfig}
\usepackage{tikz}
\usepackage{amsmath}
\usepackage{enumitem}
\usepackage{amsfonts}
\usepackage{amssymb}
\usepackage{placeins}
\usepackage[cleanup={},process=all]{pstool}
\usepackage[export]{adjustbox}

\newcommand{\x}{\cdot}

\newcommand*{\affmark}[1][*]{\textsuperscript{#1}}
\EndPreamble

\DeclareRobustCommand\full  {\tikz[baseline=-0.6ex]\draw[thick] (0,0)--(0.5,0);}
\DeclareRobustCommand\dotted{\tikz[baseline=-0.6ex]\draw[thick,dotted] (0,0)--(0.54,0);}

\newcommand{\DefineRemark}[2]{%
  \expandafter\newcommand\csname rmk-#1\endcsname{#2}%
}
\newcommand{\Remark}[1]{\csname rmk-#1\endcsname}

\DefineRemark{n2dr}{-27}
\DefineRemark{n2cfdr}{-7}
\DefineRemark{n2drnet}{-100}
\DefineRemark{n2ainc}{19.4}
\DefineRemark{n3dr}{-1}
\DefineRemark{n3cfdr}{8}
\DefineRemark{n3drnet}{-17}
\DefineRemark{n3ainc}{10.1}
\DefineRemark{n4dr}{-9}
\DefineRemark{n4cfdr}{9}
\DefineRemark{n4drnet}{-24}
\DefineRemark{n4ainc}{19.4}
\DefineRemark{n5dr}{-26}
\DefineRemark{n5cfdr}{9}
\DefineRemark{n5drnet}{-42}
\DefineRemark{n5ainc}{39.0}
\DefineRemark{n6dr}{-10}
\DefineRemark{n6cfdr}{8}
\DefineRemark{n6drnet}{-17}
\DefineRemark{n6ainc}{19.4}
\DefineRemark{n7dr}{0}
\DefineRemark{n7cfdr}{5}
\DefineRemark{n7drnet}{-1}
\DefineRemark{n7ainc}{4.7}
\DefineRemark{n8dr}{-17}
\DefineRemark{n8cfdr}{9}
\DefineRemark{n8drnet}{-24}
\DefineRemark{n8ainc}{29.4}
\DefineRemark{n9dr}{-9}
\DefineRemark{n9cfdr}{7}
\DefineRemark{n9drnet}{-11}
\DefineRemark{n9ainc}{17.2}
\DefineRemark{n10dr}{0}
\DefineRemark{n10cfdr}{4}
\DefineRemark{n10drnet}{-98}
\DefineRemark{n10ainc}{3.5}
\DefineRemark{n11dr}{9}
\DefineRemark{n11cfdr}{10}
\DefineRemark{n11drnet}{-10}
\DefineRemark{n11ainc}{1.9}
\DefineRemark{n12dr}{8}
\DefineRemark{n12cfdr}{9}
\DefineRemark{n12drnet}{-1}
\DefineRemark{n12ainc}{1.7}
\DefineRemark{n13dr}{8}
\DefineRemark{n13cfdr}{11}
\DefineRemark{n13drnet}{-10}
\DefineRemark{n13ainc}{3.5}
\DefineRemark{n14dr}{5}
\DefineRemark{n14cfdr}{8}
\DefineRemark{n14drnet}{-2}
\DefineRemark{n14ainc}{3.5}
\DefineRemark{n15dr}{3}
\DefineRemark{n15cfdr}{8}
\DefineRemark{n15drnet}{-4}
\DefineRemark{n15ainc}{4.9}
\DefineRemark{n16dr}{-10}
\DefineRemark{n16cfdr}{4}
\DefineRemark{n16drnet}{-33}
\DefineRemark{n16ainc}{14.6}
\DefineRemark{n17dr}{-3}
\DefineRemark{n17cfdr}{5}
\DefineRemark{n17drnet}{-10}
\DefineRemark{n17ainc}{8.6}
\DefineRemark{n18dr}{5}
\DefineRemark{n18cfdr}{5}
\DefineRemark{n18drnet}{-6}
\DefineRemark{n18ainc}{0.1}
\DefineRemark{n19dr}{13}
\DefineRemark{n19cfdr}{13}
\DefineRemark{n19drnet}{-88}
\DefineRemark{n19ainc}{0.9}
\DefineRemark{n20dr}{0}
\DefineRemark{n20cfdr}{3}
\DefineRemark{n20drnet}{-289}
\DefineRemark{n20ainc}{2.4}
\DefineRemark{n21dr}{3}
\DefineRemark{n21cfdr}{3}
\DefineRemark{n21drnet}{1}
\DefineRemark{n21ainc}{0.1}
\DefineRemark{n22dr}{7}
\DefineRemark{n22cfdr}{8}
\DefineRemark{n22drnet}{0}
\DefineRemark{n22ainc}{0.4}
\DefineRemark{n23dr}{12}
\DefineRemark{n23cfdr}{13}
\DefineRemark{n23drnet}{-4}
\DefineRemark{n23ainc}{0.9}
\DefineRemark{n24dr}{15}
\DefineRemark{n24cfdr}{16}
\DefineRemark{n24drnet}{-14}
\DefineRemark{n24ainc}{1.6}
\DefineRemark{n25dr}{15}
\DefineRemark{n25cfdr}{17}
\DefineRemark{n25drnet}{-32}
\DefineRemark{n25ainc}{2.4}
\DefineRemark{n26dr}{13}
\DefineRemark{n26cfdr}{16}
\DefineRemark{n26drnet}{-55}
\DefineRemark{n26ainc}{3.5}
\DefineRemark{n27dr}{1}
\DefineRemark{n27cfdr}{1}
\DefineRemark{n27drnet}{0}
\DefineRemark{n27ainc}{0.1}
\DefineRemark{n28dr}{3}
\DefineRemark{n28cfdr}{4}
\DefineRemark{n28drnet}{2}
\DefineRemark{n28ainc}{0.4}
\DefineRemark{n29dr}{6}
\DefineRemark{n29cfdr}{6}
\DefineRemark{n29drnet}{3}
\DefineRemark{n29ainc}{0.9}
\DefineRemark{n30dr}{9}
\DefineRemark{n30cfdr}{10}
\DefineRemark{n30drnet}{4}
\DefineRemark{n30ainc}{1.6}
\DefineRemark{n31dr}{9}
\DefineRemark{n31cfdr}{11}
\DefineRemark{n31drnet}{1}
\DefineRemark{n31ainc}{2.4}
\DefineRemark{n32dr}{9}
\DefineRemark{n32cfdr}{12}
\DefineRemark{n32drnet}{-3}
\DefineRemark{n32ainc}{3.5}
\DefineRemark{n33dr}{1}
\DefineRemark{n33cfdr}{1}
\DefineRemark{n33drnet}{1}
\DefineRemark{n33ainc}{0.1}
\DefineRemark{n34dr}{0}
\DefineRemark{n34cfdr}{1}
\DefineRemark{n34drnet}{0}
\DefineRemark{n34ainc}{0.4}
\DefineRemark{n35dr}{3}
\DefineRemark{n35cfdr}{4}
\DefineRemark{n35drnet}{2}
\DefineRemark{n35ainc}{0.9}
\DefineRemark{n36dr}{3}
\DefineRemark{n36cfdr}{5}
\DefineRemark{n36drnet}{1}
\DefineRemark{n36ainc}{1.6}
\DefineRemark{n37dr}{2}
\DefineRemark{n37cfdr}{5}
\DefineRemark{n37drnet}{-1}
\DefineRemark{n37ainc}{2.4}
\DefineRemark{n38dr}{2}
\DefineRemark{n38cfdr}{6}
\DefineRemark{n38drnet}{-3}
\DefineRemark{n38ainc}{3.5}
\DefineRemark{n40dr}{2}
\DefineRemark{n40cfdr}{7}
\DefineRemark{n40drnet}{-74}
\DefineRemark{n40ainc}{5.1}
\DefineRemark{n41dr}{-4}
\DefineRemark{n41cfdr}{5}
\DefineRemark{n41drnet}{-71}
\DefineRemark{n41ainc}{8.9}
\DefineRemark{n42dr}{9}
\DefineRemark{n42cfdr}{12}
\DefineRemark{n42drnet}{-6}
\DefineRemark{n42ainc}{3.5}
\DefineRemark{n43dr}{5}
\DefineRemark{n43cfdr}{6}
\DefineRemark{n43drnet}{2}
\DefineRemark{n43ainc}{1.2}
\DefineRemark{n44dr}{6}
\DefineRemark{n44cfdr}{8}
\DefineRemark{n44drnet}{2}
\DefineRemark{n44ainc}{2.3}
\DefineRemark{n45dr}{-5}
\DefineRemark{n45cfdr}{6}
\DefineRemark{n45drnet}{-23}
\DefineRemark{n45ainc}{11.0}
\DefineRemark{n46dr}{-1}
\DefineRemark{n46cfdr}{6}
\DefineRemark{n46drnet}{-10}
\DefineRemark{n46ainc}{6.8}
\DefineRemark{n47dr}{2}
\DefineRemark{n47cfdr}{2}
\DefineRemark{n47drnet}{1}
\DefineRemark{n47ainc}{0.5}
\DefineRemark{n49dr}{11}
\DefineRemark{n49cfdr}{11}
\DefineRemark{n49drnet}{-47}
\DefineRemark{n49ainc}{0.2}
\DefineRemark{n50dr}{14}
\DefineRemark{n50cfdr}{14}
\DefineRemark{n50drnet}{-6}
\DefineRemark{n50ainc}{0.4}
\DefineRemark{n51dr}{19}
\DefineRemark{n51cfdr}{19}
\DefineRemark{n51drnet}{-23}
\DefineRemark{n51ainc}{0.9}
\DefineRemark{n52dr}{19}
\DefineRemark{n52cfdr}{20}
\DefineRemark{n52drnet}{-17}
\DefineRemark{n52ainc}{1.4}
\DefineRemark{n53dr}{21}
\DefineRemark{n53cfdr}{22}
\DefineRemark{n53drnet}{-35}
\DefineRemark{n53ainc}{2.0}
\DefineRemark{n54dr}{8}
\DefineRemark{n54cfdr}{8}
\DefineRemark{n54drnet}{4}
\DefineRemark{n54ainc}{0.3}
\DefineRemark{n55dr}{17}
\DefineRemark{n55cfdr}{19}
\DefineRemark{n55drnet}{-4}
\DefineRemark{n55ainc}{1.9}
\DefineRemark{n56dr}{2}
\DefineRemark{n56cfdr}{2}
\DefineRemark{n56drnet}{1}
\DefineRemark{n56ainc}{0.1}
\DefineRemark{n57dr}{13}
\DefineRemark{n57cfdr}{14}
\DefineRemark{n57drnet}{4}
\DefineRemark{n57ainc}{1.6}
\DefineRemark{n58dr}{8}
\DefineRemark{n58cfdr}{8}
\DefineRemark{n58drnet}{5}
\DefineRemark{n58ainc}{0.6}
\DefineRemark{n60dr}{10}
\DefineRemark{n60cfdr}{12}
\DefineRemark{n60drnet}{-84}
\DefineRemark{n60ainc}{2.9}
\DefineRemark{n61dr}{7}
\DefineRemark{n61cfdr}{12}
\DefineRemark{n61drnet}{-69}
\DefineRemark{n61ainc}{4.7}
\DefineRemark{n62dr}{7}
\DefineRemark{n62cfdr}{7}
\DefineRemark{n62drnet}{2}
\DefineRemark{n62ainc}{0.6}
\DefineRemark{n63dr}{12}
\DefineRemark{n63cfdr}{14}
\DefineRemark{n63drnet}{-8}
\DefineRemark{n63ainc}{2.3}
\DefineRemark{n64dr}{7}
\DefineRemark{n64cfdr}{8}
\DefineRemark{n64drnet}{3}
\DefineRemark{n64ainc}{0.9}
\DefineRemark{n65dr}{3}
\DefineRemark{n65cfdr}{4}
\DefineRemark{n65drnet}{2}
\DefineRemark{n65ainc}{0.4}
\DefineRemark{n66dr}{6}
\DefineRemark{n66cfdr}{9}
\DefineRemark{n66drnet}{-8}
\DefineRemark{n66ainc}{4.3}
\DefineRemark{n67dr}{6}
\DefineRemark{n67cfdr}{7}
\DefineRemark{n67drnet}{2}
\DefineRemark{n67ainc}{1.8}
\DefineRemark{n68dr}{5}
\DefineRemark{n68cfdr}{5}
\DefineRemark{n68drnet}{-2}
\DefineRemark{n68ainc}{0.1}
\DefineRemark{n69dr}{19}
\DefineRemark{n69cfdr}{20}
\DefineRemark{n69drnet}{-27}
\DefineRemark{n69ainc}{0.8}
\DefineRemark{n70dr}{22}
\DefineRemark{n70cfdr}{23}
\DefineRemark{n70drnet}{-67}
\DefineRemark{n70ainc}{1.5}
\DefineRemark{n71dr}{20}
\DefineRemark{n71cfdr}{21}
\DefineRemark{n71drnet}{-15}
\DefineRemark{n71ainc}{1.1}
\DefineRemark{n72dr}{15}
\DefineRemark{n72cfdr}{15}
\DefineRemark{n72drnet}{3}
\DefineRemark{n72ainc}{0.6}
\DefineRemark{n73dr}{21}
\DefineRemark{n73cfdr}{22}
\DefineRemark{n73drnet}{-14}
\DefineRemark{n73ainc}{1.7}
\DefineRemark{n74dr}{7}
\DefineRemark{n74cfdr}{7}
\DefineRemark{n74drnet}{4}
\DefineRemark{n74ainc}{0.3}
\DefineRemark{n75dr}{9}
\DefineRemark{n75cfdr}{9}
\DefineRemark{n75drnet}{5}
\DefineRemark{n75ainc}{0.4}
\DefineRemark{n76dr}{13}
\DefineRemark{n76cfdr}{14}
\DefineRemark{n76drnet}{4}
\DefineRemark{n76ainc}{1.3}
\DefineRemark{n77dr}{3}
\DefineRemark{n77cfdr}{3}
\DefineRemark{n77drnet}{2}
\DefineRemark{n77ainc}{0.1}
\DefineRemark{n79dr}{21}
\DefineRemark{n79cfdr}{21}
\DefineRemark{n79drnet}{-17}
\DefineRemark{n79ainc}{0.3}
\DefineRemark{n80dr}{26}
\DefineRemark{n80cfdr}{26}
\DefineRemark{n80drnet}{-20}
\DefineRemark{n80ainc}{0.7}
\DefineRemark{n81dr}{7}
\DefineRemark{n81cfdr}{7}
\DefineRemark{n81drnet}{4}
\DefineRemark{n81ainc}{0.1}
\DefineRemark{n82dr}{19}
\DefineRemark{n82cfdr}{19}
\DefineRemark{n82drnet}{8}
\DefineRemark{n82ainc}{0.4}
\DefineRemark{n83dr}{1}
\DefineRemark{n83cfdr}{1}
\DefineRemark{n83drnet}{1}
\DefineRemark{n83ainc}{0.0}
\DefineRemark{n84dr}{16}
\DefineRemark{n84cfdr}{16}
\DefineRemark{n84drnet}{10}
\DefineRemark{n84ainc}{0.5}

\begin{document}

\title{Drag Reduction and Energy Saving by Spanwise Traveling Transversal Surface Waves for Flat Plate Flow}
\titlerunning{Drag reduction for flat plate flow}

\author{Marian Albers\affmark[1]  \and 
        Pascal S. Meysonnat\affmark[1] \and 
        Daniel Fernex\affmark[2] \and
        Richard Semaan\affmark[2] \and 
        Bernd R.~Noack\affmark[3,2,4,5] \and
        Wolfgang Schr\"oder\affmark[1,6]}

\institute{Marian Albers \at 
           \email{m.albers@aia.rwth-aachen.de}\\
           \and
           \affmark[1] Institute of Aerodynamics, RWTH Aachen University, W\"ullnerstrasse 5a, 52062 Aachen, Germany\\
           \affmark[2] Institut f\"{u}r Str\"{o}mungsmechanik, Technische Universit\"{a}t Braunschweig, Hermann-Blenk-Str. 37, 38108 Braunschweig, Germany\\
           \affmark[3] LIMSI, CNRS, Universit\'e Paris-Saclay, B{\^a}t 507, 
rue du Belv\'ed\`ere, Campus Universitaire,
F-91403 Orsay, France\\
           \affmark[4] Institut f\"ur Str\"omungsmechanik und Technische Akustik (ISTA), Technische Universit\"at Berlin, M\"uller-Breslau-Stra{\ss}e 8, 10623 Berlin, Germany\\
           \affmark[5] Institute for Turbulence-Noise-Vibration Interaction and Control, Harbin Institute of Technology, Shenzhen Campus,  China\\
           \affmark[6] JARA Center for Simulation and Data Science, RWTH Aachen University, Seffenter Weg 23, 52074 Aachen, Germany
         }

\authorrunning{Albers et al}

\date{\today}

\maketitle

\begin{abstract}
  Wall-resolved large-eddy simulations are performed to study the
  impact of spanwise traveling transversal surface waves in
  zero-pressure gradient turbulent boundary layer flow. Eighty
  variations of wavelength, period, and amplitude of the space- and
  time-dependent sinusoidal wall motion are considered for a boundary
  layer at a momentum thickness based Reynolds number of
  $Re_\theta = 1000$. The results show a strong decrease of friction
  drag of up to $\Remark{n80dr}\,\%$ and considerable net power saving
  of up to $\Remark{n84drnet}\,\%$. However, the highest net power
  saving does not occur at the maximum drag reduction. The drag
  reduction is modeled as a function of the actuation parameters by
  support vector regression using the LES data.
% SVR + ridge lines confuse more than they explain.
%  Using support vector
%    regression a drag reduction model is derived from sparse
%    data and a ridgeline behavior of optimum drag reduction through
%    the parameter space is identified. 
  A substantial attenuation of
  the near-wall turbulence intensity and especially a weakening of the
  near-wall velocity streaks are observed.  Similarities between the
  current actuation technique and the method of a spanwise oscillating
  wall without any normal surface deflection are reported.  In
  particular, the generation of a directional spanwise oscillating
  Stokes layer is found to be related to skin-friction reduction.
\end{abstract}


\section{Introduction}
Surface friction in turbulent wall-bounded flows is one of the major
contributors to the overall drag of flow over slender bodies in
general and passenger planes at cruise flight in particular. Lowering
turbulent friction drag is therefore essential to meet future
$\text{CO}_2$ reduction goals. Besides preventing fully turbulent flow
and benefiting from the considerably lower laminar drag
\cite{Spalart2011}, there is substantial past and ongoing research in
the field of turbulent drag reduction. Unlike active
techniques, which require energy introduction to the system, passive
techniques such as riblets
\cite{Walsh1978,Bechert1985,Walsh1989,Garcia-Mayoral2011,Garcia-Mayoral2011b}
and compliant surfaces \cite{Choi1997,Kim2014,Luhar2015,Zhang2017}
yield reduced skin friction without any added energy.  However,
compared to passive approaches, which are optimized for single
operating conditions, active techniques are adaptive and, at least for
some techniques, can achieve higher net power saving. These
results, however, hold mostly in canonical flow setups like
turbulent channel flows under laboratory conditions, i.e., at
extremely low technology readiness levels.

In the following, we briefly review active flow control techniques
that are most relevant to the current study. Particularly, we focus on
methods that use either in-plane wall motion, such as forcing parallel
to the wall or out of plane wall motion. In this study, we employ
actuation that belongs to the latter category.  We discuss its
similarities and peculiarities over existing techniques, and present
its drag reduction and net power saving potentials, which reach 26\%
and 10\% compared to the unactuated flow.

Inspired by turbulence suppression by temporary pressure gradient
variations \cite{Moin1990}, Jung~et~al.~\cite{Jung1992} performed the
first simulations of spanwise wall oscillations which resulted in
significantly lowered friction drag. The method was subsequently
investigated in detail in the following years for Poiseuille flow
\cite{Choi1998,Quadrio2004} and turbulent boundary layer flow
\cite{Ricco2004b, Yudhistira2011, Lardeau2013}. Detailed analyses
indicated an interaction of the oscillating spanwise shear with the
near-wall velocity streaks
\cite{Touber2012,Agostini2014}. Furthermore, it was found that the
maximum drag reduction in turbulent boundary layer flow is moderately
lower than in turbulent channel flow and is reached at a significantly
lower oscillation period \cite{Lardeau2013}. Motivated by this simple
but effective approach, other forms of spatio-temporal forcing have
been developed, which is excellently discussed by Quadrio
\cite{Quadrio2011}. A modified variant of the purely temporal
oscillations of spanwise velocity \cite{Touber2012} are spanwise
traveling waves of spanwise forcing \cite{Du2000,Du2002} or spanwise
traveling waves of a flexible surface \cite{Zhao2004}. Although the
techniques are different in their actuation principle, the effect of
introducing oscillating spanwise shear close to the wall is alike.

Another actuation variant is spanwise traveling transversal
surface waves \cite{Itoh2006}. Instead of directly introducing
spanwise velocity, the surface is wavily deflected in the wall-normal
direction to generate a secondary flow field of periodic wall-normal
and spanwise fluctuations. Positive drag reduction using this
technique was achieved experimentally
\cite{Itoh2006,Tamano2012,Li2018} and numerically for
channel flow \cite{Tomiyama2013}, boundary layer flow
\cite{Klumpp2011,Koh2015,Koh2015a,Ishar2019}, and airfoil flow
\cite{Albers2019}. Tomiyama and Fukagata~\cite{Tomiyama2013} observed a
possible shielding effect of quasi-streamwise vortices from the wall
by the wave-like deformations and showed that a combination of the
thickness of the Stokes layer, i.e., the actuation period, and the actuation
velocity amplitudes scales reasonably well with drag reduction.

However, the question remains what happens at higher amplitudes and
wavelengths and lower periods, especially considering the vast gap
between the mostly relatively short wavelength setups in numerical
simulations and the large wavelengths in all experimental setups
limited by mechanical actuator constraints. We will investigate if the
trend of higher drag reduction for longer wavelengths \cite{Du2002}
can be confirmed. Furthermore, an optimum forcing period $T^+$ in
inner scaling was not determined for this technique and it remains an
open question if one exists and if so if it is in the range of other
techniques, e.g., $T^+ \approx 70$ for spanwise oscillating wall in
turbulent boundary layer flow \cite{Lardeau2013}. In this study, we
address these questions. We investigate the higher reduction trends
with longer wavelengths and examine the flow sensitivities over the
space spanned by the three actuation parameters, i.e., wavelength,
wave period, and wave amplitude, using high-resolution large-eddy
simulations (LES) of turbulent boundary layer flow. In total, 80
configurations are computed. The objective is to achieve drag
reduction and net energy saving in the range of other actuation
techniques and to compare the flow response to that from pure spanwise
oscillations.

The paper has the following structure. First, the numerical method is
concisely described in section~\ref{sec::numerical}. Then the flow
setup and all flow and actuation parameters are specified in
section~\ref{sec::setup}. The results are discussed in
section~\ref{sec::results}. Finally, the essential results are
summarized in section~\ref{sec::conclusions}.

\section{Numerical method}
\label{sec::numerical}
The actuated turbulent boundary layer flow is computed by solving the
unsteady compressible Navier-Stokes equations by a large-eddy
simulation (LES) formulation. To capture the temporal variation of the
geometry, the equations are written in the Arbitrary
Lagrangian-Eulerian (ALE) formulation \cite{Hirt1997} such that the
actuated wall can be represented by an appropriate mesh
deformation. Additional volume fluxes are determined to satisfy the
Geometry Conservation Law (GCL).

The discrete solution is based on a finite-volume approximation on a
structured body-fitted mesh. A second-order accurate formulation of
the inviscid fluxes using the advection upstream splitting method
(AUSM) is applied. The cell-surface values of the flow
quantities are reconstructed from the surrounding cell-center values
using a Monotone Upstream Scheme for Conservation Laws (MUSCL) type
strategy. The viscous fluxes are discretized by a modified cell-vertex
scheme at second-order accuracy. The time integration is performed by
a second-order accurate five-stage Runge-Kutta scheme, rendering the
overall discretization second-order accurate.

The subgrid scales in the LES are implicitly modeled following the
monotonically integrated large-eddy simulation
approach~\cite{Boris1992}, i.e., the numerical dissipation of the AUSM
scheme models for the viscous dissipation of the high wavenumber
turbulence spectrum \cite{Meinke2002}. Thus, the small-scale
structures are not explicitly resolved in the whole flow domain and
the grid is used as a spatial filter resolving the large
energy-containing structures in the inertial subrange.

The numerical method has thoroughly been validated by computing a wide
variety of internal and external flow
problems~\cite{Ruetten2005,Alkishriwi2006,Renze2008,Statnikov2017}. Analyses
of drag reduction have been performed for riblet structured
surfaces~\cite{Klumpp2010a} and for traveling transversal surface
waves in canonical turbulent boundary layer
flow~\cite{Klumpp2011,Koh2015a,Koh2015,Meysonnat2016} and in
turbulent airfoil flow~\cite{Albers2019}.  The quality of the results
confirms the validity of the approach for the current flow problem.

\section{Computational Setup}
\label{sec::setup}
The zero-pressure gradient (ZPG) turbulent boundary layer flow over a
wall actuated by a sinusoidal wave motion is defined in a Cartesian
domain with the $x$-axis in the main flow direction, the
$y$-axis in the wall-normal direction, and the $z$-axis in the
spanwise direction. The velocity vector in the Cartesian frame of
reference $\mathbf{x} = (x,y,z)$ is denoted by $\mathbf{u} = (u,v,w)$,
the pressure is given by $p$, and the density by $\rho$. The flow
variables are non-dimensionalized using the flow quantities at rest,
the speed of sound $a_0$, and the momentum thickness of the boundary
layer at $x_0 = 0$ such that $\theta(x_0=0) = 1$. The momentum
thickness based Reynolds number is
$Re_\theta = u_{\infty} \theta / \nu = 1,000$ at $x_0$ where
$u_\infty$ is the freestream velocity and $\nu$ is the kinematic
viscosity. The Mach number is $M = 0.1$, i.e., the flow is nearly
incompressible. Note that unlike standard ZPG turbulent boundary layer
flow, the actuated flow is statistically three-dimensional due to the
wave propagating in the $z$-direction.

An overview of the setup is given in Fig.~\ref{fig::grid}. The
dimensions of the physical domain are $L_x = 190\,\theta$, $L_y =
105\,\theta$ in the streamwise and wall-normal direction. For the
spanwise direction, five domain widths, $L_z \in [21.65\,\theta,$
$25.98\,\theta,$ $34.64\,\theta,$ $38.97\,\theta, 64.95\,\theta]$ are
used. The mesh resolution is $\Delta x^+ = 12.0$ in the streamwise
direction, $\left. \Delta y^+\right|_{\text{wall}} = 1.0$ in the
wall-normal direction with gradual coarsening off the wall up to
$\Delta y^+ = 16.0$ at the boundary layer edge, and $\Delta z^+ = 4.0$
in the spanwise direction. This yields a DNS-like resolution near the
wall. Away from the wall, the resolution requirements are lower such
that overall an LES resolution is achieved.

At the inflow of the domain, the reformulated synthetic turbulence
generation (RSTG) method by Roidl et al.~\cite{Roidl2013} is used to prescribe a
fully turbulent inflow distribution with an adaptation length of less
than five boundary-layer thicknesses $\delta_{99}$. A fully turbulent
boundary layer is achieved at $x_0$, which marks the onset of the
actuation. Characteristic outflow conditions are applied at the
downstream and upper boundaries, whereas periodic conditions
are used in the spanwise direction. On the wall, no-slip conditions
are imposed and the wall motion is described by
\begin{equation}
  \label{eqn::actuation}
y^+|_\mathrm{wall}(z^+,t^+)  = g(x) A^+\cos\left(\frac{2\pi}{\lambda^+} z^+ + \frac{2\pi}{T^+}t^+\right),
\end{equation}
where $A^+ = A u_\tau/ \nu$ is the amplitude, $\lambda^+ = \lambda u_\tau / \nu$ is the wavelength, and $T^+ = T u_\tau^2 / \nu$ is the
period. If not otherwise stated an inner scaling is used for all wave
parameters, i.e., the quantities are scaled by the kinematic viscosity
$\nu$ and the friction velocity $u_\tau$ of the non-actuated reference
case $N_1$. 

In total, $80$ variations of $A^+ \in [0,78]$, $T^+ \in [20,120]$, and
$\lambda^+ \in [200,3000]$ are simulated. A detailed list of all
parameter combinations can be found in Tab.~\ref{tab::simulations} in
the appendix. Note that the narrowest domain has a spanwise
extent of $L_z^+ = 1000$ such that for all wavelengths $\lambda^+ <
1000$ multiple wavelengths are considered. A sketch of all wavelengths
and the respective maximum amplitude at each wavelength is illustrated in
Fig.~\ref{fig::wavelengths}. To enable a smooth spatial transition
from the stationary flat plate to the deflected wall and vice versa,
the piecewise defined function
\begin{subequations}
\label{eqn::piecewise}
\begin{eqnarray}
g(x) &=& \left \{
  \begin{array}{ll}
    0 & \text{if} \quad x < -5 \\ 
    \frac{1}{2}\left[ 1-\cos\left( \frac{\pi (x+5)}{10}\right) \right] & \text{if} \quad -5 \leq x < 5\\ 
    1  & \text{if} \quad 5 \leq x < 130\\
    \frac{1}{2}\left[ 1+\cos\left( \frac{\pi (x-130)}{10}\right) \right] & \text{if} \quad 130 \leq x < 140\\ 
    0 & \text{otherwise}
  \end{array}
    \right.
\\
\end{eqnarray}
\end{subequations}
is used in Eq.~\ref{eqn::actuation}.

The drag is integrated over the wall surface within the streamwise
interval $x \in [50.0, 100.0]$ and over the entire spanwise
extent. This area is colored in Fig.~\ref{fig::grid}. Hence,
the drag is only computed in the region where the flow is fully
influenced by the traveling wave actuation. The actuated boundary
layer is not impacted by the flow upstream and downstream of the
actuated surface.

The computing strategy is such that first, the non-actuated reference
case is simulated for $t u_\infty / \theta \approx 650$ convective
times until a quasi-steady state is observed in the integrated
drag. All actuated cases are then initialized by the flow field of the
reference case and the transition from a flat plate to an actuated
wall flow is performed via a temporal decay controlled by
$1-\cos(t)$. Once a new quasi-steady state is observed all simulations
are averaged over $t u_\infty / \theta \approx 1250$ times.
\begin{figure}
\begin{center}
  \begin{tikzpicture}[x={(0.939cm,-0.34cm)}, y={(0cm,1cm)}, z={(0.939cm,0.34cm)}]\
    \draw [->,>=stealth] (0,0,-1.5) -- (0,0,-2.0) node [left]{$z$};
    \draw [->,>=stealth] (0,0,-1.5) -- (0,0.5,-1.5) node [above]{$y$};
    \draw [->,>=stealth] (0,0,-1.5) -- (0.5,0,-1.5) node [right]{$x$};
    \draw [<->,>=stealth] (0,0,-0.5) -- (7,0,-0.5) node [pos=.5,below=1.0]{$L_x$};
    \draw [<->,>=stealth] (0,0,-0.5) -- (0,2,-0.5) node [pos=.5,left]{$L_y$};
    \draw [<->,>=stealth] (7.5,0,0) -- (7.5,0,2) node [pos=.5,below=2.0]{$L_z$};
    \draw (0,0,0) -- (1,0,0) sin (1.5,0.05,0) cos (2,0.1,0) -- (5,0.1,0) .. controls (5.3,0.1,0) and (5.7,-0.1,0) .. (7,0.0,0) -- (7,2,0) -- (7,2,0) -- (0,2,0) -- cycle;
    \draw (0,0,2) -- (1,0,2) sin (1.5,0.05,2) cos (2,0.1,2) -- (5,0.1,2) .. controls (5.3,0.1,2) and (5.7,-0.1,2) .. (7,0.0,2) -- (7,2,2) -- (7,2,2) -- (0,2,2) -- cycle;
    \draw[color=red,fill=red, fill opacity=0.2, domain=0:2, variable=\z]  (2.5,0.1,0) -- plot (2.5,{0.1*sin((1.5707+2*3.14159*\z) r)},\z)  -- (4.5,0.1,2) --  plot (4.5,{0.1*sin((1.5707+2*3.14159*\z) r)},2.0-\z)  -- (4.5,0.1,0) -- (2.5,0.1,0) -- cycle;
    \draw (7,0,0) -- (7,0,2);
    \draw (0,0,0) -- (0,0,2);
    \draw (0,2,0) -- (0,2,2);
    \draw (7,2,0) -- (7,2,2);
    \draw[opacity=0.5, variable=\z, samples at={0,0.05,...,2.05}]
    plot (5, {0.1*sin((1.5707+2*3.14159*\z) r)}, \z);      
    \draw[opacity=0.5, variable=\z, domain=0:2]
    plot (2, {0.1*sin((1.5707+2*3.14159*\z) r)}, \z);    
    \draw (5,0.1,0) -- (5,0.25,0);
    \draw (5,0.1,1) -- (5,0.25,1);
    \draw[<->] (5,0.2,0) -- (5,0.2,1) node [pos=.5,sloped,above] {$\lambda$};
    
    \node (x0) at (1.5,0,-1.3) {$x_0$};
    \draw[->,>=stealth] (x0) -- (1.5,0,-0.6);
    \node (inflow) at (-1,2,1) {Inflow};
    \node (flat) at (1, 0,1) {\small{Wall (flat)}};
    \node (flat) at (3.3, 0,1) {\small{Wall (wave)}};
    \draw[<->,>=stealth] (3,1.8,0) -- (3,1.8,-0.2) -- (3,2.2,-0.2) -- node[pos=.5,sloped,above] {Periodic BC} (3,2.2,2.2) -- (3,1.8,2.2) -- (3,1.8,2.0);
    \draw[color=blue, dashed] (0,0,1) .. controls (0.5,0.2,1) ..  (0.5,2,1) -- (0,2,1) -- (0,0,1);
    \draw[color=blue, ->,>=stealth] (0,0.2,1) -- (0.3,0.2,1);
    \draw[color=blue, ->,>=stealth] (0,0.4,1) -- (0.45,0.4,1);
    \draw[color=blue, ->,>=stealth] (0,0.6,1) -- (0.5,0.6,1);
    \draw[color=blue, ->,>=stealth] (0,0.8,1) -- (0.5,0.8,1);
    \draw[color=blue, ->,>=stealth] (0,1.0,1) -- (0.5,1.0,1);
    \draw[color=blue, ->,>=stealth] (0,1.2,1) -- (0.5,1.2,1);
    \draw[color=blue, ->,>=stealth] (0,1.4,1) -- (0.5,1.4,1);
    \draw[color=blue, ->,>=stealth] (0,1.6,1) -- (0.5,1.6,1);
    \draw[color=blue, ->,>=stealth] (0,1.8,1) -- (0.5,1.8,1);
  \end{tikzpicture}
  \caption{Overview of the physical domain of the actuated turbulent
    boundary layer flow. The quantities $L_x, L_y,$ and $L_z$ are the
    dimensions of the domain in the Cartesian directions, $\lambda$ is
    the wavelength of the spanwise traveling wave, and $x_0$ marks the
    onset of the actuation. The surface area $A_\mathrm{surf}$ for the
    integration of the wall-shear stress $\tau_w$ is shaded red.}
  \label{fig::grid}

  \begin{tikzpicture}[scale=0.8]
    \draw [dashed](6.5,1) -- (6.5,-5.5);
    \draw [dashed](6.5,-4) -- (13,-4);

    % Lambda = 200
    \def\dwidth{2.0}
    \def\wlength{0.4}
    \def\offsetx{0.0}
    \def\offsety{0.0}
    \def\amp{0.09}
    \def\thickness{0.1}
    \draw[<->,>=stealth] (\offsetx,\offsety+0.5) -- (\offsetx+\wlength,\offsety+0.5) node [pos=0.5,sloped, above] {$\lambda^+ = 200$};
    \draw (\offsetx,\offsety+0.5-0.1) -- (\offsetx,\offsety+0.5+0.1);
    \draw (\offsetx+\wlength,\offsety+0.5-0.1) -- (\offsetx+\wlength,\offsety+0.5+0.1);
    \node at (\offsetx-0.1,\offsety) [left] {$A^+_{\mathrm{max}} = 45$};
    \draw[domain=0:\dwidth, variable=\x, samples=200]  (\offsetx,\offsety+0) -- plot (\offsetx+\x,{\offsety+\amp*sin((2*3.14159*\x/\wlength) r)});
    \fill[fill=black,opacity=0.2,domain=0:\dwidth, variable=\x, samples=200]  (\offsetx,\offsety+0) -- plot (\offsetx+\x,{\offsety+\amp*sin((2*3.14159*\x/\wlength) r)})  -- (\offsetx+\dwidth,\offsety+0) -- (\offsetx+\dwidth,\offsety-\thickness) -- plot (\offsetx+\dwidth-\x,{\amp*sin((2*3.14159*(\wlength-\x)/\wlength) r)-\thickness+\offsety}) -- cycle;

    % Lambda = 500
    \def\dwidth{2.0}
    \def\wlength{1.0}
    \def\offsetx{0.0}
    \def\offsety{-1.7}
    \def\amp{0.128}
    \def\thickness{0.1}
    \draw[<->,>=stealth] (\offsetx,\offsety+0.5) -- (\offsetx+\wlength,\offsety+0.5) node [pos=0.5,sloped, above] {$\lambda^+ = 500$};
    \draw (\offsetx,\offsety+0.5-0.1) -- (\offsetx,\offsety+0.5+0.1);
    \draw (\offsetx+\wlength,\offsety+0.5-0.1) -- (\offsetx+\wlength,\offsety+0.5+0.1);
    \node at (\offsetx-0.1,\offsety) [left] {$A^+_{\mathrm{max}} = 64$};
    \draw[domain=0:\dwidth, variable=\x, samples=100]  (\offsetx,\offsety+0) -- plot (\offsetx+\x,{\offsety+\amp*sin((2*3.14159*\x/\wlength) r)});
    \fill[fill=black,opacity=0.2,domain=0:\dwidth, variable=\x, samples=100]  (\offsetx,\offsety+0) -- plot (\offsetx+\x,{\offsety+\amp*sin((2*3.14159*\x/\wlength) r)})  -- (\offsetx+\dwidth,\offsety+0) -- (\offsetx+\dwidth,\offsety-\thickness) -- plot (\offsetx+\dwidth-\x,{\amp*sin((2*3.14159*(\wlength-\x)/\wlength) r)-\thickness+\offsety}) -- cycle;

    % Lambda = 600
    \def\dwidth{2.4}
    \def\wlength{1.2}
    \def\offsetx{9.0}
    \def\offsety{0.0}
    \def\amp{0.132}
    \def\thickness{0.1}
    \draw[<->,>=stealth] (\offsetx,\offsety+0.5) -- (\offsetx+\wlength,\offsety+0.5) node [pos=0.5,sloped, above] {$\lambda^+ = 600$};
    \draw (\offsetx,\offsety+0.5-0.1) -- (\offsetx,\offsety+0.5+0.1);
    \draw (\offsetx+\wlength,\offsety+0.5-0.1) -- (\offsetx+\wlength,\offsety+0.5+0.1);
    \node at (\offsetx-0.1,\offsety) [left] {$A^+_{\mathrm{max}} = 66$};
    \draw[domain=0:\dwidth, variable=\x, samples=100]  (\offsetx,\offsety+0) -- plot (\offsetx+\x,{\offsety+\amp*sin((2*3.14159*\x/\wlength) r)});
    \fill[fill=black,opacity=0.2,domain=0:\dwidth, variable=\x, samples=100]  (\offsetx,\offsety+0) -- plot (\offsetx+\x,{\offsety+\amp*sin((2*3.14159*\x/\wlength) r)})  -- (\offsetx+\dwidth,\offsety+0) -- (\offsetx+\dwidth,\offsety-\thickness) -- plot (\offsetx+\dwidth-\x,{\amp*sin((2*3.14159*(\wlength-\x)/\wlength) r)-\thickness+\offsety}) -- cycle;

    % Lambda = 900
    \def\dwidth{3.6}
    \def\wlength{1.8}
    \def\offsetx{9.0}
    \def\offsety{-1.7}
    \def\amp{0.126}
    \def\thickness{0.1}
    \draw[<->,>=stealth] (\offsetx,\offsety+0.5) -- (\offsetx+\wlength,\offsety+0.5) node [pos=0.5,sloped, above] {$\lambda^+ = 900$};
    \draw (\offsetx,\offsety+0.5-0.1) -- (\offsetx,\offsety+0.5+0.1);
    \draw (\offsetx+\wlength,\offsety+0.5-0.1) -- (\offsetx+\wlength,\offsety+0.5+0.1);
    \node at (\offsetx-0.1,\offsety) [left] {$A^+_{\mathrm{max}} = 63$};
    \draw[domain=0:\dwidth, variable=\x, samples=50]  (\offsetx,\offsety+0) -- plot (\offsetx+\x,{\offsety+\amp*sin((2*3.14159*\x/\wlength) r)});
    \fill[fill=black,opacity=0.2,domain=0:\dwidth, variable=\x, samples=50]  (\offsetx,\offsety+0) -- plot (\offsetx+\x,{\offsety+\amp*sin((2*3.14159*\x/\wlength) r)})  -- (\offsetx+\dwidth,\offsety+0) -- (\offsetx+\dwidth,\offsety-\thickness) -- plot (\offsetx+\dwidth-\x,{\amp*sin((2*3.14159*(\wlength-\x)/\wlength) r)-\thickness+\offsety}) -- cycle;

    % Lambda = 1000
    \def\dwidth{2.0}
    \def\wlength{2.0}
    \def\offsetx{0.0}
    \def\offsety{-3.4}
    \def\amp{0.12}
    \def\thickness{0.1}
    \draw[<->,>=stealth] (\offsetx,\offsety+0.5) -- (\offsetx+\wlength,\offsety+0.5) node [pos=0.5,sloped, above] {$\lambda^+ = 1000$};
    \draw (\offsetx,\offsety+0.5-0.1) -- (\offsetx,\offsety+0.5+0.1);
    \draw (\offsetx+\wlength,\offsety+0.5-0.1) -- (\offsetx+\wlength,\offsety+0.5+0.1);
    \node at (\offsetx-0.1,\offsety) [left] {$A^+_{\mathrm{max}} = 60$};
    \draw[domain=0:\dwidth, variable=\x, samples=50]  (\offsetx,\offsety+0) -- plot (\offsetx+\x,{\offsety+\amp*sin((2*3.14159*\x/\wlength) r)});
    \fill[fill=black,opacity=0.2,domain=0:\dwidth, variable=\x, samples=50]  (\offsetx,\offsety+0) -- plot (\offsetx+\x,{\offsety+\amp*sin((2*3.14159*\x/\wlength) r)})  -- (\offsetx+\dwidth,\offsety+0) -- (\offsetx+\dwidth,\offsety-\thickness) -- plot (\offsetx+\dwidth-\x,{\amp*sin((2*3.14159*(\wlength-\x)/\wlength) r)-\thickness+\offsety}) -- cycle;

    % Lambda = 1800
    \def\dwidth{3.6}
    \def\wlength{3.6}
    \def\offsetx{9.0}
    \def\offsety{-3.4}
    \def\amp{0.150}
    \def\thickness{0.1}
    \draw[<->,>=stealth] (\offsetx,\offsety+0.5) -- (\offsetx+\wlength,\offsety+0.5) node [pos=0.5,sloped, above] {$\lambda^+ = 1800$};
    \draw (\offsetx,\offsety+0.5-0.1) -- (\offsetx,\offsety+0.5+0.1);
    \draw (\offsetx+\wlength,\offsety+0.5-0.1) -- (\offsetx+\wlength,\offsety+0.5+0.1);
    \node at (\offsetx-0.1,\offsety) [left] {$A^+_{\mathrm{max}} = 75$};
    \draw[domain=0:\dwidth, variable=\x, samples=50]  (\offsetx,\offsety+0) -- plot (\offsetx+\x,{\offsety+\amp*sin((2*3.14159*\x/\wlength) r)});
    \fill[fill=black,opacity=0.2,domain=0:\dwidth, variable=\x, samples=50]  (\offsetx,\offsety+0) -- plot (\offsetx+\x,{\offsety+\amp*sin((2*3.14159*\x/\wlength) r)})  -- (\offsetx+\dwidth,\offsety+0) -- (\offsetx+\dwidth,\offsety-\thickness) -- plot (\offsetx+\dwidth-\x,{\amp*sin((2*3.14159*(\wlength-\x)/\wlength) r)-\thickness+\offsety}) -- cycle;

        % Lambda = 1600
    \def\dwidth{3.2}
    \def\wlength{3.2}
    \def\offsetx{9.0}
    \def\offsety{-5.1}
    \def\amp{0.142}
    \def\thickness{0.1}
    \draw[<->,>=stealth] (\offsetx,\offsety+0.5) -- (\offsetx+\wlength,\offsety+0.5) node [pos=0.5,sloped, above] {$\lambda^+ = 1600$};
    \draw (\offsetx,\offsety+0.5-0.1) -- (\offsetx,\offsety+0.5+0.1);
    \draw (\offsetx+\wlength,\offsety+0.5-0.1) -- (\offsetx+\wlength,\offsety+0.5+0.1);
    \node at (\offsetx-0.1,\offsety) [left] {$A^+_{\mathrm{max}} = 72$};
    \draw[domain=0:\dwidth, variable=\x, samples=50]  (\offsetx,\offsety+0) -- plot (\offsetx+\x,{\offsety+\amp*sin((2*3.14159*\x/\wlength) r)});
    \fill[fill=black,opacity=0.2,domain=0:\dwidth, variable=\x, samples=50]  (\offsetx,\offsety+0) -- plot (\offsetx+\x,{\offsety+\amp*sin((2*3.14159*\x/\wlength) r)})  -- (\offsetx+\dwidth,\offsety+0) -- (\offsetx+\dwidth,\offsety-\thickness) -- plot (\offsetx+\dwidth-\x,{\amp*sin((2*3.14159*(\wlength-\x)/\wlength) r)-\thickness+\offsety}) -- cycle;

    % Lambda = 3000
    \def\dwidth{6.0}
    \def\wlength{6.0}
    \def\offsetx{0.0}
    \def\offsety{-5.1}
    \def\amp{0.156}
    \def\thickness{0.1}
    \draw[<->,>=stealth] (\offsetx,\offsety+0.5) -- (\offsetx+\wlength,\offsety+0.5) node [pos=0.5,sloped, above] {$\lambda^+ = 3000$};
    \draw (\offsetx,\offsety+0.5-0.1) -- (\offsetx,\offsety+0.5+0.1);
    \draw (\offsetx+\wlength,\offsety+0.5-0.1) -- (\offsetx+\wlength,\offsety+0.5+0.1);
    \node at (\offsetx-0.1,\offsety) [left] {$A^+_{\mathrm{max}} = 78$};
    \draw[domain=0:\dwidth, variable=\x, samples=50]  (\offsetx,\offsety+0) -- plot (\offsetx+\x,{\offsety+\amp*sin((2*3.14159*\x/\wlength) r)});
    \fill[fill=black,opacity=0.2,domain=0:\dwidth, variable=\x, samples=50]  (\offsetx,\offsety+0) -- plot (\offsetx+\x,{\offsety+\amp*sin((2*3.14159*\x/\wlength) r)})  -- (\offsetx+\dwidth,\offsety+0) -- (\offsetx+\dwidth,\offsety-\thickness) -- plot (\offsetx+\dwidth-\x,{\amp*sin((2*3.14159*(\wlength-\x)/\wlength) r)-\thickness+\offsety}) -- cycle;
  \end{tikzpicture}
  \caption{Overview of the sinusoidal wall function with all used
    wavelengths and the corresponding maximum amplitude. The spanwise extent is varied to fit an
    integer number of wavelengths into the domain.}
  \label{fig::wavelengths}
\end{center}
\end{figure}

\section{Results}
\label{sec::results}
In the following, the results of the parameter study will be
investigated in detail. First, a grid convergence study is performed
in Sec.~\ref{sec::gridconvergence}. Then, the wall-shear stress
reductions as a function of the wave parameters are thoroughly
discussed in Sec.~\ref{sec::wall_shear_stress}. The findings are
compared with data from the literature for the same and similar drag
reduction techniques. This analysis is followed in
Sec.~\ref{sec::drag_reduction} by a discussion of the variation of the
total drag, i.e., the wall-shear stress multiplied by the wetted
surface, since the wetted surface changes for different parameter
setups. Support vector regression is used to predict drag reductions
and to examine the drag reduction sensitivities for varying actuation
settings is concisely presented in Sec.~\ref{sec::ml}.  The statistics
of the non-actuated and actuated flow field are compared, and links to
drag reduction mechanisms are drawn in
Sec.~\ref{sec::flow_statistics}. The spanwise shear distribution in
the near-wall region with special focus on the periodic Stokes shear
and its relevance for drag reduction is analyzed in
Sec.~\ref{sec::spanwiseshear}. Finally, the net energy balance results
are presented in Sec.~\ref{sec::energy_saving}.

%%%%%%%%%%%%%%%%%%%%%%%%%%%%
%%%%%%% GRID STUDY %%%%%%%%%
%%%%%%%%%%%%%%%%%%%%%%%%%%%%
\subsection{Grid convergence}
\label{sec::gridconvergence}
To ensure a sufficient grid resolution for the large-eddy simulation
of the actuated turbulent boundary layer flow a grid convergence study
is conducted. The data of the three meshes that are compared are
listed in Tab.~\ref{tab::grid}. Besides the mesh data, the drag
reduction $\Delta c_d$, which is defined in
Sec.~\ref{sec::drag_reduction}, is given. The actuation parameters in
inner coordinates are $\lambda^+ = 1000$, $T^+ = 40$, and $A^+ =
40$. They define case $N_{24}$ in Tab.~\ref{tab::simulations} in
  the appendix for which a moderate drag reduction is achieved. It is
evident from the results in Tab.~\ref{tab::grid} that the drag
reduction values on the standard and the fine grid are nearly
identical. On the coarse grid, however, a clear deviation is
determined.

\begin{figure}
  \begin{center}
    \subfloat[~]{\includegraphics[width=0.45\textwidth]{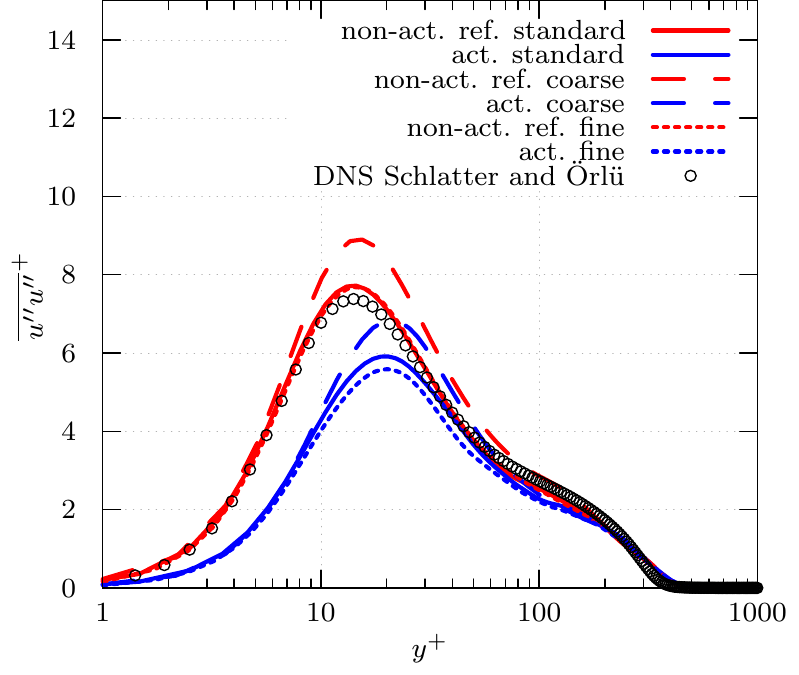}}~
    \subfloat[~]{\includegraphics[width=0.45\textwidth]{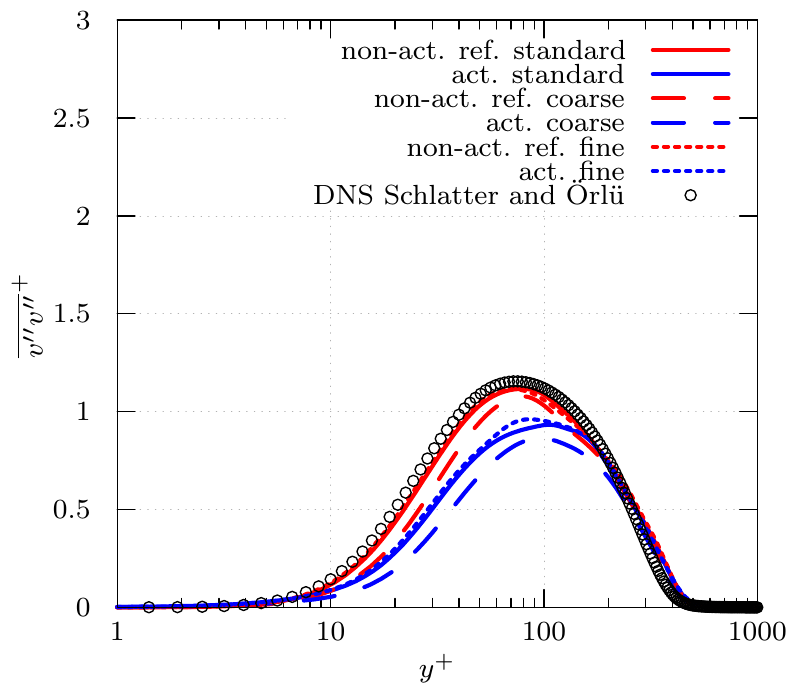}}

    \subfloat[~]{\includegraphics[width=0.45\textwidth]{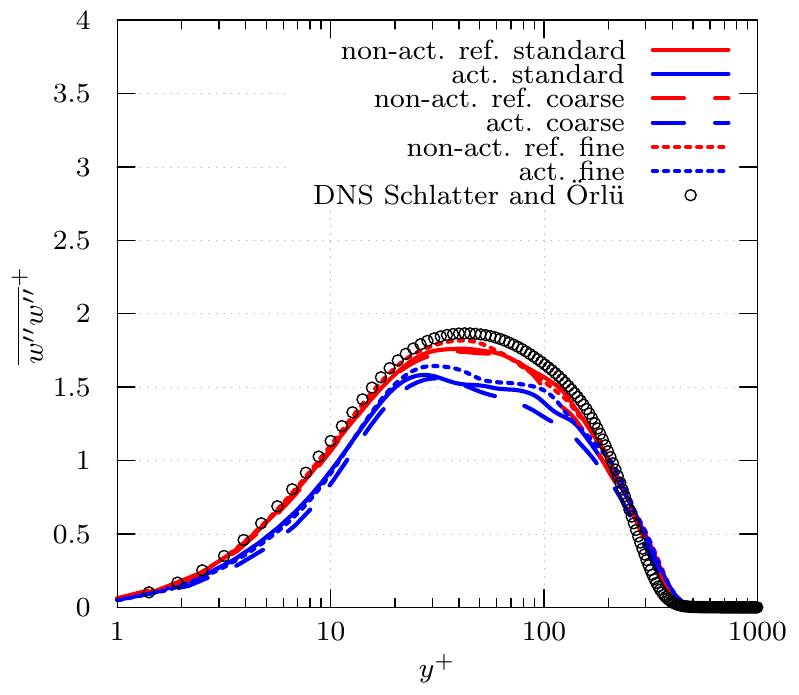}}~
    \subfloat[~]{\includegraphics[width=0.45\textwidth]{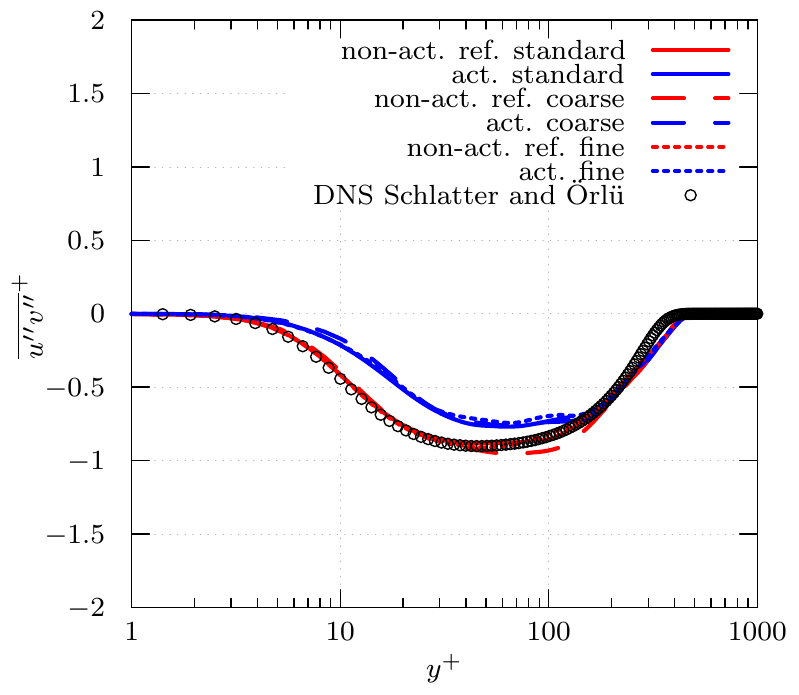}}
    \caption{Comparison of the wall-normal distributions of the
      symmetric stresses and the shear-stress component of the
      Reynolds stress tensor on the coarse, standard, and fine grids
      for the non-actuated reference case and the actuated case
      $N_{24}$ with DNS data \cite{Schlatter2010}.}
    \label{fig::gridstudy}
  \end{center}
\end{figure}

A comparison of the symmetric stresses and the shear-stress component
of the Reynolds stress tensor at the streamwise location of $x/\theta
= 50$, i.e., on the non-actuated surface at $Re_\theta = 1033$, is
shown in Fig.~\ref{fig::gridstudy}. The distributions of the standard
and the fine mesh nearly collapse for the non-actuated reference case
and for the actuated case $N_{24}$ for all four components. Note that
similar results were determined for other actuation parameter
configurations. Only on the coarse mesh larger deviations are
obtained. Furthermore, the data of the non-actuated reference case for
the standard and the fine mesh shows good agreement with DNS data
\cite{Schlatter2010} of a turbulent boundary layer at a similar
Reynolds number, i.e., $Re_\theta = 1006$.

In conclusion, the analysis of the data shows that the resolution of
the standard grid can be considered sufficient to accurately predict
actuated turbulent boundary layer flow.
\begin{table}
  \centering
  \begin{tabularx}{1.0\textwidth}{r@{\quad}r@{\quad}r@{\quad}r@{\quad}r@{\quad}r@{\quad}r@{\quad}r}
    Case & $\Delta x^+$ & $\Delta y^+|_\mathrm{wall}$ & $\Delta z^+$ & $N_\mathrm{BL}$ & $N_i \times N_j \times N_k$ & $N_\mathrm{total}$ & $\Delta c_d$\\
    \midrule
  coarse & 20.0 & 1.5 & 8.0 & 78 & $438 \times 118 \times 125$  & $6.5\cdot10^6$ & $19.28\,\%$\\
standard & 12.0 & 1.0 & 4.0 & 89 &$732 \times 131 \times 250$   & $24.0\cdot10^6$ & $15.17\,\%$\\
    fine & 10.0 & 0.7 & 2.0 & 100 & $877 \times 142 \times 500$ & $62.3\cdot10^6$ & $15.18\,\%$\\
    \bottomrule
  \end{tabularx}
  \caption{Summary of the grid spacings $\Delta x^+$, $\Delta
    y^+_\mathrm{wall}$, and $\Delta z^+$, the number of cells inside
    the boundary layer $N_{BL}$, the number of cells in the coordinate
    directions, the total number of cells, and the computed drag reduction $\Delta
    c_d$ for the coarse, standard, and fine grid.}
  \label{tab::grid}
\end{table}

% =============================================================
% 															SUBSECTION
% =============================================================
\subsection{Wall-shear stress reductions}
\label{sec::wall_shear_stress}
\begin{figure}
  \begin{center}
    \subfloat[~]{\includegraphics[width=0.5\textwidth]{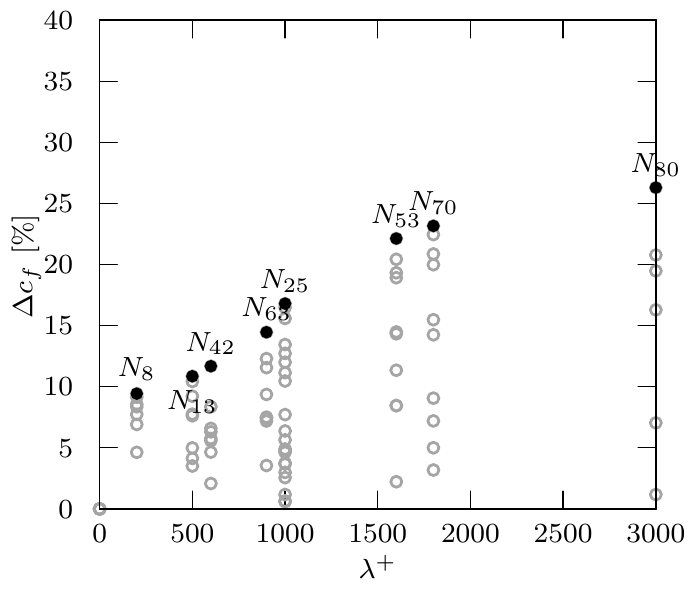}\label{fig::cfdep::wavelength}}~
    \subfloat[~]{\includegraphics[width=0.5\textwidth]{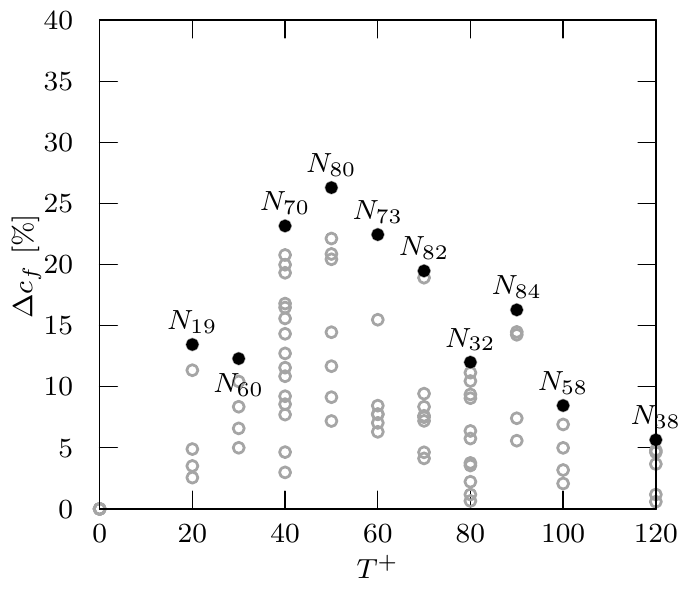}\label{fig::cfdep::period}}

    \subfloat[~]{\includegraphics[width=0.5\textwidth]{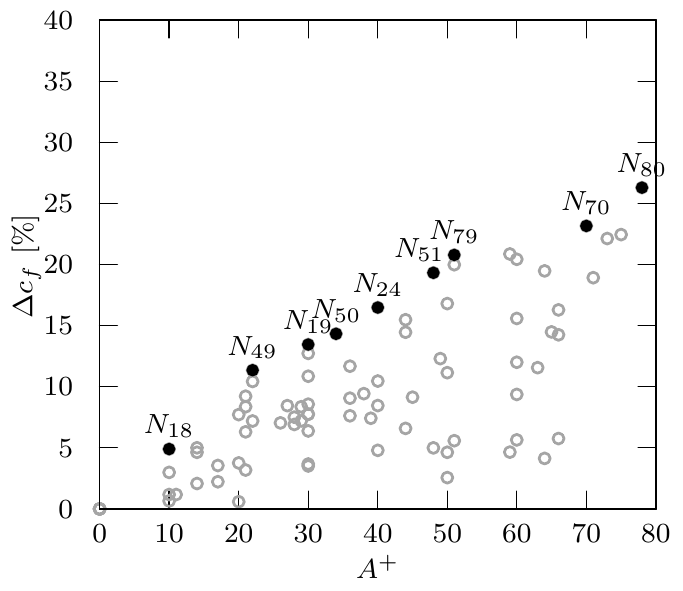}\label{fig::cfdep::amp}}~
    \subfloat[~]{\includegraphics[width=0.5\textwidth]{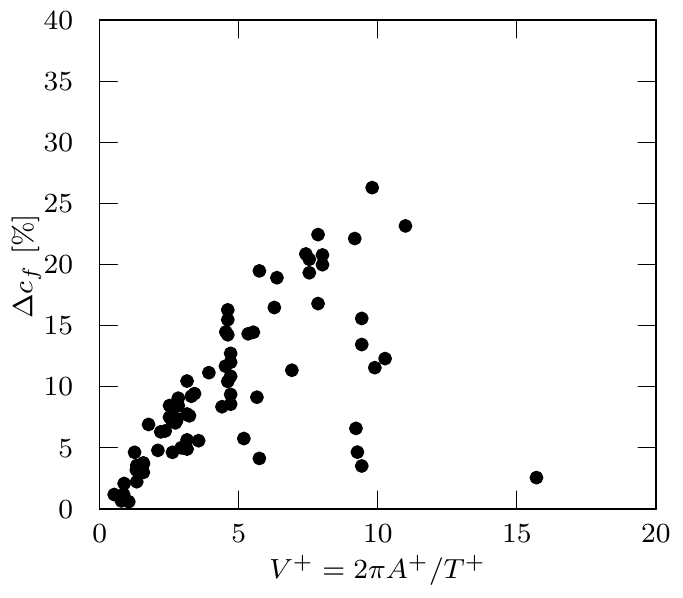}\label{fig::cfdep::vel}}
    \caption{Dependence of the skin-friction reduction $\Delta c_f$ on
      \protect\subref{fig::cfdep::wavelength} the wavelength
      $\lambda^+$, \protect\subref{fig::cfdep::period} the period
      $T^+$, \protect\subref{fig::cfdep::amp} the amplitude $A^+$, and
      \protect\subref{fig::cfdep::vel} the actuation velocity $V^+ =
      2\pi A^+/T^+$. For clarity, only the cases that define the upper
      envelope of the skin-friction reduction $\Delta c_f$ are shown
      in \protect\subref{fig::cfdep::wavelength},
      \protect\subref{fig::cfdep::period}, and
      \protect\subref{fig::cfdep::amp}.}
    \label{fig::cfdep}
  \end{center}
\end{figure}

The skin-friction reduction $\Delta c_f$ is defined in percent by
\begin{align*}
  \Delta c_f = \frac{\tau_{w,\mathrm{ref}} - \tau_{w,\mathrm{act}}}{\tau_{w,\mathrm{ref}}} \cdot 100\>,
\end{align*}
where the wall-shear stress $\tau_w$ is averaged over the shaded
surface $A_\mathrm{surf}$ in Fig.~\ref{fig::grid}. The values for $\Delta
c_f$ of the $80$ cases are listed in Tab.~\ref{tab::simulations} in
the appendix. The dependence of $\Delta c_f$ on the various
parameters, i.e., the wavelength, period, amplitude, and amplitude
velocity, is summarized in Fig.~\ref{fig::cfdep}. The highlighted and
numbered distributions are mainly from cases of the upper envelope of
the wall-shear stress reduction, i.e., the maximum $\Delta c_f$ values
for the wavelength, the period, and the amplitude are emphasized. The
discussion and illustration in Fig.~\ref{fig::cfdep} summarize the
pronounced varying dependence of the wall-shear stress on the
different actuation parameters.

In Fig.~\subref*{fig::cfdep::wavelength} a quasi-linear increase of
the skin-friction reduction $\Delta c_f$ as a function of the
wavelength $ \lambda^+$ is observed. Note, however, that this
quasi-linear distribution is achieved by changing simultaneously the
amplitude $A^+$ and the period $T^+$. Especially the latter has to
undergo quite a non-linear variation to obtain such an approximately
linear $\Delta c_f$ growth.

The dependence of $\Delta c_f$ on the wave period $T^+$ at various
$A^+$ and $\lambda^+$ is presented in
Fig.~\subref*{fig::cfdep::period}. Due to the coupling between the
forcing strength and the actuation period, which is in contrast to
other actuation methods like spanwise traveling waves of spanwise
forcing \cite{Du2002} and traveling waves of flexible wall
\cite{Zhao2004}, the optimum period $T^+$ is determined by an
internal, i.e., fluid mechanical, and an external, i.e., actuator,
related condition. The ideal $T^+$ is defined by the streak formation
time scale \cite{Touber2012}, i.e., for oscillatory spanwise
  forcing the period must be small enough to disrupt the reorganization
  of the streaks, and a sufficient strength of the forcing, which
increases with decreasing period, is required. The dependence of the
skin-friction reduction on the period in
Fig.~\subref*{fig::cfdep::period} shows that the optimum period among
all $80$ cases is on the order of $T^+ = \mathcal{O}\left( 50
\right)$, which is slightly lower than the optimum period $T^+ \approx
70$ of a spanwise oscillating wall in turbulent boundary layer flow
\cite{Lardeau2013}.

Note that likewise tendencies can be found for spanwise traveling
oscillatory forcing \cite{Du2002} such as increased drag reduction
with higher wavelengths
(cf.~Fig.\subref*{fig::cfdep::wavelength}). The longest wavelength
considered in this study ($\lambda^+ = 3000$) is comparable to that
used in the experimental setups by Tamano and Itoh \cite{Tamano2012}
and Li et al. \cite{Li2018}. Although their lowest investigated period
is $T^+ \approx 110$ and thus considerably higher than the
optimum found in this study, their results corroborate the
tendency of higher wall-shear stress reduction at lower periods in the
regime $110 \leq T^+ \leq 302.5$.

The distribution of the skin-friction decrease as a function of the
amplitude in Fig.~\subref*{fig::cfdep::amp} shows that the maximum
skin-friction reduction is directly coupled to the amplitude. This is
to some extent expected since the velocity and thus, the strength of
the actuation $V^+ = 2\pi A^+ / T^+$ is directly related to the
amplitude. That is, at a given period the strength of the actuation is
determined by the amplitude. This is confirmed by the experimental
findings of Li et al. \cite{Li2018}. They obtain in a lower amplitude
range a monotonic increase of the skin-friction reduction for
increasing amplitude. Figure \subref*{fig::cfdep::vel} presents the
skin-friction reductions as a function of the velocity amplitude of
the actuation $V^+$, where the scaling shows a quasi-linear behavior
for larger wavelengths $\lambda^+ > 1000$.

\begin{figure}
  \begin{center}
    \subfloat[$\lambda^+ \leq 1000$]{\includegraphics[width=0.5\textwidth]{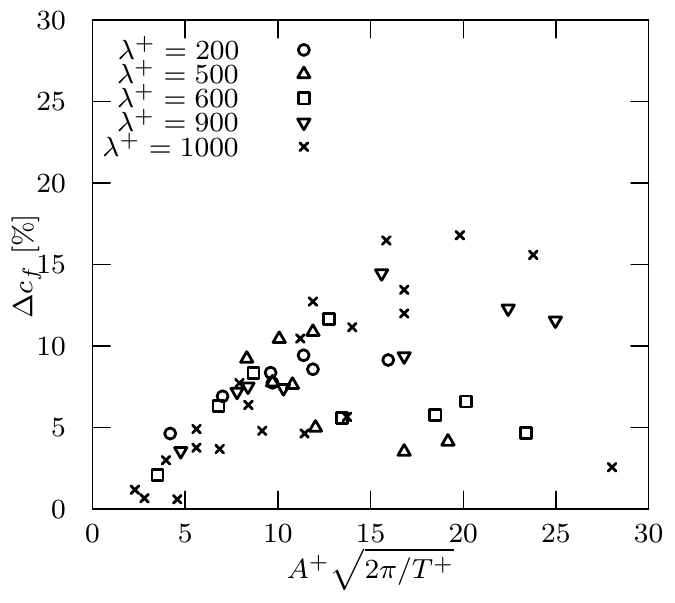}\label{fig::dr::small}}~
    \subfloat[$\lambda^+ > 1000$]{\includegraphics[width=0.5\textwidth]{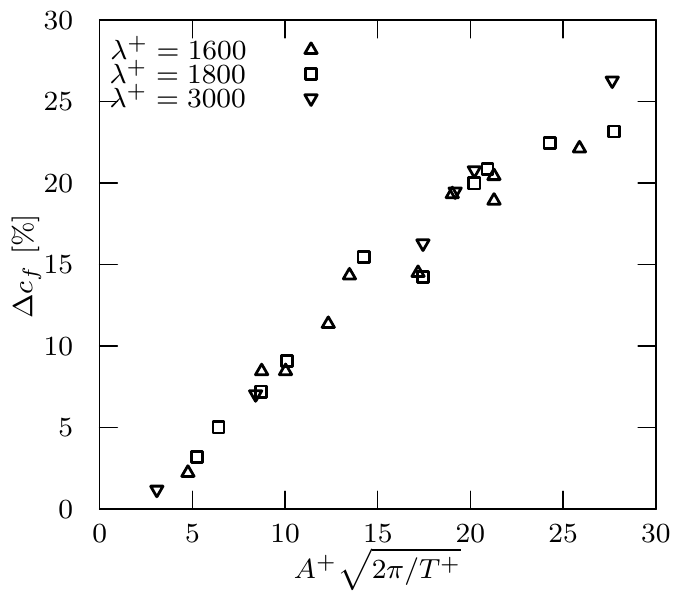}\label{fig::dr::large}}
    \caption{Relative skin-friction reduction of all cases as a
      function of the scaling parameter $A^+\sqrt{2 \pi/ T^+}$ for \protect\subref{fig::dr::small}
      $\lambda^+ \leq 1000$ and \protect\subref{fig::dr::large} $\lambda^+ > 1000$.}
    \label{fig::dr}
  \end{center}
\end{figure}

A similar scaling for $\Delta c_f$ was proposed by Tomiyama and
Fukagata~\cite{Tomiyama2013} by combining the amplitude of the
actuation velocity $V^+$ (cf. Fig.~\subref*{fig::cfdep::vel}) and the
thickness of the Stokes layer $\sqrt{T^+/(2 \pi)}$ such that $\Delta
c_f = f(A^+\sqrt{2\pi /T^+})$ which is plotted in
Fig.~\ref{fig::dr}. For shorter wavelengths
(cf. Fig.~\subref*{fig::dr::small}), i.e., $\lambda^+ \leq 1000$, a
linear scaling is only observed for small values $A^+\sqrt{2\pi / T^+}
< 10$. Note that the current overall reductions are lower than those
in \cite{Tomiyama2013} which is likely due to the higher Reynolds
number in this study ($Re_\tau = 360$ vs.  $Re_\tau = 180$ in
\cite{Tomiyama2013}) and due to a generally lower skin-friction
reduction efficiency in turbulent boundary layers compared to
turbulent channel flow \cite{Ricco2004}. For higher scaling factor
values, the distribution is more scattered. For larger wavelengths
$\lambda^+ > 1000$ (cf. Fig.~\subref*{fig::dr::large}), the
skin-friction reduction scales almost linearly over the entire
range. Above a certain value $A^+\sqrt{2\pi / T^+} \gtrsim 20$, however, the linear behavior of the
skin-friction reduction deteriorates. We believe the reason for this
degradation is the large momentum injection into the boundary layer
via too high a velocity amplitude. This increases the spanwise
velocity component which leads to an amplified turbulent exchange.
% =============================================================
% 															SUBSECTION
% =============================================================
\subsection{Drag reduction}
\label{sec::drag_reduction}
Introducing a wave motion of the surface means that the area of the
moving wall increases with the amplitude and the wavelength of the
wave. The data in Tab.~\ref{tab::simulations} in the appendix shows that this change
of the wetted surface $\Delta A_\mathrm{surf}$ can be quite
substantial, especially at small wavelengths and high amplitudes. At
high wavelengths, this variation becomes rather small. Since the
friction drag is defined by the product of the wall-shear stress and
the surface interacting with the fluid, the variation of the wetted
surface has to be taken into account leading to a non-linear relation
between wall-shear stress and friction drag reduction.

The averaged drag reduction is defined as
\begin{align*}
  \Delta c_d = \frac{c_{d,\mathrm{ref}} - c_{d,\mathrm{act}}}{c_{d,\mathrm{ref}}} \cdot 100
\end{align*}
where $c_d$ is the drag coefficient computed by an integration over the shaded surface $A_\mathrm{surf}$ in Fig.~\ref{fig::grid},
\begin{align*}
  c_d &= \frac{2}{\rho_\infty u_\infty^2 A_{\mathrm{ref}}} \int_{A_\mathrm{surf}}  \tau_w \mathbf{e}_y\cdot\mathbf{n}dA~.
\end{align*}
The quantity $\mathbf{n}$ denotes the unit normal vector of the surface,
$\mathbf{e}_y$ is the unit vector in the $y$-direction, and
$A_\mathrm{ref} = 1$ is the reference surface.

The data in Tab.~\ref{tab::simulations} in the appendix evidences the differences
between $\Delta c_f$ and $\Delta c_d$, especially at small
wavelengths. The highest drag reduction is $\Delta c_d =
\Remark{n80dr}\,\%$ for a wavelength of $\lambda^+ = 3000$, a period
of $T^+ = 50$, and an amplitude of $A^+ = 78$. Due to the large
wavelength, the increase of the wetted surface is only $\Delta
A_\mathrm{surf} = \Remark{n80ainc}\,\%$. The highest drag increase $\Delta
c_d = \Remark{n2dr}\,\%$ with the highest corresponding skin-friction
coefficient increase $\Delta c_f = \Remark{n2cfdr}\,\%$ is observed
for $\lambda^+ = 200$, $T^+ = 20$, and $A^+ = 30$. As stated before,
this configuration with a small wavelength suffers considerably from a
drastic increase of the wetted surface $\Delta A_\mathrm{surf} =
\Remark{n2ainc}\,\%$.

\begin{figure}
  \begin{center}
    \includegraphics[width=1.0\textwidth]{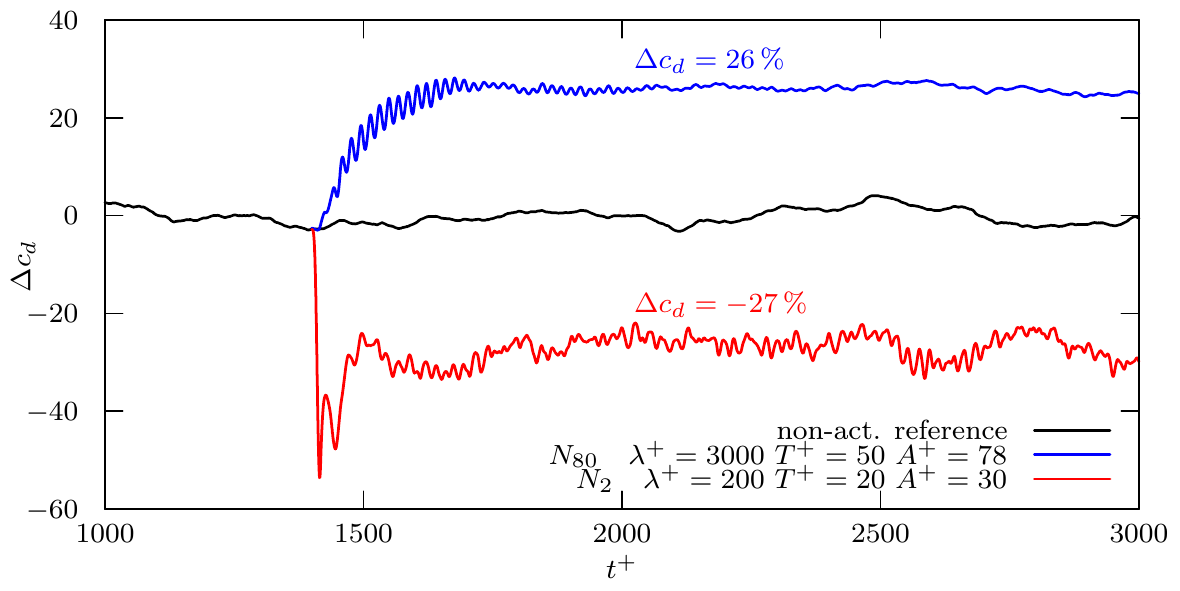}
    \caption{Temporal evolution of the instantaneous drag reduction
      $\Delta c_d$ for the non-actuated reference case, the actuated
      case with the highest drag reduction ($N_{80}$,
      Tab.~\ref{tab::simulations} in the appendix), and the actuated case with the
      highest drag increase ($N_2$, Tab.~\ref{tab::simulations} in the appendix).}
    \label{fig::cd}
  \end{center}
\end{figure}

The temporal distributions of the instantaneous drag of the ''best''
and ''worst'', i.e., highest and lowest drag reduction $N_{80}$
$\Delta c_d = \Remark{n80dr}\,\%$ and $N_2$ $\Delta c_d =
\Remark{n2dr}\,\%$, which is a massive drag increase, are compared
exemplarily in Fig.~\ref{fig::cd}, with the instantaneous drag of the
reference case. The temporal fluctuations of the drag appear stronger
for the ''worst'' case. Note that due to the larger wavelength the
drag of the ''best'' case is numerically integrated over a three times
larger spanwise extent. Due to the spanwise averaging, this leads to
the temporally smoother distribution of the $N_{80}$ ($\Delta c_d =
\Remark{n80dr}\,\%$) case compared to the $N_2$ ($\Delta c_d =
\Remark{n2dr}\,\%$) case.

% =============================================================
% SUBSECTION
% =============================================================
\subsection{Drag reduction modeling and sensitivity analysis}
\label{sec::ml}
\begin{figure}
  \begin{center}
    \includegraphics[width=0.8\textwidth]{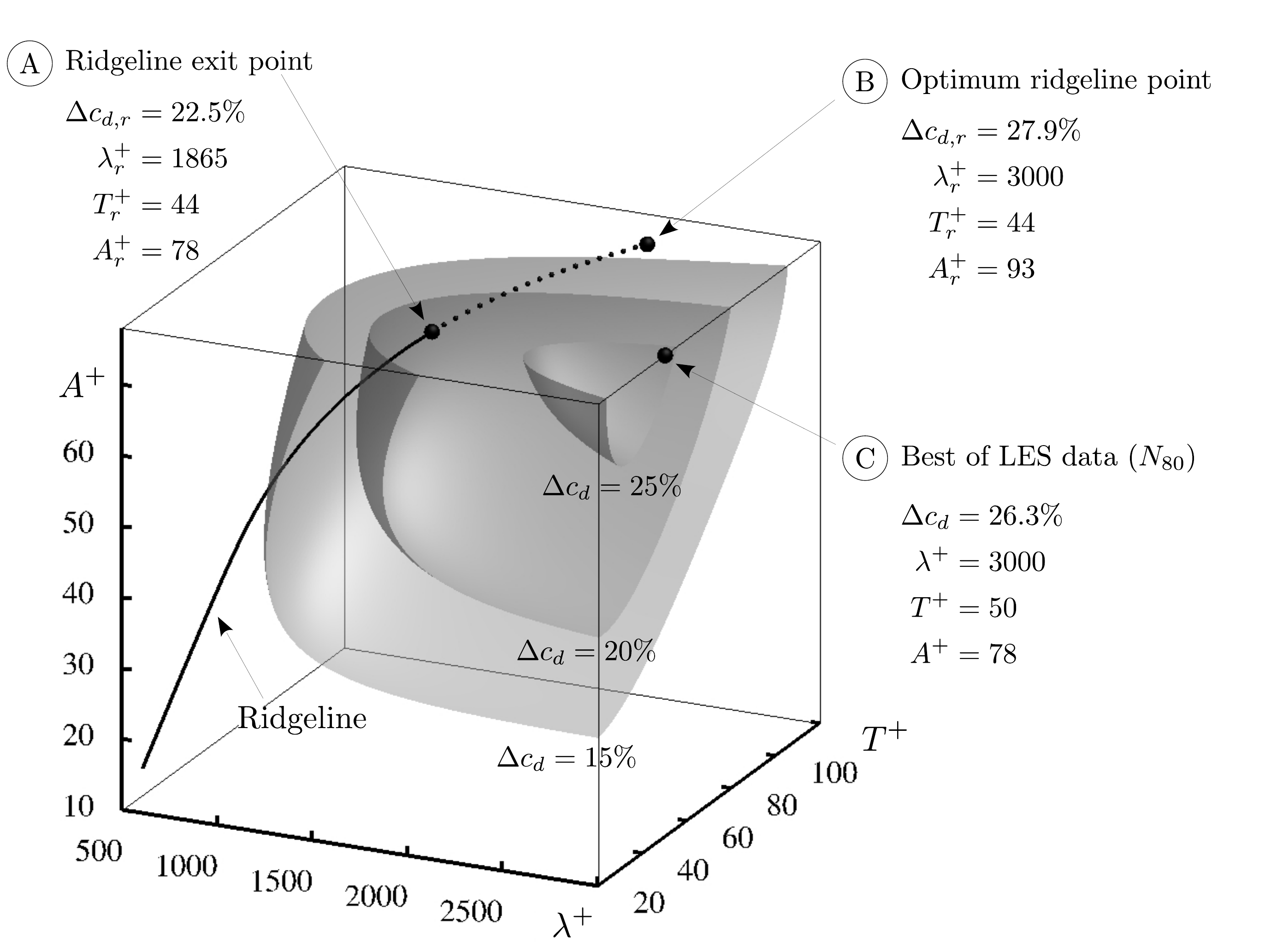}
    \caption{Actuation response surface.
      The gray surfaces represent drag reduction at three levels: $\Delta c_d = 15$, $20$, and $25\,\%$.
      The thick black line illustrates the ridgeline. 
      Its solid portion is interpolated using all LES data.
      The dotted ridgeline between points $A$ and $B$ extrapolates a better actuation response at given $\lambda^+$
      for amplitudes $A^+>78$, i.e.\ beyond the investigated parameter range. 
      Point $C$ corresponds to the actuation parameters of the LES simulation $N_{80}$ with the largest drag reduction.
    }
    \label{fig::IsoSurfaces}
  \end{center}
\end{figure}
In the following, 
the drag reduction $\Delta c_d$ is modeled 
as a function of the actuation parameters $\lambda^+$, $T^+$, and $A^+$.
For this task, 
LES simulations provide only a sparse data set comprising 80 points.
This amount would roughly correspond to a Cartesian discretization 
of a three-dimensional data space with only  $3 \times 3 \times 3$ points.
However, the modeling is further complicated by the fact 
that these points are far from regularly distributed.
A dense coverage of the actuation space using expensive LES simulations is hardly feasible. 

The modeling is performed using
a powerful regression solver from machine learning:
support vector regression (SVR) \cite{Cortes1995}. 
The algorithm is chosen for its prediction accuracy
and its smooth response distribution. 
SVR is a supervised learning algorithm 
that constructs a mapping between features or inputs and a known response.
Here, SVR maps the actuation parameters $\lambda^+$, $T^+$, and $A^+$ on the averaged drag reduction $\Delta c_d$. 
Note that due to the highly non-linear response behavior and the scarcity of data points at very low wavelengths, 
cases with wavelengths $\lambda^+ < 500$ are ignored during the modeling process.
This range is also of little interest 
as the best drag reduction is found at larger wavelengths.
This data exclusion effectively yields 71 data points instead of 80.
Overfitting is prevented with a 5-fold cross-validation.
The SVR model has a coefficient of determination of $R^2 = 0.93$ 
indicating an excellent prediction accuracy. 

The SVR model from the LES data 
is employed to visualize a continuous actuation response
in the investigated parameter range 
of $A^+ \in [0,78]$, $T^+ \in [20,120]$, and $\lambda^+ \in [500,3000]$.
Fig.~\ref{fig::IsoSurfaces} shows the isosurfaces 
of three drag reduction levels: $\Delta c_d=15$, $20$, and $25$\%.
  Within this parameter range, the best performance of 26.5 \% is achieved at 
  $\lambda^+=3000$, $T^+=38$ and $A^+=78$, which is slightly higher than the best
  simulated LES configuration $N_{80}$ with $\Delta c_d=26.3$ \%.
This location indicates that better
drag reduction could be achieved by increasing amplitude and wavelength.

%The best performance of 26.3 \% is achieved at point $C$
%on the intersection line of the maximum amplitude $A^+=78$ 
%and maximum wavelength $\lambda^+=3000$ plane.
%This location indicates that better drag reduction could be achieved
%by increasing amplitude and wavelength.

An extrapolation of better performance
outside the investigated parameter range
is obtained with a ridgeline.
In every $\lambda^+=\text{const}$ plane, 
the drag reduction $\Delta c_d$ features a single maximum $(A_r^+,T_r^+)$ 
with respect to the actuation amplitude $A^+$ and period $T^+$.
The curve of $(A^+,T^+, \lambda^+)$ 
connecting all these $\lambda^+$-dependent $\Delta c_d$ maxima is the \emph{ridgeline}, 
which is illustrated as a thick black curve in Fig.~\ref{fig::IsoSurfaces}.
Variables on this ridgeline are denoted by the subscript `r'.

In the range $\lambda^+ \in [560,1865]$,
this maximum is inside the  modeled $T^+$ and $A^+$ data range.
It is illustrated by the solid black curve.
However, the ridgeline leaves this modeled data range 
through the exit point $A$ at the top surface $A^+ =78$ near $\lambda^+ \approx 1865$.
The rigdeline is extrapolated outside the data range 
and depicted as dotted curve between points $A$ and $B$.
The extrapolation method is detailed in \cite{Fernex2019jfm}.
Along the ridgeline, 
$\Delta c_d$ monotonously 
increases from 7\% at $\lambda^+=560$ 
to the maximum of 27.9\% at $\lambda^+=3000$ (point $B$).
Note that this point is outside the current LES parameter range.

\begin{figure}
  \begin{center}
    \subfloat[~]{%
      \includegraphics[width=0.48\textwidth,valign=t]{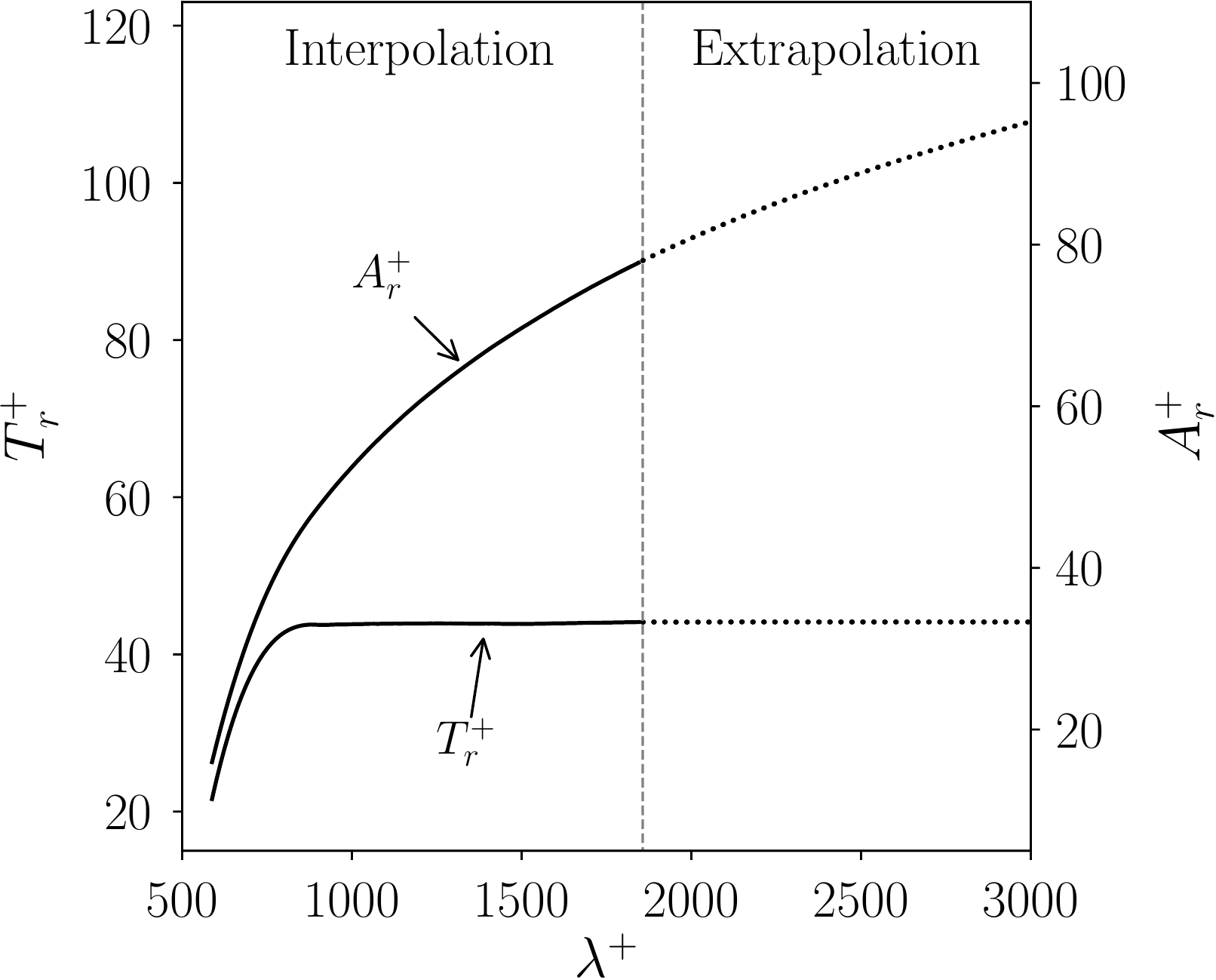}\label{fig::Ridgeline}%
      \vphantom{\includegraphics[width=0.44\textwidth,valign=t]{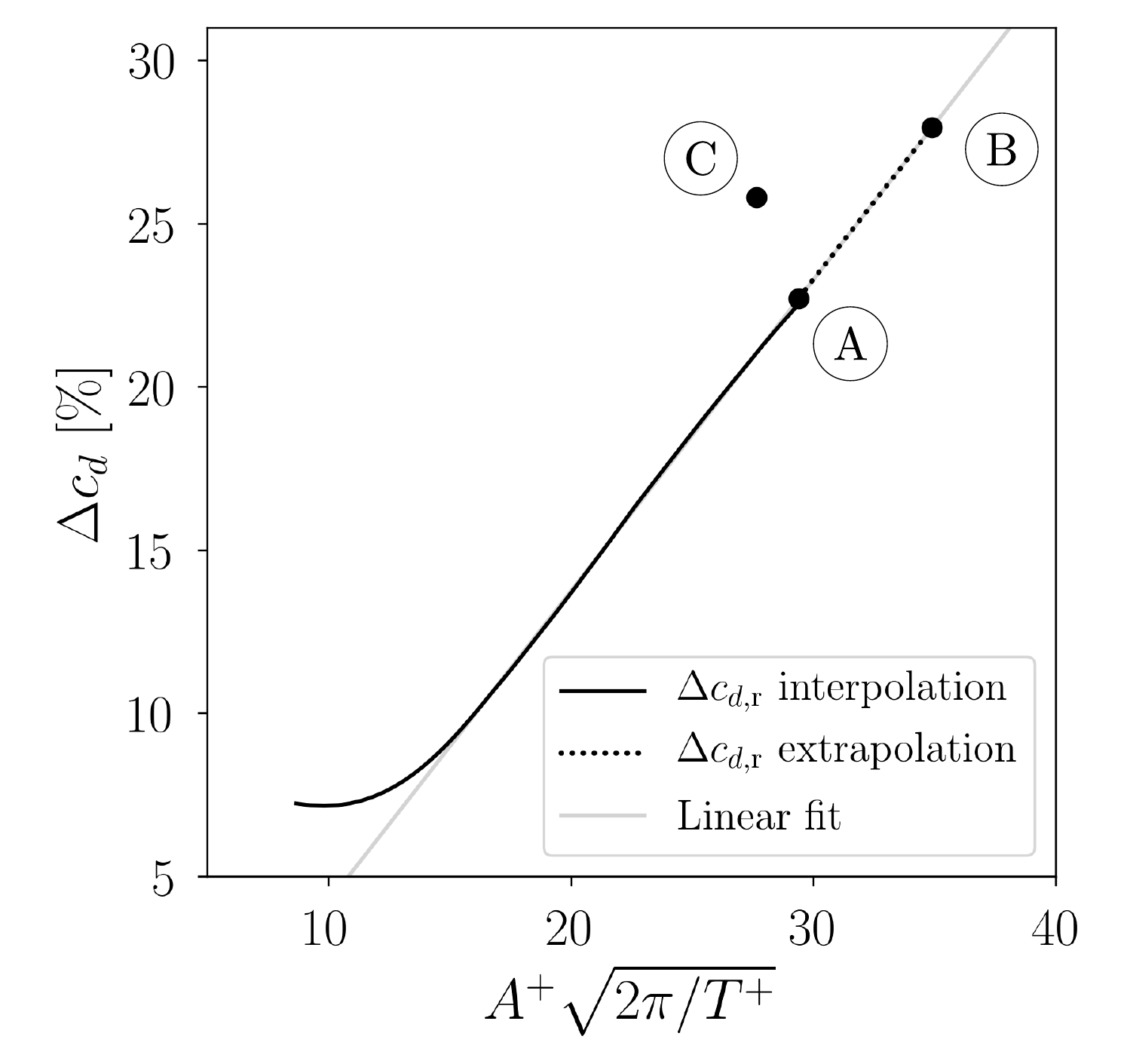}}%
    } \quad
    \subfloat[~]{%
      \includegraphics[width=0.44\textwidth,valign=t]{./figures/RidgeTomiyama.pdf}\label{fig::RidgeTomiyama}
    }
    \caption{(a) Projection of the ridgeline in the $\lambda^+$--$T^+$ and
      $\lambda^+$--$A^+$ planes.  (b) Drag reduction along the
      ridgeline as a function of the scaling proposed by Tomiyama and
      Fukagata \cite{Tomiyama2013}.  The solid circles marked $A$, $B$,
      and $C$ correspond to the likewise marked points in
      Fig.~\ref{fig::IsoSurfaces}.  }
    \label{fig::RidgelineAll}
  \end{center}
\end{figure}

The ridgeline defines a `skeleton' of the parametric behavior 
as illustrated in Fig.~\ref{fig::RidgelineAll}. 
Fig.~\ref{fig::Ridgeline} shows its projection 
in the $\lambda^+-T^+$ and $\lambda^+-A^+$ planes.
Like in Fig.~\ref{fig::IsoSurfaces}, 
the dotted sections correspond 
to the extrapolated ridgeline between points $A$ to $B$.
The amplitude $A_r^+$ and period $T_r^+$ 
along the ridgeline
monotonously increase with the wavelength $\lambda^+$.
The period asymptotes rapidly towards $44$.
The amplitude continually increases with the wavelength
although at a decreasing rate.

An intriguing physical insight about the drag-reduction mechanism
is revealed in Fig.~\ref{fig::RidgeTomiyama} complementing Fig.~\ref{fig::dr::large}.
The relative drag reduction $\Delta c_{d}$ along the ridgeline 
is shown as a function of the 
scaling parameter proposed by Tomiyama and Fukagata \cite{Tomiyama2013}
based on a Stokes layer of a transverse wall oscillation.
 $\Delta c_{d}$ clearly exhibits a linear behavior 
along the ridgeline in the scaling parameter range between 15 and 30. %$\lambda^+>1000$.
% with $\lambda^+ \in [560, 1865]$. 
%The extrapolated section between $A$ and $B$ slightly departs from the linear fit.
%along the interpolated ridgeline with $\lambda^+ \in [560, 1865]$. 
%The extrapolated section between $A$ and $B$ slightly departs from the linear fit.
Away from the ridgeline, 
the scaling shows scatter on the order of that observed in
Fig.~\subref*{fig::dr::large}. 

%This ridgeline model can be used to predict drag reduction outside the
%current actuation space defined by the actuation parameters. To better
%understand the physical mechanisms that cause drag reduction in the
%spanwise traveling transversal wave system a thorough analysis of the
%near-wall flow structures will be presented next.

% =============================================================
% 															SUBSECTION
% =============================================================
\subsection{Turbulent flow statistics}
\label{sec::flow_statistics}
In the following, the mean statistics of a drag reduced flow will be
investigated in detail. For this analysis, the case with the highest
drag reduction, i.e., the case $N_{80}$, will be considered. If data
from other cases is used, it is explicitly indicated. All presented
wall-normal distributions are considered at the streamwise position
$x = 90\, \theta$, which is located in the actuated region. The
actuated fully developed turbulent flow possesses a momentum based
Reynolds number $\mathrm{Re}_\theta = 1077$. The flow field of the
actuated cases is phase averaged in the spanwise direction.
Therefore, a triple decomposition \cite{Hussain1970} of the flow
variables is used
\begin{equation}
\label{eq:1}
\phi = \underbrace{\overline{\phi} + \tilde{\phi}}_{\langle \phi \rangle} + \phi'',
\end{equation}
where $\overline{\phi}$ is the temporal and spanwise average,
$\tilde{\phi}$ are periodic fluctuations, $\langle \phi \rangle =
\overline{\phi} + \tilde{\phi}$ are phase averaged quantities, and
$\phi''$ are stochastic fluctuations. Using this decomposition,
$\overline{\phi} = f(x,y)$ represents phase independent quantities,
$\tilde{\phi} = f(x,y,z)$ are the periodic fluctuations generated
through the actuation, i.e., the secondary flow field, and $\phi'' =
f(x,y,z,t)$ are turbulent fluctuations. Spanwise averages are obtained
along lines of constant distance from the wall, i.e., along the curved
mesh lines. This calculation of the spanwise average suffers from some
uncertainty for short wavelengths with high amplitudes, where the
traveling wave massively intrudes into the boundary layer.  For
spanwise averages of long wavelengths as in the $N_{80}$ case, where
the local perturbation of the viscous sublayer and the buffer layer is
less drastic, this problem does not occur.

\begin{figure}
  \begin{center}
    \subfloat[~]{\adjincludegraphics[width=0.4\textwidth,trim={0 {.05\height} {.6\width} 0},clip]{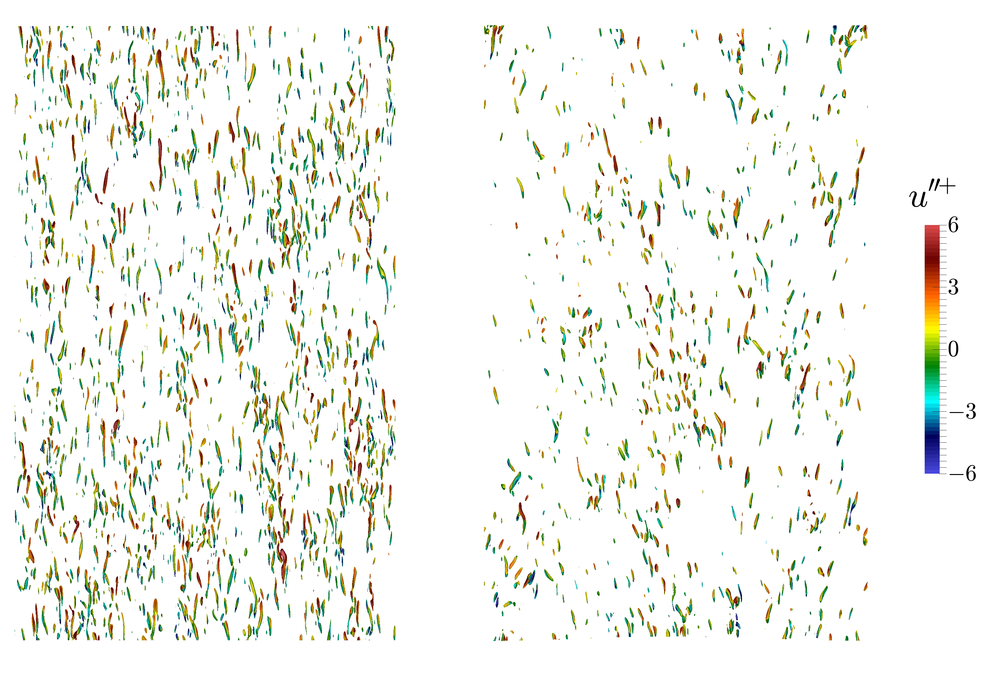}\label{fig::lambda2::compare::ref}}%
    \subfloat[~]{\adjincludegraphics[width=0.6\textwidth,trim={{.4\width} {.05\height} 0 0},clip]{./figures/lambda2_w3000_t50_a78_top_white_uplus_scale_small.png}\label{fig::lambda2::compare::act}}
    \caption{Contours of the $\lambda_2$-criterion \cite{Jeong1995}
      colored by the random velocity fluctuations $u''$ in the
      near-wall region $y^+ < 20$ of
      \protect\subref{fig::lambda2::compare::ref} the non-actuated
      reference case and \protect\subref{fig::lambda2::compare::act}
      the actuated case with the highest drag reduction $N_{80}$,
      i.e., $\lambda^+ = 3000$, $T^+ = 50$, and $A^+ = 78$.}
    \label{fig::lambda2::compare}
  \end{center}
\end{figure}

A first overall impression of the impact of the wave actuation on the
turbulent coherent structures is given in
Fig.~\ref{fig::lambda2::compare} by comparing contours of the
$\lambda_2$-criterion \cite{Jeong1995} for the random velocity
fluctuations $u''_i$. It is evident that the total number of vortical
structures in the near-wall region is significantly reduced for the
actuated flow. Extended regions of little to hardly any structures
occur in Fig.~\subref*{fig::lambda2::compare::act}. A closer look
evidences that unlike the structures of the non-actuated reference
flow in Fig.~\subref*{fig::lambda2::compare::ref}, the structures of
the actuated flow are inclined to the left and right depending on the
phase angle of the traveling wave. It will be discussed in
Sec.~\ref{sec::spanwiseshear} that this wave determined orientation of
the flow structures is an important feature related to drag
reduction \cite{Touber2012}.

\begin{figure}
  \begin{center}
    \subfloat[~]{\adjincludegraphics[width=0.5\textwidth,trim={0 0 {.5\width} 0},clip]{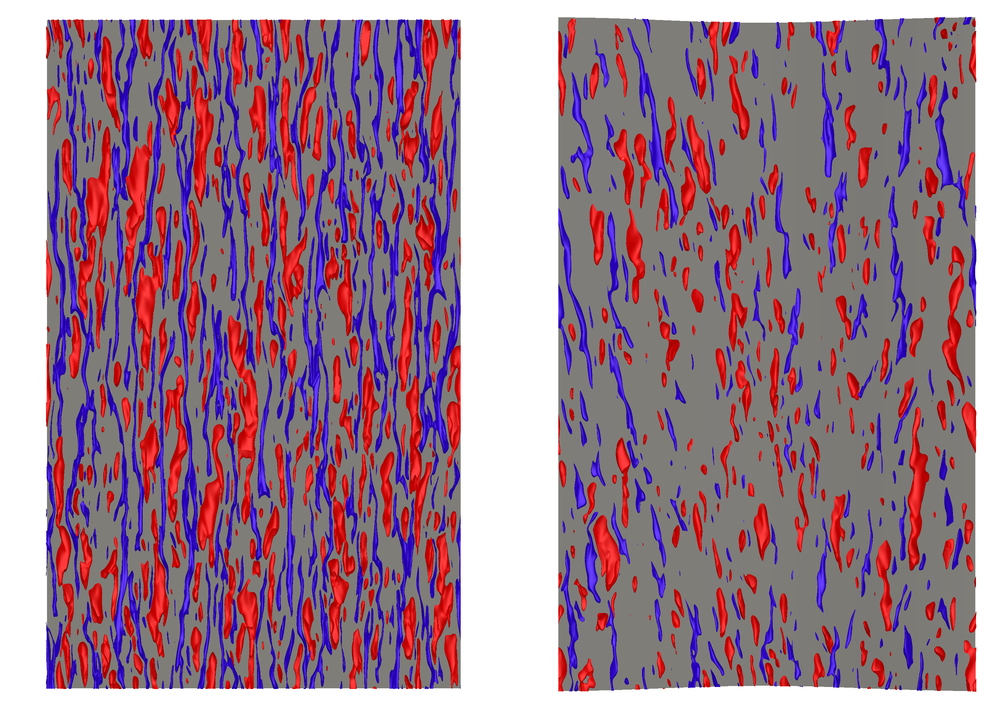}\label{fig::uprimeprime::ref}}%
    \subfloat[~]{\adjincludegraphics[width=0.5\textwidth,trim={{.5\width} 0 0 0},clip]{./figures/uprimeprime_w3000_lowspeed_highspeed_uplus_small.png}\label{fig::uprimeprime::act}}
    \caption{Contours of the random streamwise velocity fluctuations
      for $u''^+ = -3$ (blue) and $u''^+ = 3$ (red) in the near-wall
      region $0 < y^+ < 20$ of \protect\subref{fig::uprimeprime::ref} the
      non-actuated reference case and
      \protect\subref{fig::uprimeprime::act} the actuated case with
      the highest drag reduction $N_{80}$, i.e., $\lambda^+ = 3000$,
      $T^+ = 50$, and $A^+ = 78$.}
    \label{fig::uprimeprime}
  \end{center}
\end{figure}

To highlight the influence of the actuation on the instantaneous flow
field, Fig.~\ref{fig::uprimeprime} shows contours of the instantaneous
random fluctuations of the velocity component in the streamwise
direction $u''$ in a region $0 < y^+ < 20$ above the wall. For the
non-actuated flow, regions of localized high-speed and low-speed
fluid, i.e., streaks, are illustrated whose length and width are on
the order of $\mathcal{O}(10^3)$ and $\mathcal{O}(10^2)$ in inner
units. The actuated flow field shows much less pronounced regions of
high- and low-speed fluid. That is, the distinctive structure of thin
meandering streaks is considerably alleviated compared to the
non-actuated reference case.

\begin{figure}
  \begin{center}
     \subfloat[~]{\includegraphics[width=0.5\textwidth]{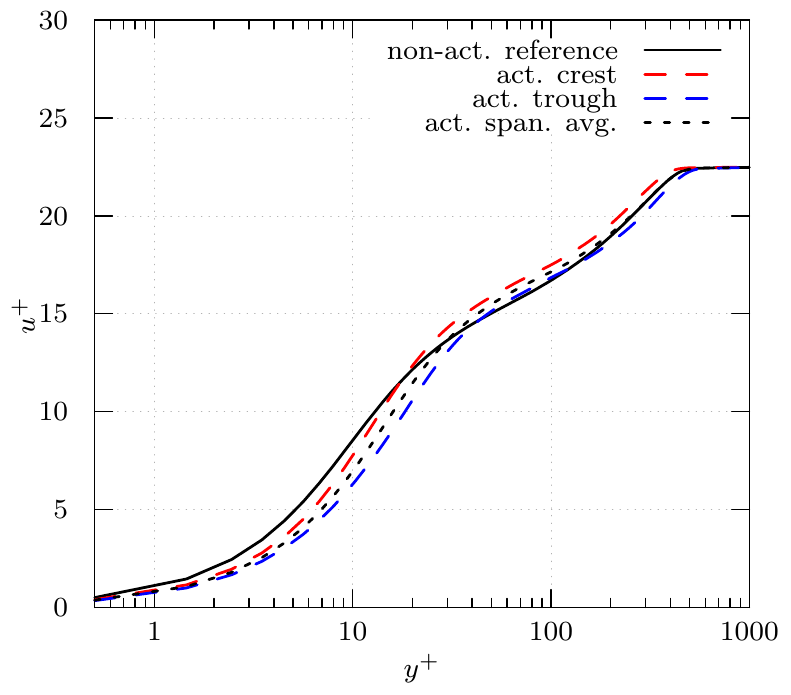}\label{fig::vel::ref}}~
     \subfloat[~]{\includegraphics[width=0.5\textwidth]{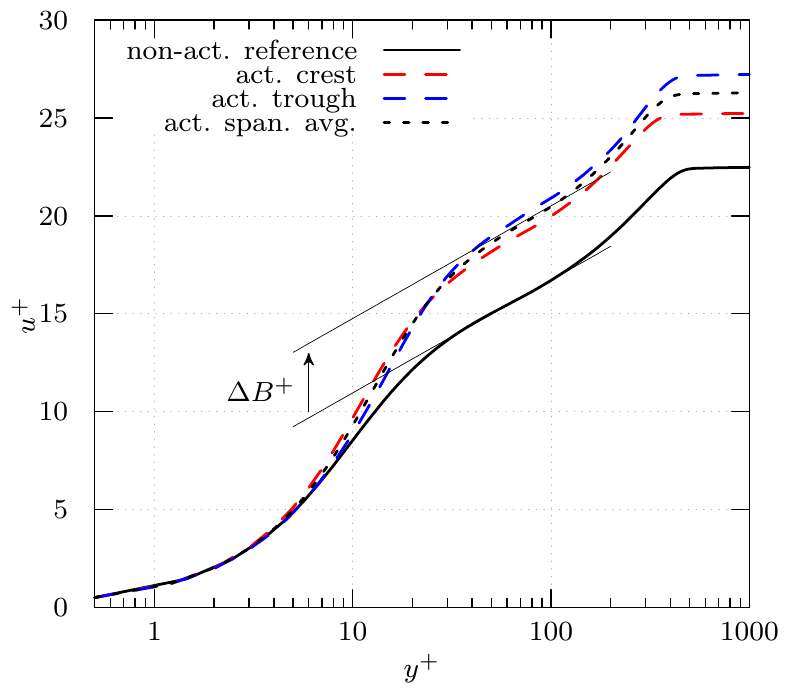}\label{fig::vel::nonref}}
     \caption{Wall-normal distributions of the phase averaged
       streamwise velocity $\langle u \rangle$ above the crest and in
       the trough and the spanwise averaged mean velocity
       $\overline{u}$ for the non-actuated reference case and the
       actuated case $N_{80}$, in \protect\subref{fig::vel::ref} the
       $N_{80}$ distributions are non-dimensionalized by the friction
       velocity of the non-actuated reference case and in
       \protect\subref{fig::vel::nonref} the $N_{80}$ distributions
       are non-dimensionalized by the friction velocity of the
       actuated case $N_{80}$.}
    \label{fig::vel}
  \end{center}
\end{figure}

The wall-normal distributions of the phase averaged streamwise
velocity $\langle u \rangle$ above the wave crest and in the wave
trough and the mean velocity $\overline{u}$ are shown in
Fig.~\ref{fig::vel}. The scaling with the friction velocity of the
non-actuated reference case $u_{\tau,\mathrm{ref}}$ in
Fig.~\subref*{fig::vel::ref} illustrates the decrease of the velocity
in the near-wall region. The wall-normal gradient at the wall is
lowered, which results in drag reduction. Scaling the velocities with
the friction velocity $u_\tau$ of the actuated wall in
Fig.~\subref*{fig::vel::nonref} leads to an offset of the velocity
profiles in the logarithmic region with respect to the non-actuated
reference case by $\Delta B^+ \approx 3.8$. Based on the idea of the
analysis of the impact of roughness on fully turbulent flow
\cite{Nikuradse1933, Clauser1956}, Gatti~and~Quadrio~\cite{Gatti2016}
suggested the offset $\Delta B^+$ to predict drag reduction at higher
Reynolds numbers
  \begin{equation} \label{Eq:BPlus}
    \Delta B^+ = \sqrt{\frac{2}{c_{f,0}}} \left[ \left(
1-\Delta c_f \right)^{-1/2} - 1\right] -
\frac{1}{\kappa}\left(\frac{\mathrm{Re}_\tau}{\mathrm{Re}_{\tau,0}}\right)\, .
  \end{equation}
  Note, however, that this equation cannot be further simplified since
  for actuated turbulent boundary layer flow the term
  $\frac{\mathrm{Re}_\tau}{\mathrm{Re}_{\tau,0}}$ is neither constant,
  as for constant pressure gradient turbulent channel flow, nor can it
  be substituted by the drag reduction rate, as for constant flow rate
  turbulent channel flow. Thus, $\Delta c_f$ cannot be directly
  determined by equation~\eqref{Eq:BPlus}. Nevertheless, using the
  local values of $c_{f,0}$, $\Delta c_f$, $\mathrm{Re}_\tau$, and
  $\mathrm{Re}_{\tau,0}$ at $x = 90\theta$ the calculated offset from
  equation~\eqref{Eq:BPlus} is $\Delta B^+ = 4.07$, which reasonably
  agrees with the result $\Delta B^+ = 3.8$ shown in
  Fig.~\subref*{fig::vel::nonref}.  The velocity profiles in
  Fig.~\ref{fig::vel} show that for the current actuation neither the
  non-actuated nor the actuated friction velocity scaling
  ---regardless from crest, trough, or spanwise averaged wall shear
  scaling--- result in a collapsed distribution over the entire
  boundary layer.  In other words, an inner scaling does not hold over
  the entire boundary layer.

\begin{figure}
  \begin{center}
     \subfloat[stresses]{\includegraphics[width=0.5\textwidth]{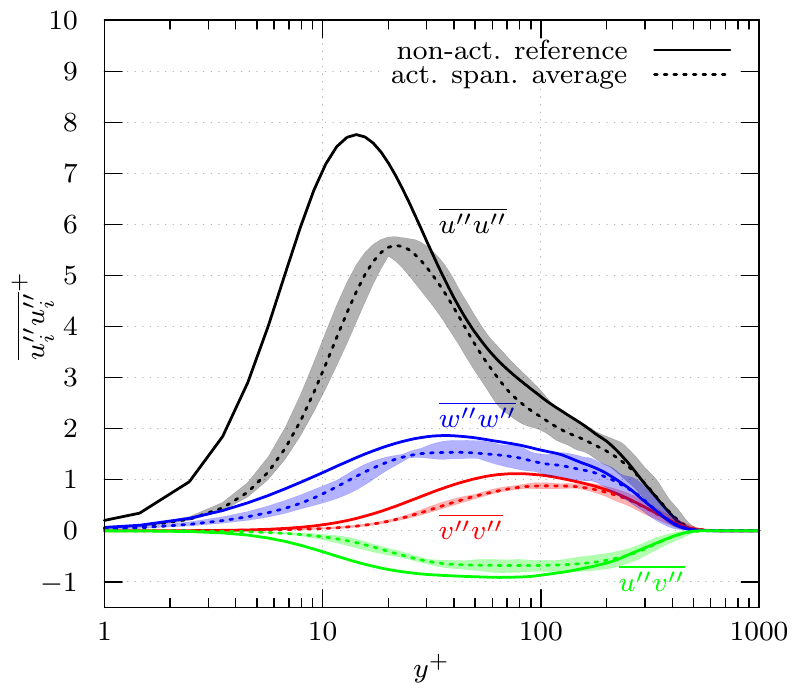}\label{fig::rst::rst}}
     \subfloat[anisotropy map]{\includegraphics[width=0.5\textwidth]{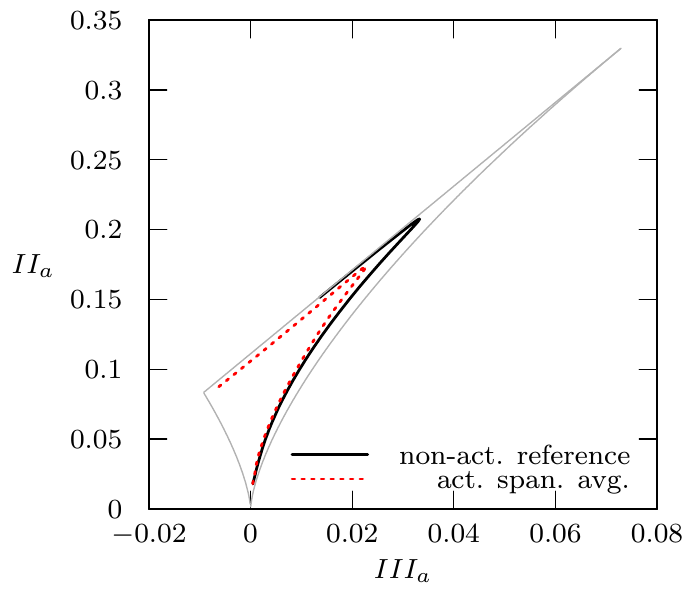}\label{fig::rst::lumley}}
     \caption{Wall-normal distributions of the stochastic components
       of the Reynolds stress tensor scaled by the non-actuated reference friction velocity; \protect\subref{fig::rst::rst}
       symmetric components $\overline{u''_iu''_i}$ and Reynolds shear
       stress $\overline{u''v''}$ for the non-actuated reference case and the actuated case $N_{80}$.
       Spanwise averaged values are shown as lines and the shaded
       regions illustrate phase variations of the depicted
       quantity; \protect\subref{fig::rst::lumley} Lumley anisotropy
       map of the Reynolds stress tensor.}
    \label{fig::rst}
  \end{center}
\end{figure}

Next, the components of the Reynolds stress tensor are depicted in
Fig.~\ref{fig::rst}. Through the actuation, the symmetric Reynolds
stresses $\overline{u''_iu''_i}$ and the Reynolds shear stress
$\overline{u''v''}$ shown in Fig.~\subref*{fig::rst::rst} are
significantly lowered with only minor phase variations. Considering
all cases for $\lambda^+ > 1000$, a good correlation of the decrease
of the skin-friction with the decrease of the peak of the streamwise
velocity fluctuations is computed ($R = 0.90$). For the case $N_{80}$,
the reductions at $y^+ = 14.2$, which defines the location of the peak
of the streamwise fluctuations and the location of the maximum
streamwise velocity streak intensity, are approx. $39\,\%$ for the
streamwise component and $62\,\%$ for the shear-stress component. This
suggests that the turbulent motion close to the wall is massively
damped. Touber and Leschziner~\cite{Touber2012} have reported
similarly large reductions in this region for spanwise wall
oscillations without normal deflection. They emphasize the importance
of the reduced near-wall Reynolds shear stress and drag, as
characterized by the Fukagata, Iwamoto, Kasagi (FIK) identity
\cite{Fukagata2002}, i.e., for the shear-stress contribution
$c_{f,\mathrm{RSS}} \sim \int_0^{\delta_{99}}
(1-y)(-\overline{u''v''})dy$. The structural property of the turbulent
motion is evidenced by the anisotropy invariant map \cite{Lumley1977}
in Fig.~\subref*{fig::rst::lumley}. The stronger suppression of the
streamwise fluctuations compared to the other components is
illustrated by the shift of the actuated distribution away from
one-dimensional turbulence in the upper right vertex to isotropic
turbulence in the lower vertex.

\begin{figure}
  \begin{center}
    \includegraphics[width=0.5\textwidth]{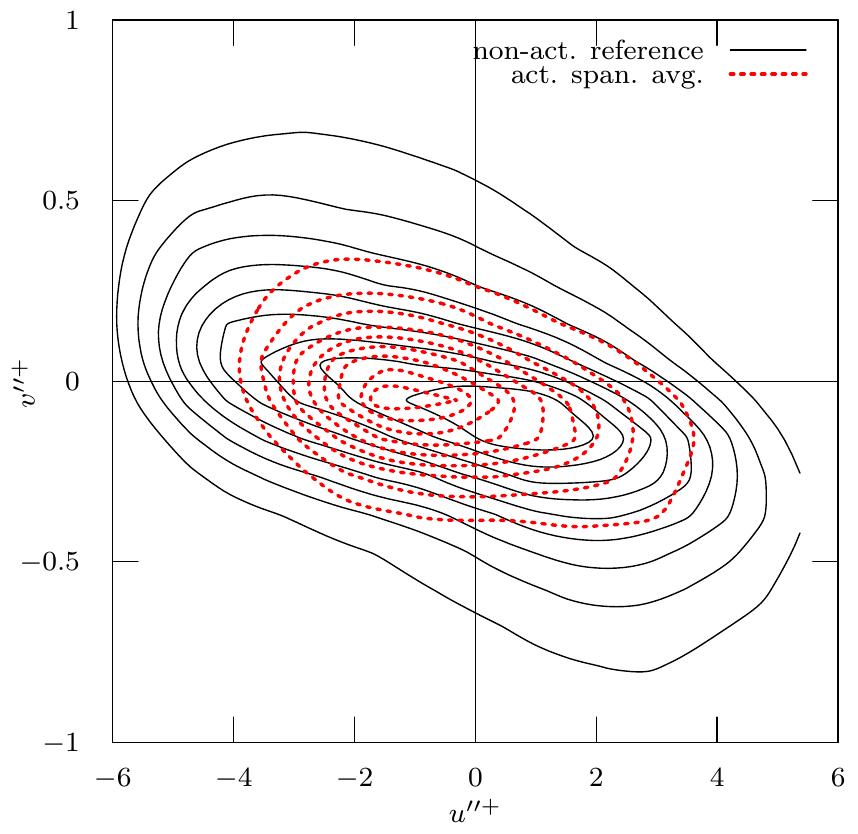}
    \caption{Joint PDF of $u''$ and $v''$ at $y^+ = 12$ for the
      non-actuated reference case and the actuated case
      $N_{80}$.}
    \label{fig::jointpdf}
  \end{center}
\end{figure}

The distributions of the joint probability density function (PDF) of
the streamwise and the wall-normal stochastic velocity fluctuations
$u''$ and $v''$ are presented in Fig.~\ref{fig::jointpdf}. High values
in the upper left quadrant (negative $u''$ and positive $v''$) denote
ejections of fluid from the near-wall region towards the outer flow,
whereas high values in the lower right quadrant (positive $u''$ and
negative $v''$) indicate sweeps of high-speed fluid from the outer
flow towards the near-wall region. As can be seen in
Fig.~\ref{fig::jointpdf}, an overall attenuation of the fluctuations
is observed with a strong damping of the sweeps and ejections in the
second and the fourth quadrant. Again, this is in agreement with the
results from spanwise oscillating walls without normal deflection
\cite{Agostini2014,Touber2012}.

\begin{figure}
  \begin{center}
    \subfloat[]{\includegraphics[width=0.5\textwidth]{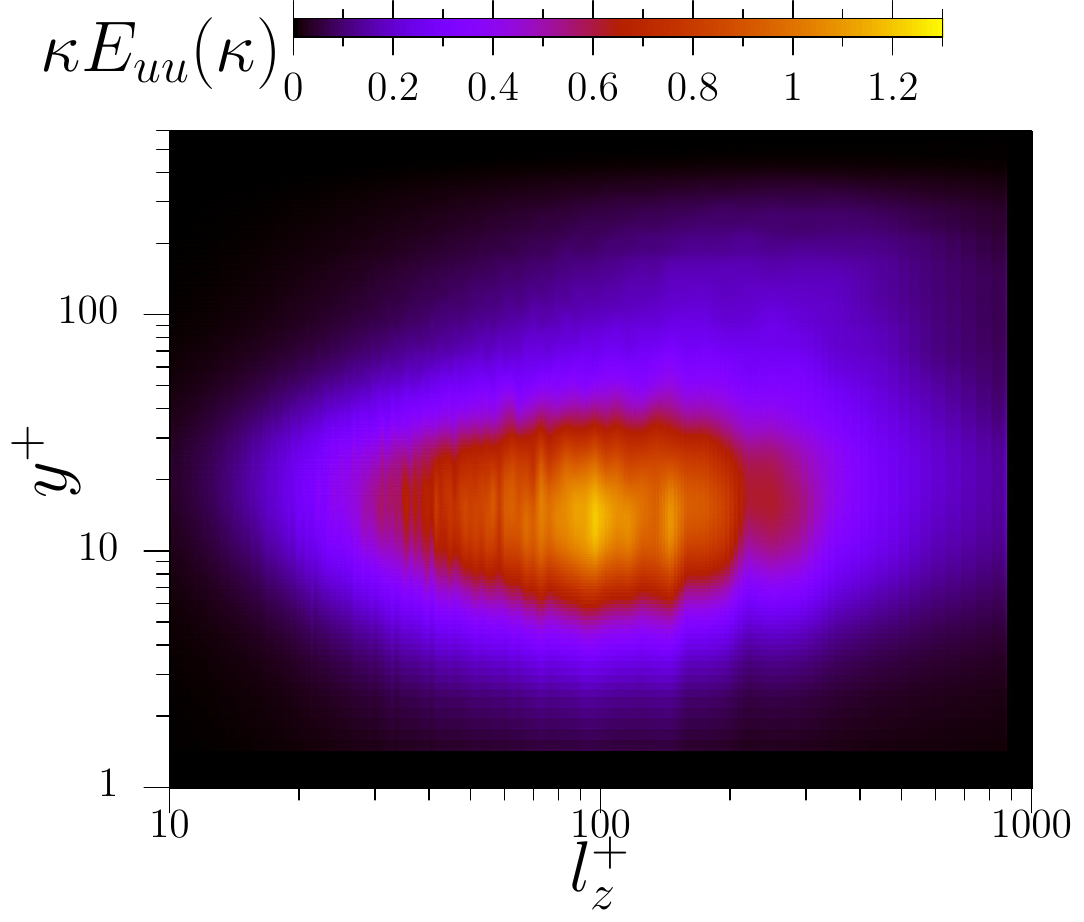}\label{fig::spectra::u::ref}}~
    \subfloat[]{\includegraphics[width=0.5\textwidth]{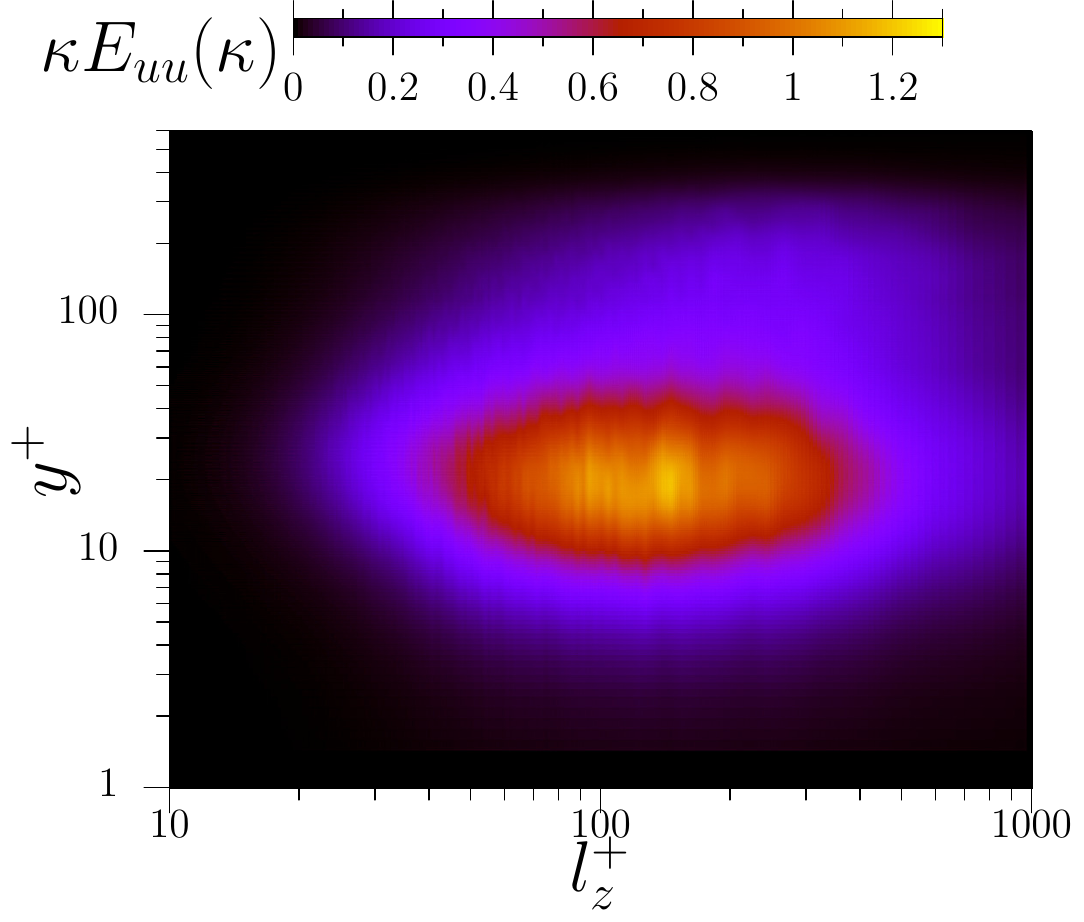}\label{fig::spectra::u::act}}

    \subfloat[]{\includegraphics[width=0.5\textwidth]{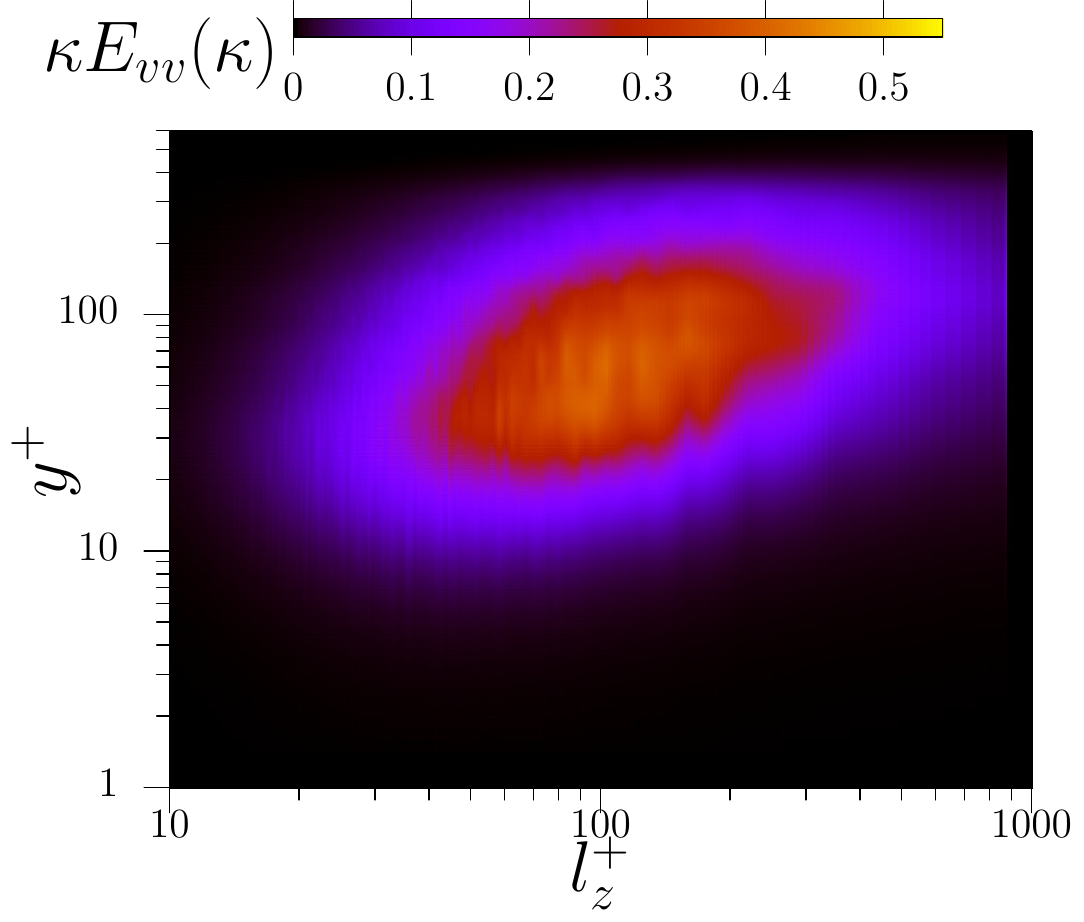}\label{fig::spectra::v::ref}}~
    \subfloat[]{\includegraphics[width=0.5\textwidth]{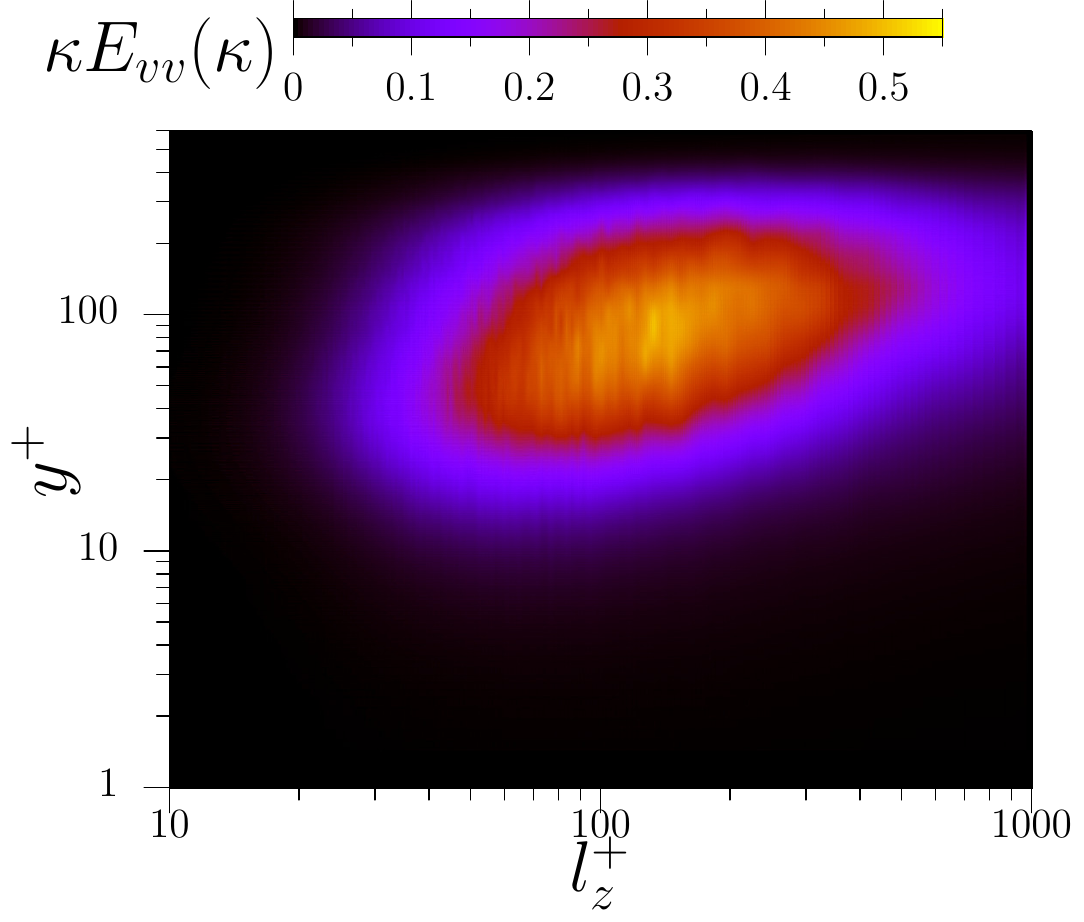}\label{fig::spectra::v::act}}

    \subfloat[]{\includegraphics[width=0.5\textwidth]{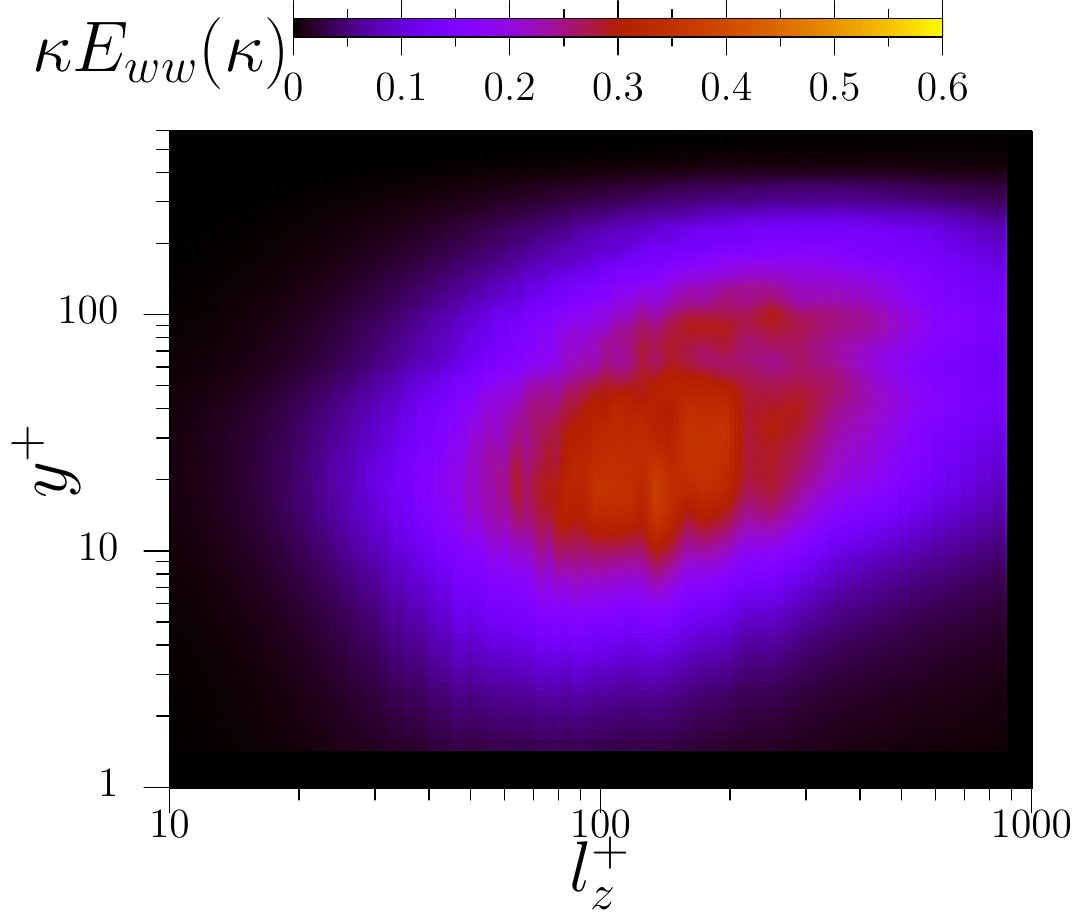}\label{fig::spectra::w::ref}}~
    \subfloat[]{\includegraphics[width=0.5\textwidth]{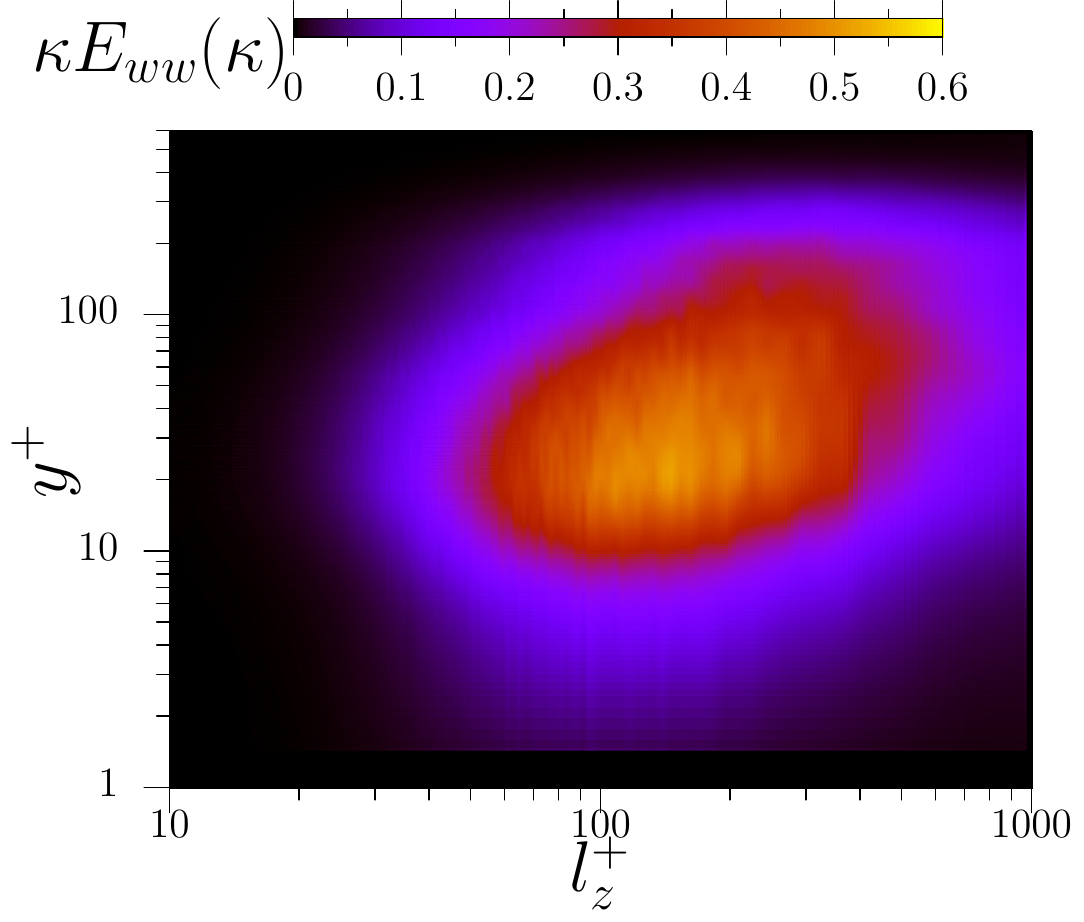}\label{fig::spectra::w::act}}
  \end{center}
  \caption{Spanwise premultiplied energy spectra of the velocity
    fluctuations $u_i''u_i''$, normalized by the total resolved
    energy of the related component and the related case for (left)
    the non-actuated reference case and (right) the actuated
    case with the highest drag reduction $N_{80}$;
    \protect\subref{fig::spectra::u::ref},\protect\subref{fig::spectra::u::act}
    streamwise,
    \protect\subref{fig::spectra::v::ref},\protect\subref{fig::spectra::v::act}
    wall-normal, and
    \protect\subref{fig::spectra::w::ref},\protect\subref{fig::spectra::w::act}
    spanwise velocity component.}
  \label{fig::spectra}
\end{figure}

Spanwise premultiplied energy spectra of the velocity fluctuations
$\kappa E_{u_i''u_i''}$, where $\kappa = 2\pi / l_z$ is the
wavenumber, are presented in Fig.~\ref{fig::spectra}. Each spectrum is
normalized by the total resolved energy of the corresponding velocity
component and the related case, i.e., the non-actuated reference case
and $N_{80}$. No general decrease in the energy peak of the actuated
case is observed, only a shift in the energy distribution as a
function of the structural wavelength and wall-normal coordinate can
be seen.  A comparison between the two differently normalized spectra
for the streamwise component (cf. Fig.~\subref*{fig::spectra::u::ref}
and Fig.~\subref*{fig::spectra::u::act}) shows an energy decrease
especially for the small scales and in the near-wall region. In other
words, for the actuated case $N_{80}$ the energy is accumulated
further off the wall in the larger scale turbulent structures. The
peak of the non-actuated reference case at $\lambda^+ \approx 100$,
which is associated with the typical spacing of the near-wall streaks
of $l_z^+ = \mathcal{O}(100)$, becomes less pronounced for the
actuated case $N_{80}$ and is shifted off the wall. This observation
corroborates the visual impression from Fig.~\ref{fig::uprimeprime} of
a strong reduction of the near-wall streaks for the actuated
case. Similar tendencies are observed for the wall-normal
(cf. Fig.~\subref*{fig::spectra::v::ref} and
Fig.~\subref*{fig::spectra::v::act}) and the spanwise velocity
component (cf. Fig.~\subref*{fig::spectra::w::ref} and
Fig.~\subref*{fig::spectra::w::act}). Additionally, a stronger
concentration of the energy in the length scale range of the near-wall
streaks is observed for the spanwise velocity component. There is a
sharper peak for the actuated case in comparison to a broader energy
distribution in the non-actuated case.

\begin{figure}
  \begin{center}
    \subfloat[streamwise]{\includegraphics[width=0.3\textwidth]{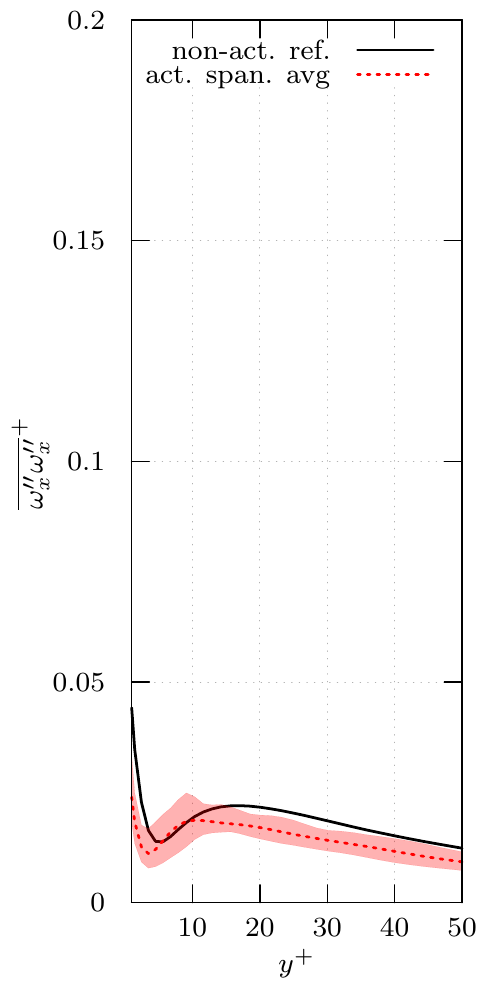}\label{fig::vorticity::x}}~
    \subfloat[wall-normal]{\includegraphics[width=0.3\textwidth]{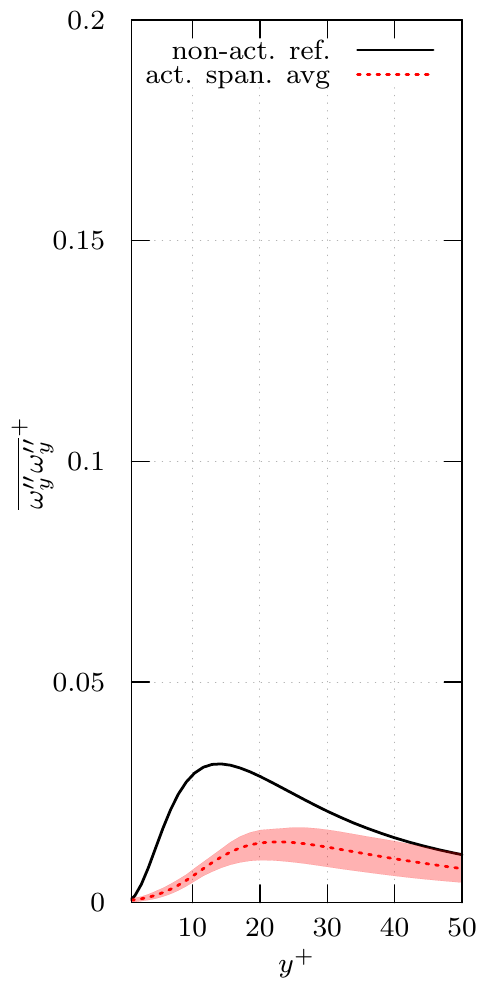}\label{fig::vorticity::y}}~
    \subfloat[spanwise]{\includegraphics[width=0.3\textwidth]{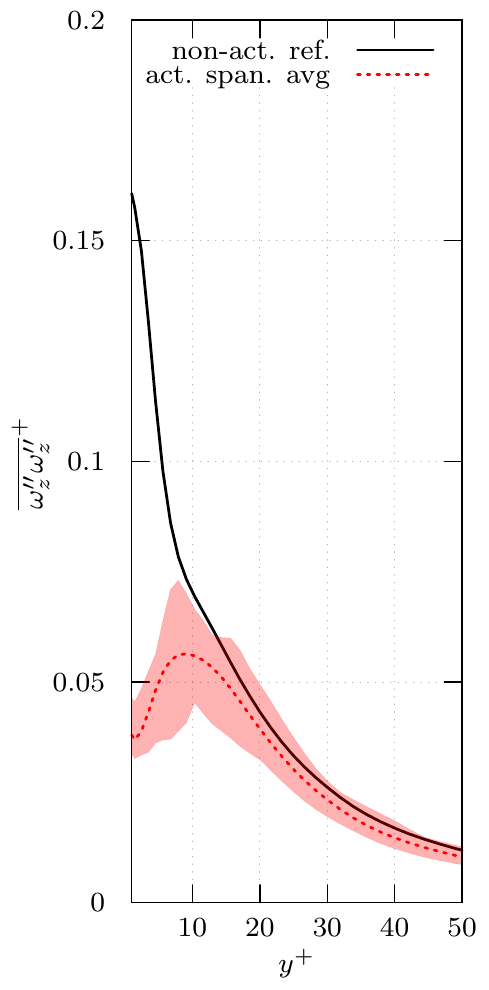}\label{fig::vorticity::z}}~
    \caption{Wall-normal distributions of the phase and spanwise
      averaged vorticity fluctuations
      $\overline{\omega''_i\omega''_i}$ for the non-actuated reference
      case and the actuated case $N_{80}$;
      \protect\subref{fig::vorticity::x} streamwise,
      \protect\subref{fig::vorticity::y} wall-normal, and
      \protect\subref{fig::vorticity::z} spanwise component.}
    \label{num::vorticity}
  \end{center}
\end{figure}

In Fig.~\ref{num::vorticity} the phase averaged and spanwise averaged
distributions of the vorticity fluctuations $\omega_i'' \omega_i''$
are presented as a function of the wall-normal distance. The
comparison of the profiles of each component shows that the major
attenuation is observed in the wall-normal and the spanwise
components, whereas the streamwise component shows only minor
changes. Generally, for all cases with $\lambda^+ > 1000$ a good
correlation between the decrease of the skin-friction and the decrease
of the peak of the wall-normal ($R = 0.96$) and spanwise ($R = 0.98$)
vorticity fluctuations was found. Again, similar vorticity trends were
reported for spanwise oscillating walls \cite{Touber2012}. The drag
reduction was discussed to be caused by the weakening of velocity
streaks near $y^+ \approx 10$ \cite{Agostini2014}. That is, at least
for actuation with large wavelengths $\lambda^+ > 1000$, a direct
interference with quasi-streamwise vortices \cite{Tomiyama2013} has a
minor effect on drag reductions.

\begin{figure}
  \begin{center}
   \subfloat[~]{\includegraphics[width=0.33\textwidth]{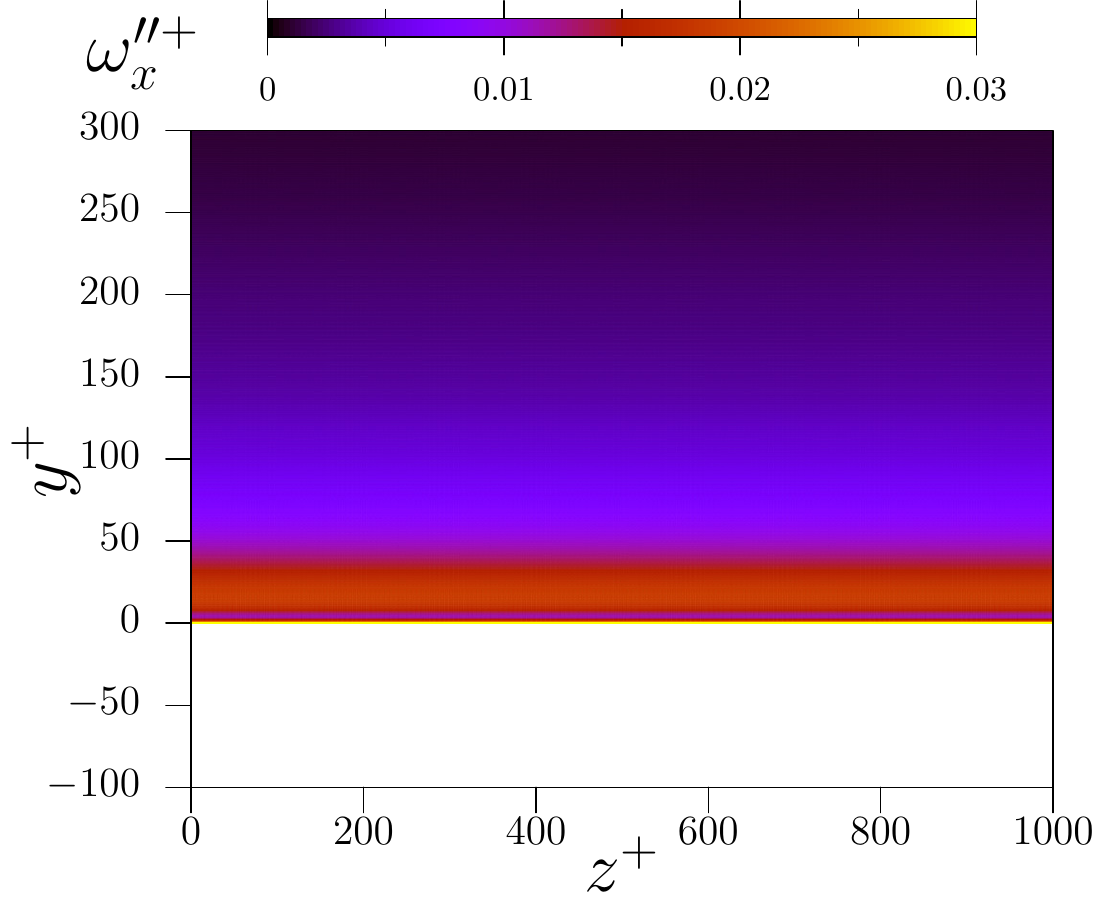}\label{fig::comp_om::1::omx}}
   \subfloat[~]{\includegraphics[width=0.33\textwidth]{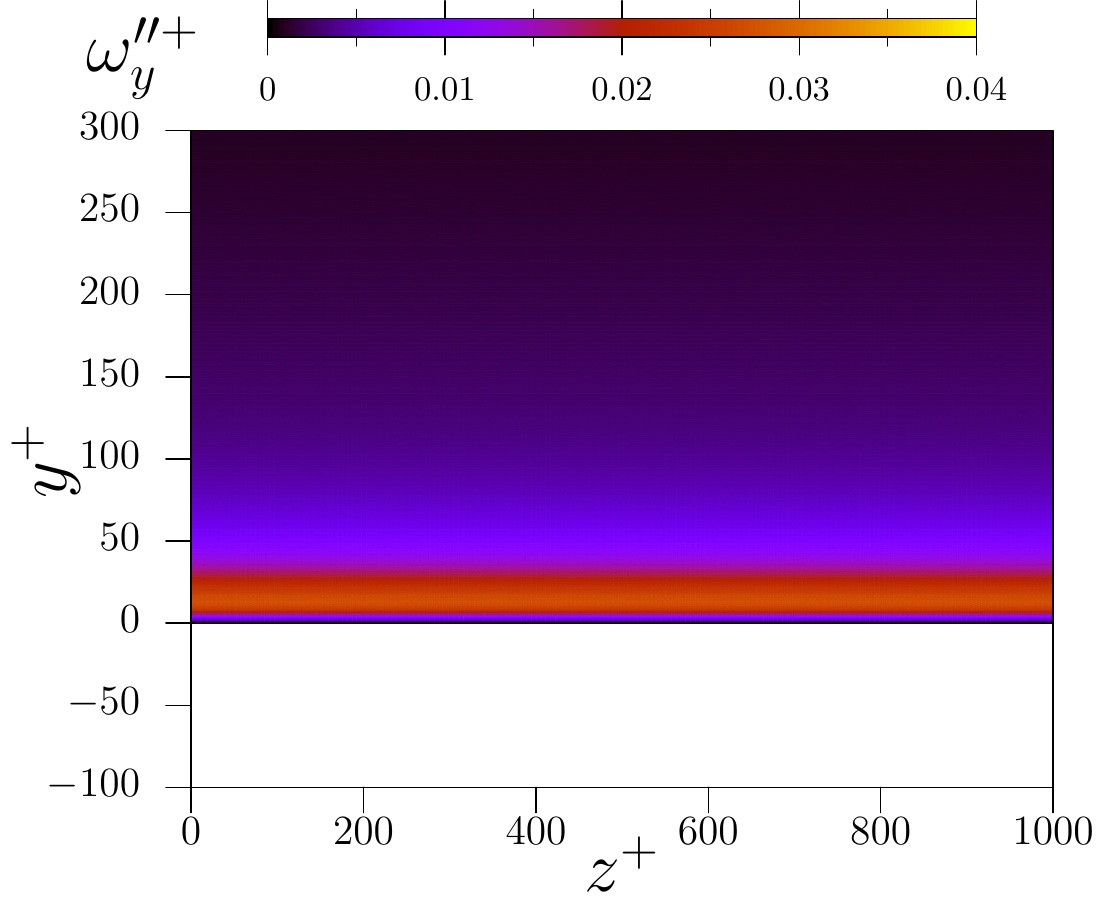}\label{fig::comp_om::1::omy}}
   \subfloat[~]{\includegraphics[width=0.33\textwidth]{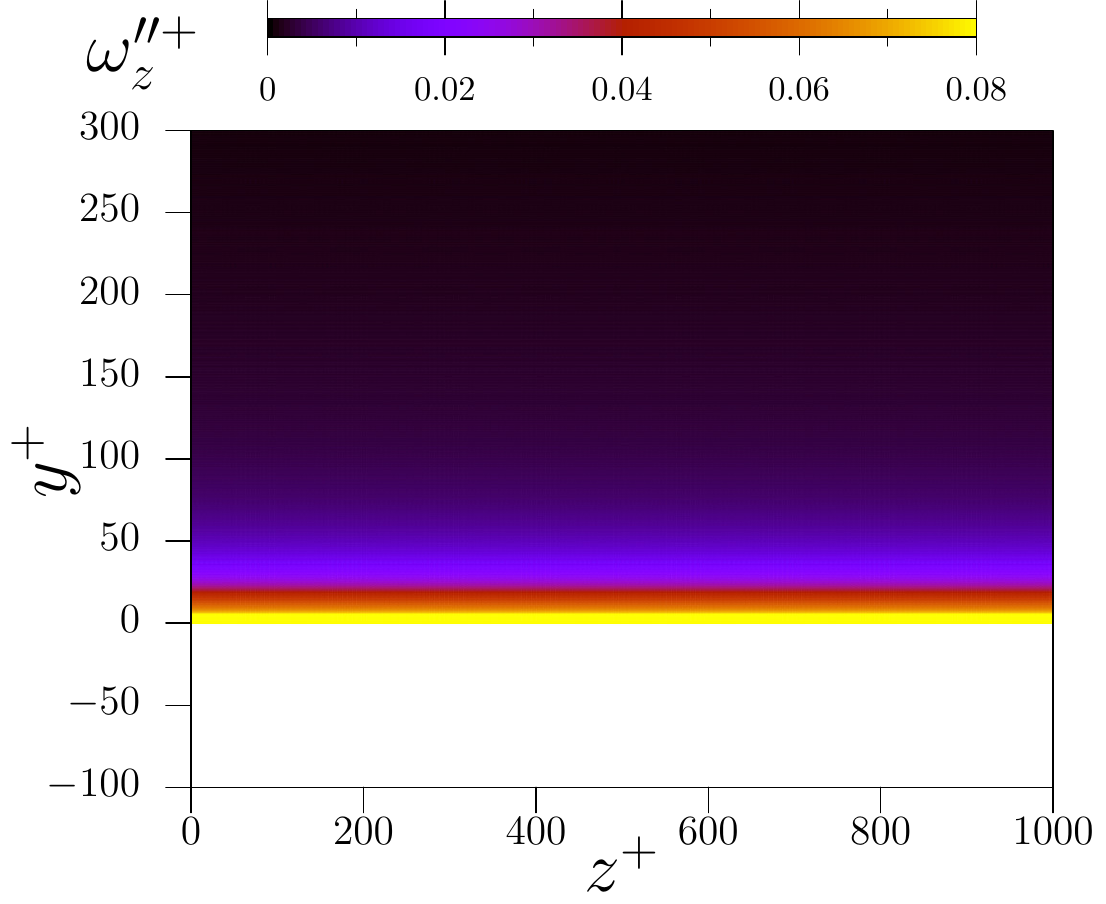}\label{fig::comp_om::1::omz}}

   \subfloat[~]{\includegraphics[width=0.33\textwidth]{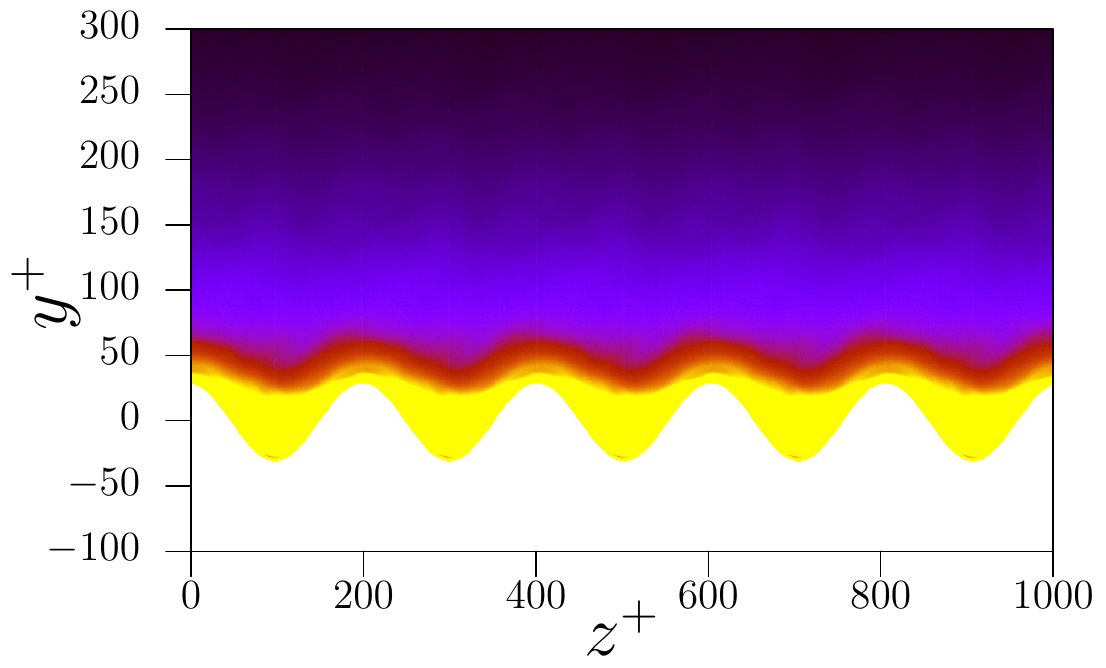}\label{fig::comp_om::2::omx}}
   \subfloat[~]{\includegraphics[width=0.33\textwidth]{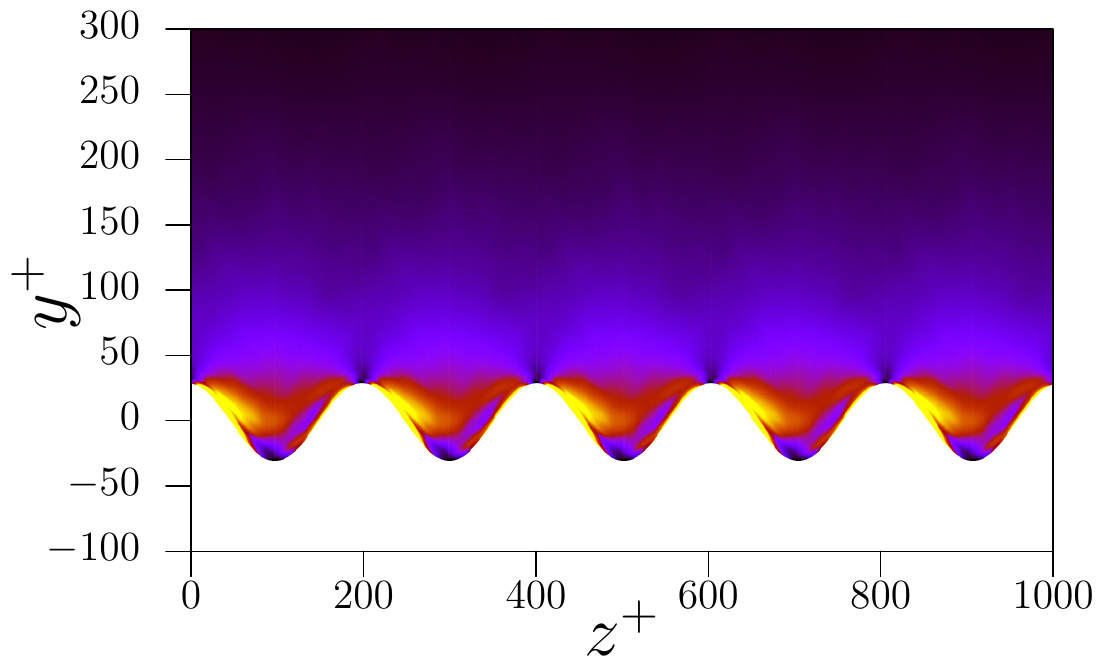}\label{fig::comp_om::2::omy}}
   \subfloat[~]{\includegraphics[width=0.33\textwidth]{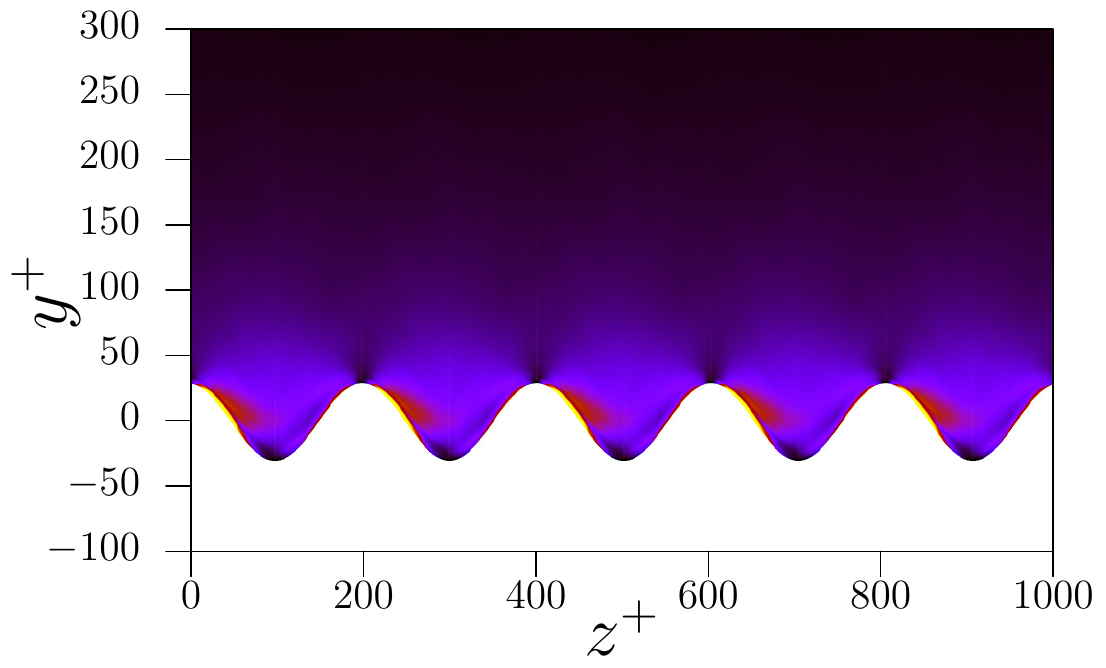}\label{fig::comp_om::2::omz}}

   \subfloat[~]{\includegraphics[width=0.33\textwidth]{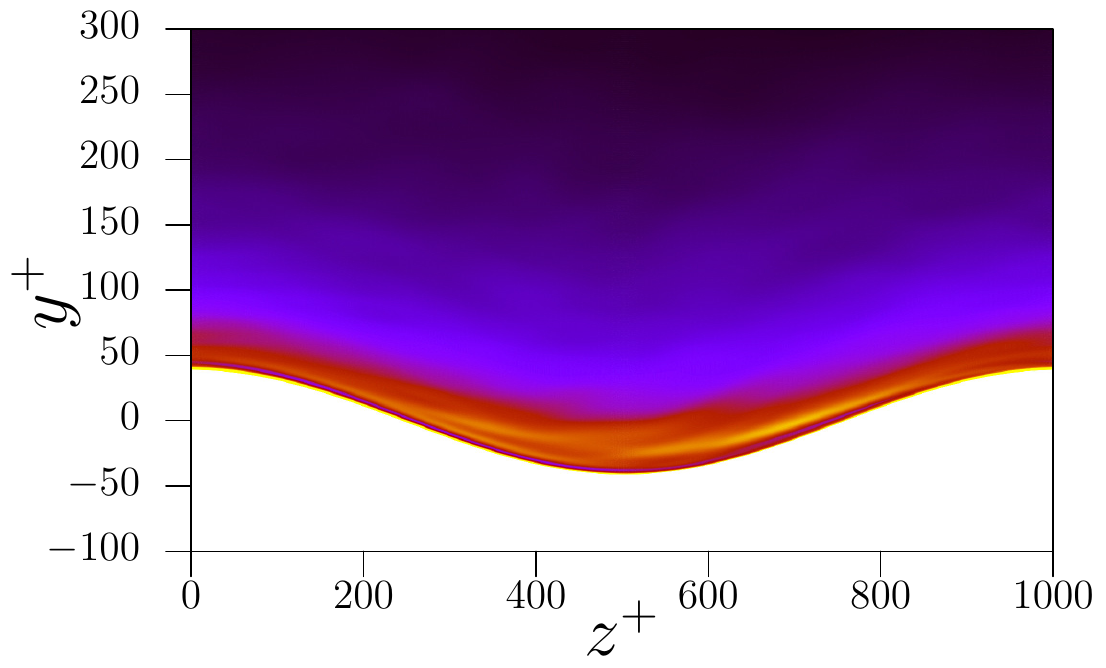}\label{fig::comp_om::24::omx}}
   \subfloat[~]{\includegraphics[width=0.33\textwidth]{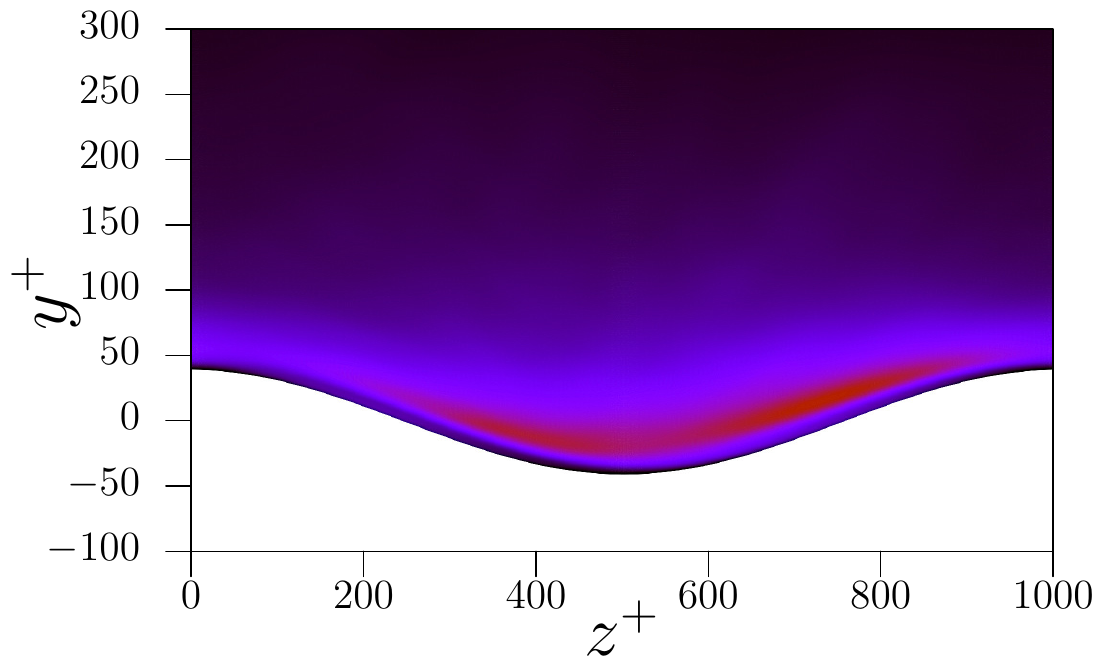}\label{fig::comp_om::24::omy}}
   \subfloat[~]{\includegraphics[width=0.33\textwidth]{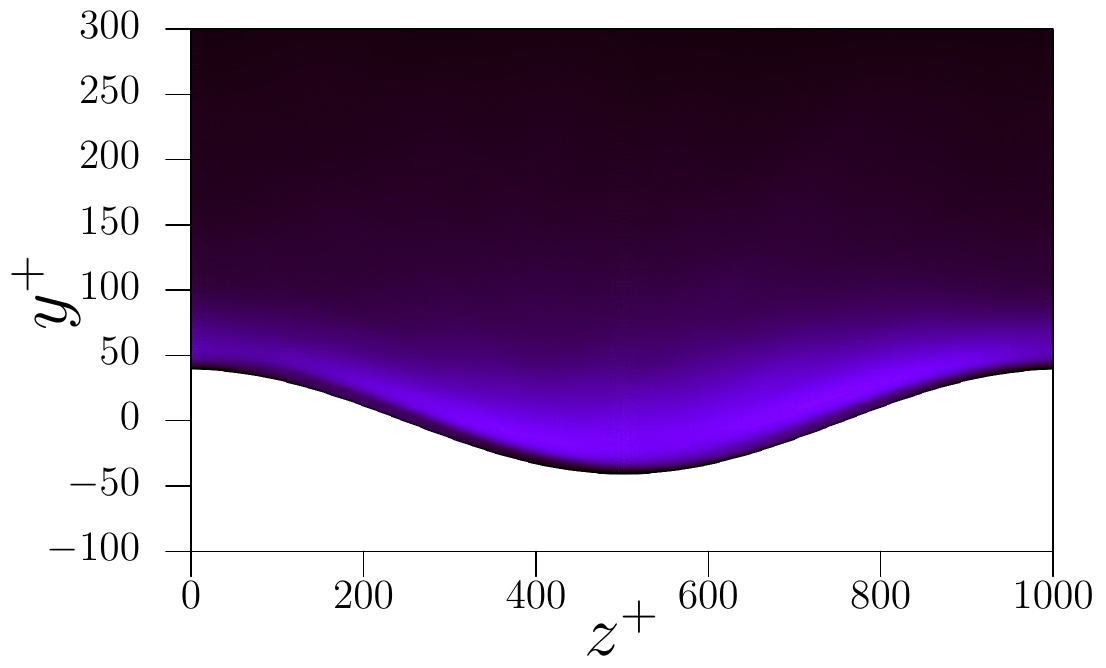}\label{fig::comp_om::24::omz}}

   \subfloat[~]{\includegraphics[width=0.33\textwidth]{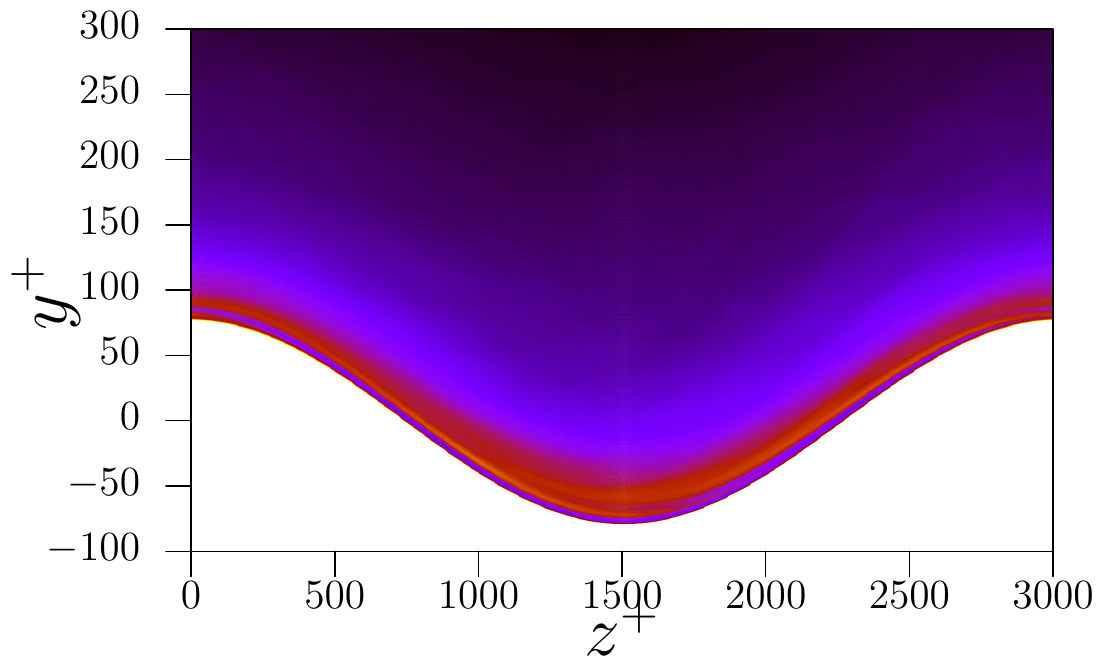}\label{fig::comp_om::80::omx}}
   \subfloat[~]{\includegraphics[width=0.33\textwidth]{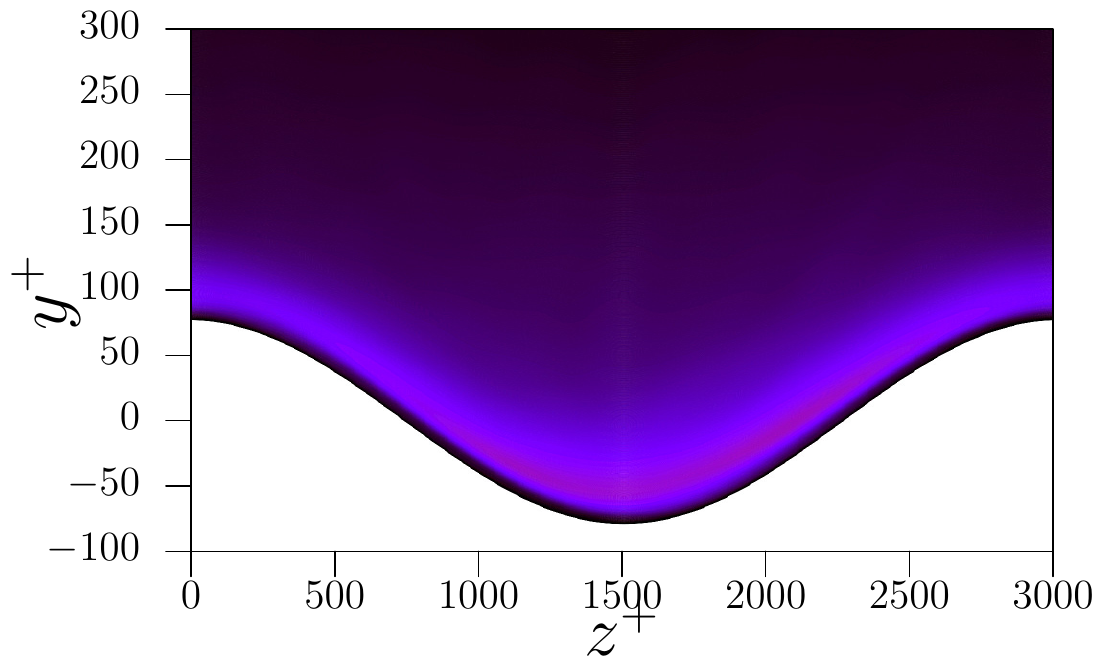}\label{fig::comp_om::80::omy}}
   \subfloat[~]{\includegraphics[width=0.33\textwidth]{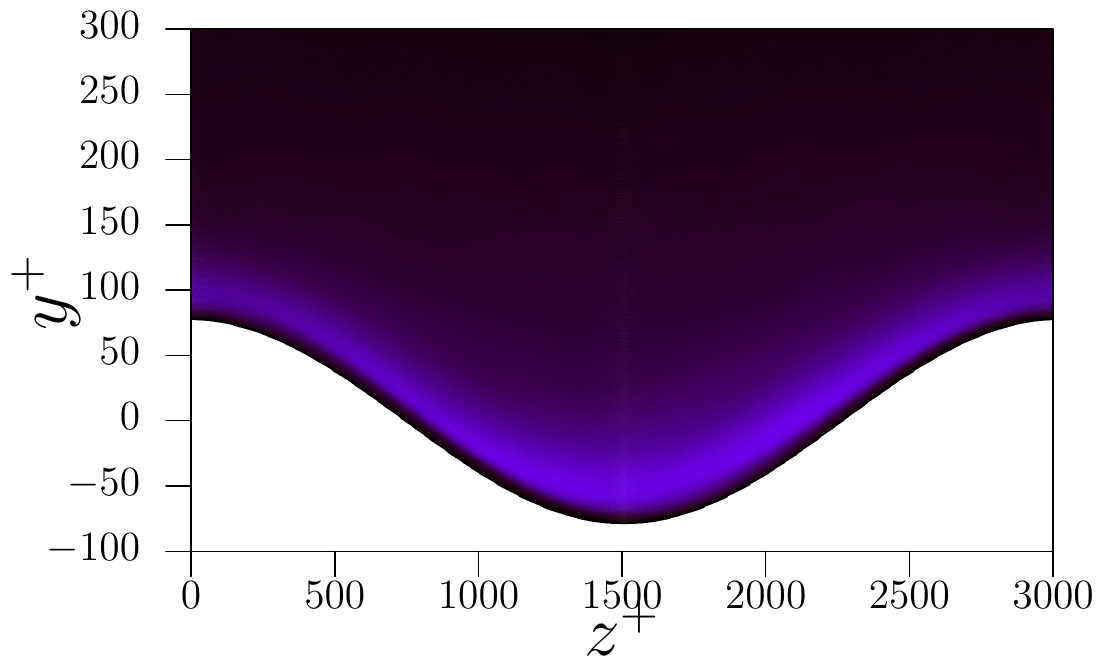}\label{fig::comp_om::80::omz}}
   \caption{Contours of the vorticity fluctuations
     $\overline{\omega''_i\omega''_i}$ in the $y$-$z$-plane at $x =
     90\,\theta$ for (first row \protect\subref{fig::comp_om::1::omx},\protect\subref{fig::comp_om::1::omy},\protect\subref{fig::comp_om::1::omz}) the non-actuated reference case, (second row \protect\subref{fig::comp_om::2::omx},\protect\subref{fig::comp_om::2::omy},\protect\subref{fig::comp_om::2::omz}) the case with the highest drag increase
     $N_{2}$, (third row \protect\subref{fig::comp_om::24::omx},\protect\subref{fig::comp_om::24::omy},\protect\subref{fig::comp_om::24::omz}) a case with moderate drag reduction
     $N_{24}$, and (fourth row \protect\subref{fig::comp_om::80::omx},\protect\subref{fig::comp_om::80::omy},\protect\subref{fig::comp_om::80::omz}) the case with the highest drag
     reduction $N_{80}$; (left) streamwise, (center) wall-normal, and
     (right) spanwise vorticity component. Note that case $N_{80}$ has
     a wavelength of $\lambda^+ = 3000$, the images for this case are
     thus compressed for lack of space.}
    \label{fig::comp_om}
  \end{center}
\end{figure}

The comparison of the vorticity fluctuation contours for four cases,
i.e., the non-actuated reference case, the case with the highest drag
increase $N_2$, a case with moderate drag reduction $N_{24}$, and the
case with the highest drag reduction $N_{80}$, in
Fig.~\ref{fig::comp_om} shows details about the phase dependence of
the overall structure of the vorticity field. It is obvious that the
drag increase is associated with strongly enlarged and highly
phase dependent vorticity contours. Due to the high amplitude and
short wavelength sinusoidal wall motion, the boundary layer flow is
massively perturbed. This is completely different for the two drag
reduction cases, where the overall boundary layer structure is
maintained but with reduced values of the wall-normal and spanwise
vorticity component. Phase variations occur especially for the
wall-normal component with the highest decrease above the wave crest
and the lowest decrease in the wave trough. Note, however, that
despite the clear phase variations, the overall lowered vorticity
levels are maintained throughout the entire actuation period, as shown
in Fig.~\ref{num::vorticity}.

% =============================================================
% 															SUBSECTION
% =============================================================
\subsection{Spanwise shear}
\label{sec::spanwiseshear}
\begin{figure}
  \begin{center}
    \subfloat[$\lambda^+ = 200$]{\includegraphics[width=0.5\textwidth]{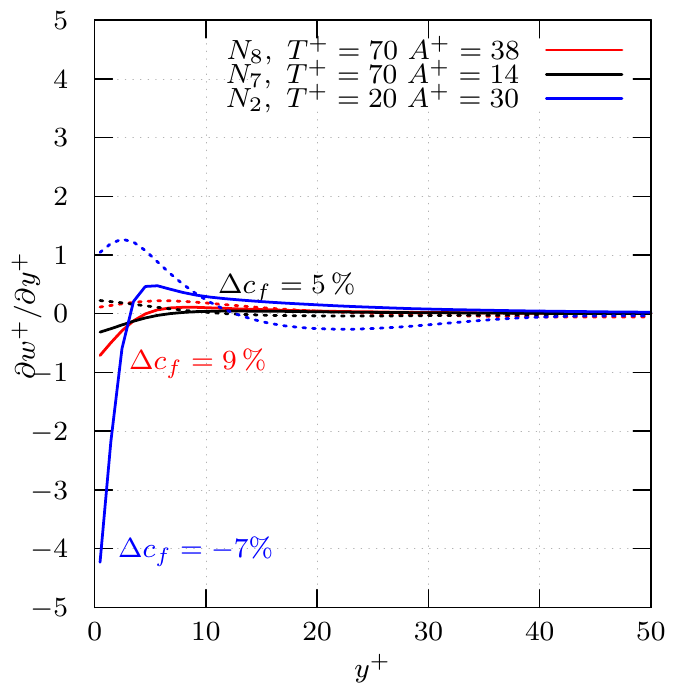}\label{fig::results::shear200}}~
    \subfloat[$\lambda^+ = 1000$]{\includegraphics[width=0.5\textwidth]{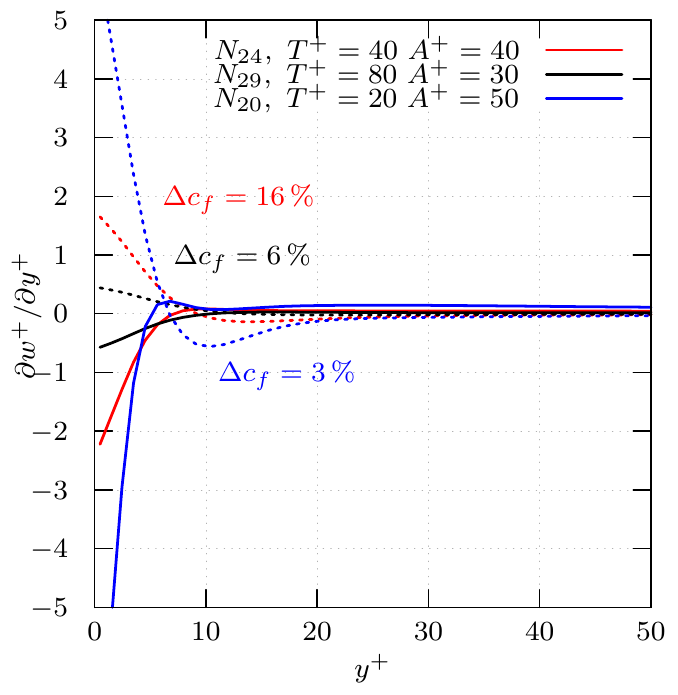}\label{fig::results::shear1000}}

    \subfloat[$\lambda^+ = 1800$]{\includegraphics[width=0.5\textwidth]{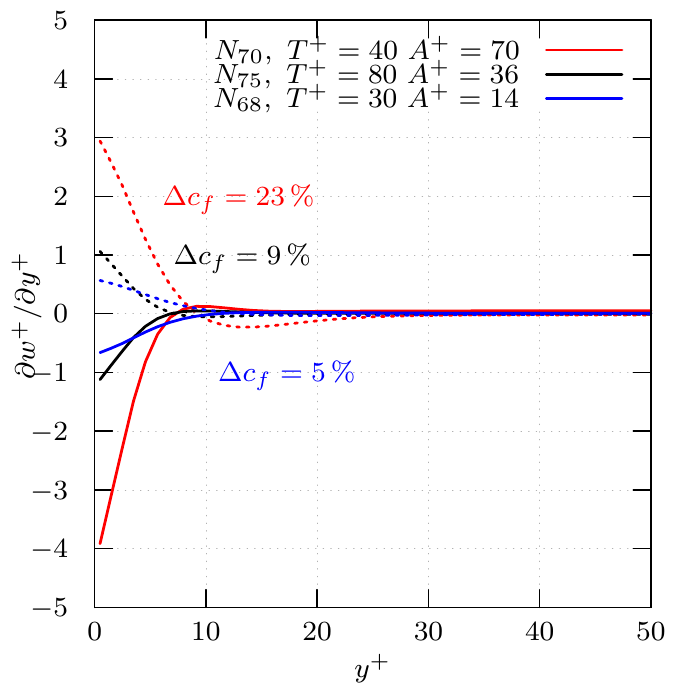}\label{fig::results::shear1800}}~
    \subfloat[$\lambda^+ = 3000$]{\includegraphics[width=0.5\textwidth]{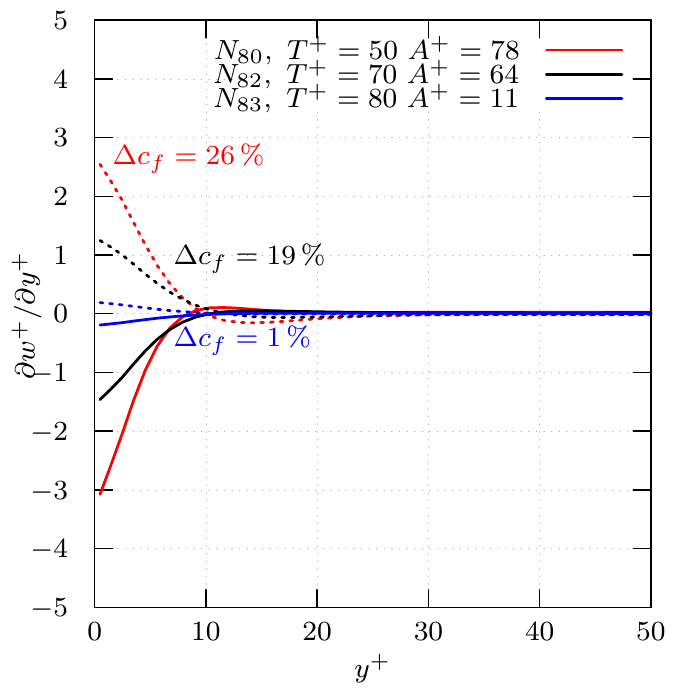}\label{fig::results::shear3000}}
    \caption{Phase averaged spanwise shear above the wave crest
      (\full) and in the wave trough (\dotted) for wavelengths
      \protect\subref{fig::results::shear200} $\lambda^+ = 200$,
      \protect\subref{fig::results::shear1000} $\lambda^+ = 1000$,
      \protect\subref{fig::results::shear1800} $\lambda^+ = 1800$, and
      \protect\subref{fig::results::shear3000} $\lambda^+ = 3000$. The
      cases listed in Tab.~\ref{tab::simulations} in the appendix are representative
      for high ($N_8$, $N_{24}$, $N_{70}$, $N_{80}$), medium ($N_7$,
      $N_{29}$, $N_{75}$, $N_{82}$), and low ($N_{20}$, $N_{68}$,
      $N_{83}$) skin-friction reduction or skin-friction increase ($N_{2}$) at each
      wavelength.}
    \label{fig::shear}
  \end{center}
\end{figure}

To investigate the secondary flow strength and its effect on the
near-wall turbulent structures, the Stokes strain
$\partial \tilde{w}^+ / \partial y^+$, i.e., the rate of change in the
wall-normal direction of the periodic fluctuations of the spanwise
velocity component, is considered in Fig.~\ref{fig::shear} for cases
with wavelengths $\lambda^+ = 200$, $\lambda^+ = 1000$,
$\lambda^+ = 1800$, and $\lambda^+ = 3000$. Based on the data
summarized in Tab~\ref{tab::simulations} in the appendix, for each
$\lambda^+ = \text{const}$ set the cases with the highest, medium, and
lowest skin-friction reduction are shown. The Stokes strain is
used to characterize the Stokes layer that develops above an
oscillating wall without any wall-normal deflection. However, similar
to a configuration with pure spanwise oscillating walls
\cite{Jung1992,Touber2012} a Stokes-like layer is also generated by a
transversal wave motion of the surface. Through the introduction of a
periodic wall-normal velocity $\tilde{v}$, a periodic spanwise
velocity component $\tilde{w}$ is induced via mass conservation
resulting in a wall-normal shear distribution defining a Stokes
layer. Cases with a high drag reduction generally show a high level of
symmetric, i.e., positive and negative, spanwise shear very close to
the wall, whereas less symmetric shear distributions yield lower drag
reduction. When the Stokes drag significantly increases near the wall,
i.e., a singular-like distribution occurs, the drag reduction reduces
drastically. This observation supports the assumption that the drag
reduction mechanism is strongly linked to spanwise oscillations which
are generated either by wave oscillations \cite{Touber2012}, traveling
waves of spanwise forcing \cite{Du2002}, spanwise velocity at the wall
\cite{Zhao2004}, or spanwise transversal surface waves
\cite{Klumpp2010b, Koh2015}.

The assumption of the importance of the oscillating spanwise shear for
skin-friction reduction also yields an explanation for the increasing
skin-friction reduction with growing wavelength, which agrees with an
observation of Du et al.~\cite{Du2002}. For short wavelengths, e.g.,
$\lambda^+ = 200$, the integration of the spanwise shear distribution
over the spanwise width of a near-wall streak, i.e.,
$l_z^+ = \mathcal{O}(100)$, results in only a minor force in the
spanwise direction acting on the streaks. For wavelengths
$\lambda^+ > 1000$, however, the spanwise shear varies only slightly
over the width of a streak, therefore a considerable spanwise force is
determined by the integration over the spanwise streak width. Note
that this behavior does not occur for spanwise wall oscillations,
since the periodic spanwise shear does not depend on the spanwise
coordinate.

\begin{figure}
  \begin{center}
    \subfloat[]{\includegraphics[width=0.5\textwidth]{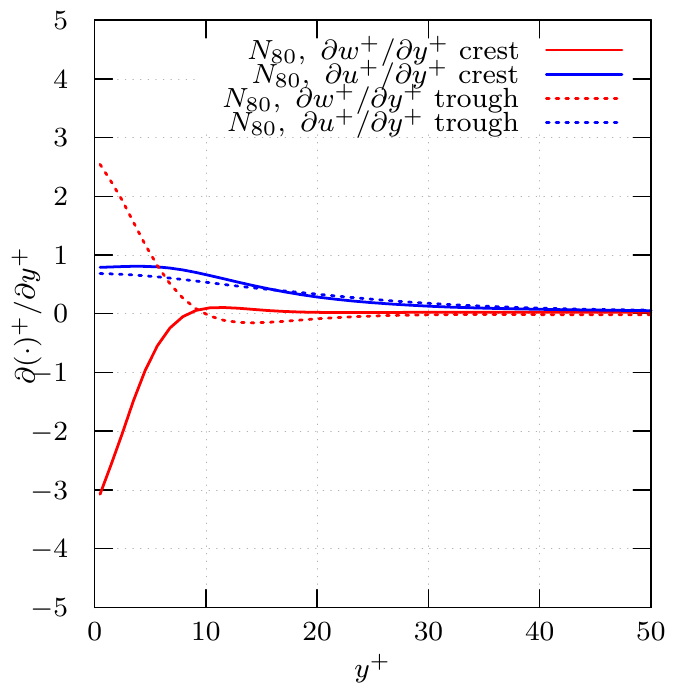}\label{fig::shearuw::split}}~
    \subfloat[]{\includegraphics[width=0.5\textwidth]{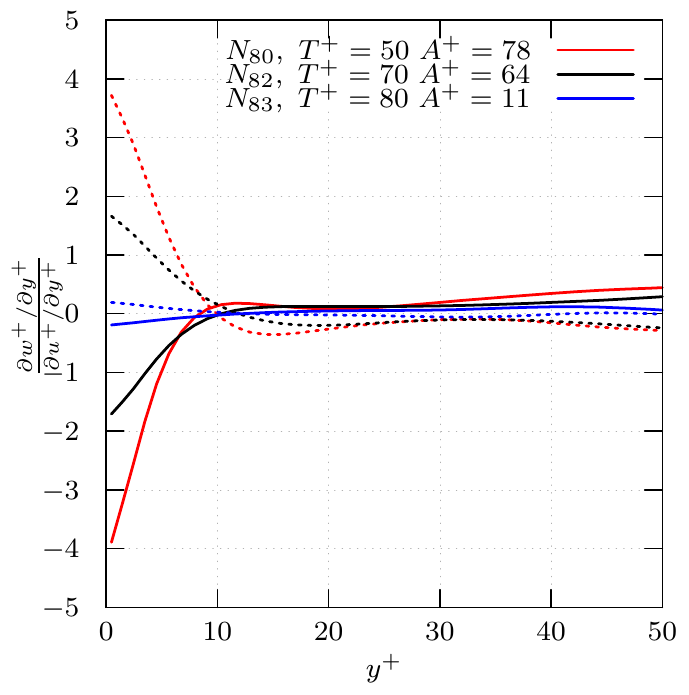}\label{fig::shearuw::ratio}}
    \caption{Wall-normal distributions of
      \protect\subref{fig::shearuw::split} the phase averaged
      streamwise and spanwise shear for the actuated case $N_{80}$ and
      \protect\subref{fig::shearuw::ratio} the ratio of the phase
      averaged spanwise and streamwise shear for cases with high
      ($N_{80}$), medium ($N_{82}$), and low ($N_{83}$) drag
      reduction.}
    \label{fig::shearuw}
  \end{center}
\end{figure}

A comparison of spanwise and streamwise shear is shown in
Fig.~\ref{fig::shearuw}. Touber and Leschziner~\cite{Touber2012}
discuss a certain optimal scenario for oscillatory forcing in
turbulent channel flow, where the ratio of spanwise to streamwise
shear reaches values of up to
$\frac{\partial \tilde{w}^+ / \partial y^+}{\partial u^+ / \partial
  y^+} \approx 3$.  Fig.~\subref*{fig::shearuw::ratio} shows that a
similar value of this ratio is obtained for the case with the highest
skin-friction reduction $N_{80}$, whereas a lower ratio is obtained
for the cases with medium ($N_{82}$) and low ($N_{83}$) skin-friction
reduction. For all cases, the change of the skin-friction is well
  correlated, i.e., $R = 0.84$ at the wave crest and $R = 0.85$ in the wave
  trough, with a spanwise-to-streamwise shear ratio of
  $\frac{\partial \tilde{w}^+ / \partial y^+}{\partial u^+ / \partial
    y^+} = 3.1$.
Overall, the results for the spanwise traveling transversal waves in
Fig.~\ref{fig::shear} underline the similarities to other drag
reduction techniques based on periodic spanwise forcing.  The results
of the cases with lower wavelength in combination with high amplitude
and high frequency deviate from this observation due to the increased
wetted surface.

% =============================================================
% 															SUBSECTION
% =============================================================
\subsection{Energy saving analysis}
\label{sec::energy_saving}
\begin{figure}
  \begin{center}
    \subfloat[~]{\includegraphics[width=0.5\textwidth]{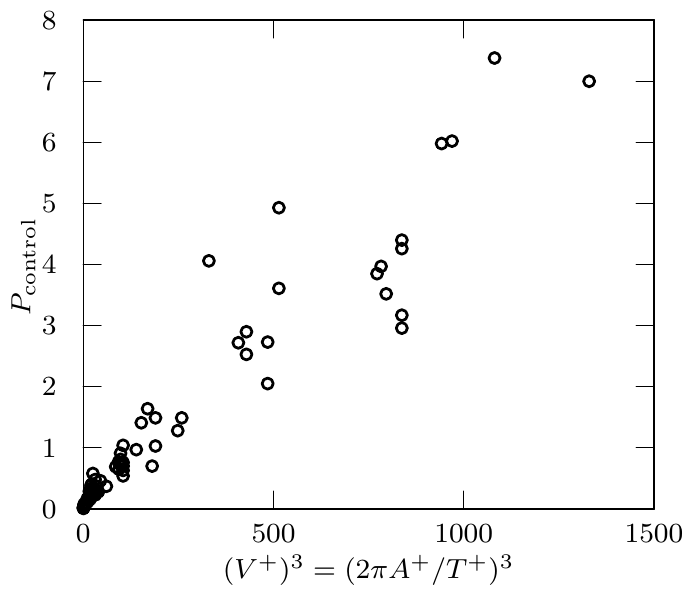}\label{fig::power::control}}~
    \subfloat[~]{\includegraphics[width=0.5\textwidth]{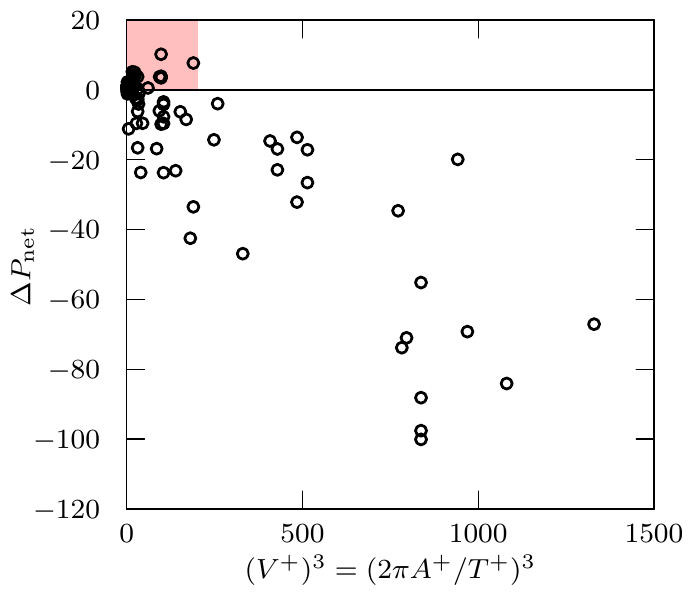}\label{fig::power::net}}

    \subfloat[~]{\includegraphics[width=1.0\textwidth]{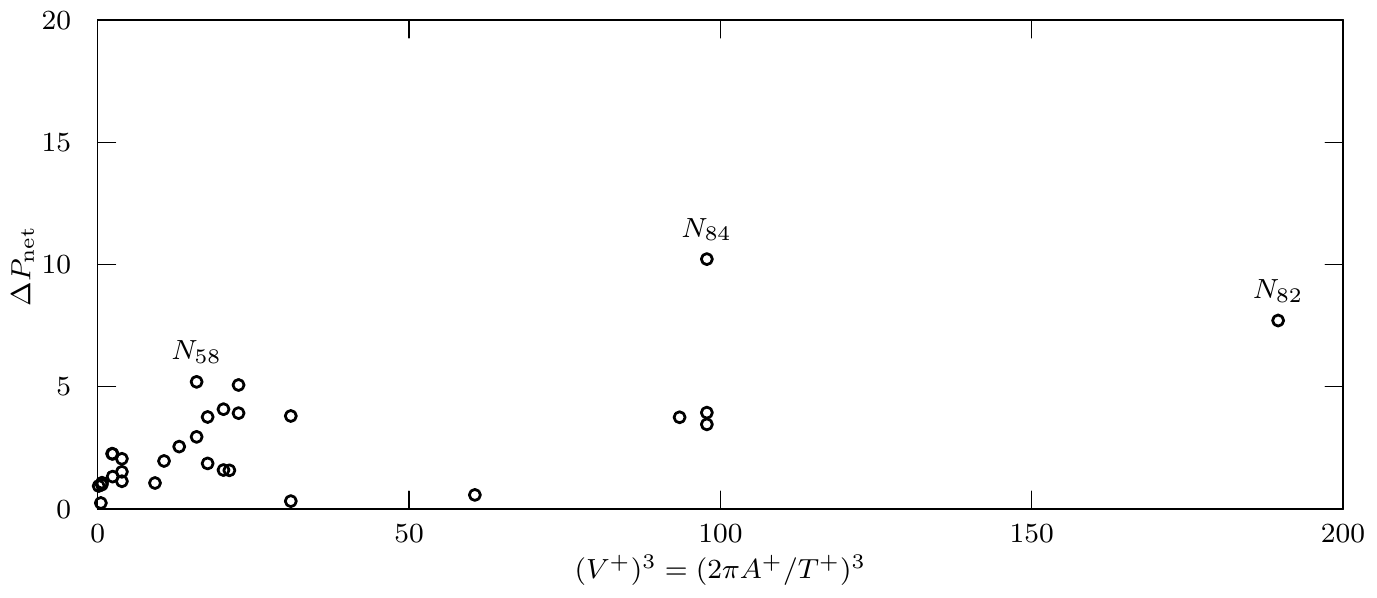}\label{fig::power::netdetail}}
    \caption{Dependence of \protect\subref{fig::power::control} the
      power spent $P_{\mathrm{control}}$ and
      \protect\subref{fig::power::net} the net power saving
      $\Delta P_{\mathrm{net}}$ on the cube of the actuation velocity
      amplitude $(V^+)^3 = (2\pi A^+/T^+)^3$; a zoom of the red
      rectangle in \protect\subref{fig::power::net} is shown in
      \protect\subref{fig::power::netdetail}. To indicate which cases
      possess net power saving the notation for three selected cases
      is given in \protect\subref{fig::power::netdetail}.}
    \label{fig::power}
  \end{center}
\end{figure}

The previous discussion has shown that considerable drag reduction
rates have been obtained. However, drag reduction is not the only metric
of interest. From a prospective application point of view, the
question of net energy saving and its relation to drag reduction must
be addressed.  The ideal net energy saving is defined as
\begin{equation}
\label{eq:2}
\Delta P_{\mathrm{net}} = \frac{P_{d,\mathrm{ref}} - (P_{d,\mathrm{act}} + P_{\mathrm{control}})}{P_{d,\mathrm{ref}}} \cdot 100~,
\end{equation}
where $P_{d,\mathrm{ref}/\mathrm{act}} = u_\infty (F_{f} + F_{p})$ is the power necessary to
overcome the friction $F_f$ and pressure forces $F_p$ of the non-actuated $P_{d,\mathrm{ref}}$ and
actuated surface $P_{d,\mathrm{act}}$ in the streamwise direction. The power spent on
control $P_{\mathrm{control}}$, i.e., on deflecting the surface for the traveling wave, is computed by
\begin{equation}
\label{eq:3}
P_{\mathrm{control}}  = \int_{A_\mathrm{surf}} v(x,z) (\tau_w \mathbf{e}_{xz} + p \mathbf{e}_y)\cdot\mathbf{n}dA~,
\end{equation}
where $e_{xz} = (1,0,1)^T$ is a combination of the unit vectors in the streamwise and wall-normal direction.
Hence, $P_{\mathrm{control}}$ is a combination of the viscous and
pressure forces effective in the $y$-direction multiplied by the speed
of the wall motion in the $y$-direction. The values of
$P_{\mathrm{control}}$ and $\Delta P_\mathrm{net}$ are depicted in
Fig.~\ref{fig::power} for all cases. The data for $\Delta
P_\mathrm{net}$ are also listed in
Tab.~\ref{tab::simulations} in the appendix. Fig.~\subref*{fig::power::control} shows the
expected approximately linear dependence between the power spent
$P_\mathrm{control}$ and the actuation velocity cubed $(V^+)^3 =
\left(2\pi A^+/T^+\right)^3$.

It is evident from Fig.~\subref*{fig::power::net} that net power
saving is only obtained for a few cases with a maximum of $\Delta
P_{\mathrm{net}} = \Remark{n84drnet}\,\%$ for case $N_{84}$. Most
cases clearly show no net power saving but net power loss. For
instance, for $N_{20}$ $\Delta P_{\mathrm{net}}$ is $\Delta
P_{\mathrm{net}} = \Remark{n20drnet}\,\%$, i.e., almost the fourfold
$P_{d,\mathrm{ref}}$ has to be invested.

A closer look at the net-power-saving cases in
Fig.~\subref*{fig::power::netdetail} shows that high drag reduction
rates are no indicator for high net power saving. That is, there is no
linear relation between drag reduction and net power saving. Instead,
a high value of the scaling parameter $A^+ \sqrt{2\pi / T^+}$ obtained
by a low amplitude speed $2 \pi A^+ / T^+$ leads to positive net power
saving. Thus, as expected, there is a trade-off between the minimum
power input to effectively influence the turbulent boundary layer and
a maximum power input above which the energy costs grow
tremendously. In the current parameter range, the optimum energy
saving solution, i.e., $\Delta P_\mathrm{net} =
\Remark{n84drnet}\,\%$, is achieved for the $N_{84}$ case with
$\lambda^+ = 3000$, $T^+ = 90$, and $A^+ = 66$, which possesses just a
medium drag reduction of $\Delta c_d = \Remark{n84dr}\,\%$. Note that
the parameters that result in high net energy saving are in the upper
range of the interval. Furthermore, the data in
Tab.~\ref{tab::simulations} in the appendix indicates that the sensitivity of $\Delta
P_\mathrm{net}$ is less pronounced for larger wavelength and above a
wave period of $60$.

\section{Conclusions}
\label{sec::conclusions}
To analyze drag reducing effects and the net energy saving potential
of spanwise traveling transversal surface waves high-resolution
large-eddy simulations were conducted. The parameter space defined by
the wave amplitude, wave period, and wavelength was investigated based
on $80$ wave parameter setups for purely spanwise traveling waves. The
variation of skin-friction reduction, i.e., mean wall-shear stress
alteration, drag reduction, i.e., surface integrated wall-shear
stress, and net energy saving was analyzed. In brief, a maximum drag
reduction and net energy saving of $\Remark{n80dr}\,\%$ and
$\Remark{n84drnet}\,\%$ was found.

The highest skin-friction reduction was achieved for a period of
$T^+ \approx 50$, which is lower than the one reported for spanwise
oscillating wall and within the range of the streak formation time
scale. Larger wavelengths and amplitudes yielded higher skin-friction
reduction.  For wavelengths larger than 1000 plus units, a scaling
with the Stokes layer height and the velocity amplitude was found to
predict skin-friction reduction reasonably well. Additionally, the
difference between skin-friction reduction and drag reduction, i.e.,
the increase of the wetted surface was taken into account, was found
to be substantial for short wavelengths in combination with high
amplitudes. A drag-reduction model was derived from the sparse dataset
using optimized support vector regression. From the model, a tendency
to an asymptotic behavior of amplitude and period could be identified,
supporting the assumption of an optimum period in the range
$40 \leq T^+ \leq 50$ for large wavelengths. Moreover, a ridgeline
behavior of optimum drag reduction in the high wavelength regime was
extracted from the model.

The statistical results of the turbulent flow field confirmed this
result for high wavelength configurations, where similar effects of
the actuation on the near-wall region compared to spanwise oscillating
walls were observed. That is, considerable reductions of the near-wall
velocity streak strength were found for the cases with high drag
reduction. For the highest drag reduction case, the smaller wall-shear
stress was coupled to a substantial decrease of the Reynolds shear
stress in the near-wall region. Generally, for large wavelength cases
$\lambda^+ > 1000$ the decrease of the wall-normal and spanwise
vorticity fluctuations strongly correlated with skin-friction
reduction and drag reduction. A comparison among several
configurations revealed that for unfavorable combinations of short
wavelength and high amplitude, a considerable increase of the
turbulent exchange resulting in skin-friction and drag increase was
observed, whereas large wavelengths circumvented this effect and led
to drag reduction. The periodic secondary flow field generated by the
wavy surface motion approximated that of Stokes flow. Similar
oscillating spanwise shear distributions were observed for many drag
reducing cases, although no perfectly symmetrical oscillatory
excitation of the near-wall structures is achieved.

No linear relationship between drag reduction and net energy saving
was determined. That is, due to the non-linear response of the
near-wall flow to the actuation the highest drag reduction does not
result in the highest net energy saving. The maximum net energy saving
$\Delta P_\mathrm{net} = \Remark{n84drnet}\,\%$ was achieved for a
drag reduction of $\Delta c_d = \Remark{n84dr}\,\%$, which is clearly
lower than the maximum drag reduction of
$\Delta c_d = \Remark{n80dr}\,\%$. A high value of the product of the
actuation amplitude speed and the thickness of the Stokes layer at low
amplitude speed results in positive net energy saving. The
susceptibility of $\Delta P_\mathrm{net}$ is less pronounced for
larger wavelength, which is a promising observation for prospective
applications.

\section*{Acknowledgements}
  The research was funded by the Deutsche Forschungsgemeinschaft (DFG)
  in the framework of the research projects SCHR 309/52 and SCHR
  309/68. The authors gratefully acknowledge the Gauss Centre for
  Supercomputing e.V. (www.gauss-centre.eu) for funding this project
  by providing computing time on the GCS Supercomputers Hazelhen at
  HLRS Stuttgart and JURECA at J\"ulich Supercomputing Centre (JSC).
  BRN acknowledges support from the French National Research Agency (ANR) 
  under grant ANR-17-ASTR-0022 (FlowCon).

\section*{Compliance with Ethical Standards}

\section*{Conflict of Interests}
The authors declare that they have no conflicts of interest.

\appendix

\FloatBarrier
\setcounter{table}{0}
\renewcommand{\thetable}{A\arabic{table}}
\section{Appendix}
\begin{table}[!htbp]
  \centering
  \begin{tabularx}{1.0\textwidth}{r@{\qquad}r@{\qquad}r@{\qquad}r@{\qquad}r@{\qquad}r@{\qquad}r@{\qquad}r@{\qquad}r}
    $N$ & $L_z^+$ & $\lambda^+$ & $T^+$ & $A^+$ & $\Delta c_d~[ \% ]$ & $\Delta c_f~[ \% ]$  & $\Delta A_\mathrm{surf}~[ \% ]$ & $\Delta P~[ \% ]$\\
    \midrule
1 & 1000 & 0 & 0 & 0 & 0 & 0 & 0.0 & 0\\
2 & 1000 & 200 & 20 & 30 & -27 & -7 & 19.4 & -100\\
3 & 1000 & 200 & 30 & 21 & -1 & 8 & 10.1 & -17\\
4 & 1000 & 200 & 40 & 30 & -9 & 9 & 19.4 & -24\\
5 & 1000 & 200 & 50 & 45 & -26 & 9 & 39.0 & -42\\
6 & 1000 & 200 & 60 & 30 & -10 & 8 & 19.4 & -17\\
7 & 1000 & 200 & 70 & 14 & 0 & 5 & 4.7 & -1\\
8 & 1000 & 200 & 70 & 38 & -17 & 9 & 29.4 & -24\\
9 & 1000 & 200 & 100 & 28 & -9 & 7 & 17.2 & -11\\
10 & 1000 & 500 & 20 & 30 & 0 & 4 & 3.5 & -98\\
11 & 1000 & 500 & 30 & 22 & 9 & 10 & 1.9 & -10\\
12 & 1000 & 500 & 40 & 21 & 8 & 9 & 1.7 & -1\\
13 & 1000 & 500 & 40 & 30 & 8 & 11 & 3.5 & -10\\
14 & 1000 & 500 & 60 & 30 & 5 & 8 & 3.5 & -2\\
15 & 1000 & 500 & 70 & 36 & 3 & 8 & 4.9 & -4\\
16 & 1000 & 500 & 70 & 64 & -10 & 4 & 14.6 & -33\\
17 & 1000 & 500 & 100 & 48 & -3 & 5 & 8.6 & -10\\
18 & 1000 & 1000 & 20 & 10 & 5 & 5 & 0.1 & -6\\
19 & 1000 & 1000 & 20 & 30 & 13 & 13 & 0.9 & -88\\
20 & 1000 & 1000 & 20 & 50 & 0 & 3 & 2.4 & -289\\
21 & 1000 & 1000 & 40 & 10 & 3 & 3 & 0.1 & 1\\
22 & 1000 & 1000 & 40 & 20 & 7 & 8 & 0.4 & 0\\
23 & 1000 & 1000 & 40 & 30 & 12 & 13 & 0.9 & -4\\
24 & 1000 & 1000 & 40 & 40 & 15 & 16 & 1.6 & -14\\
25 & 1000 & 1000 & 40 & 50 & 15 & 17 & 2.4 & -32\\
26 & 1000 & 1000 & 40 & 60 & 13 & 16 & 3.5 & -55\\
27 & 1000 & 1000 & 80 & 10 & 1 & 1 & 0.1 & 0\\
28 & 1000 & 1000 & 80 & 20 & 3 & 4 & 0.4 & 2\\
29 & 1000 & 1000 & 80 & 30 & 6 & 6 & 0.9 & 3\\
30 & 1000 & 1000 & 80 & 40 & 9 & 10 & 1.6 & 4\\
31 & 1000 & 1000 & 80 & 50 & 9 & 11 & 2.4 & 1\\
32 & 1000 & 1000 & 80 & 60 & 9 & 12 & 3.5 & -3\\
33 & 1000 & 1000 & 120 & 10 & 1 & 1 & 0.1 & 1\\
34 & 1000 & 1000 & 120 & 20 & 0 & 1 & 0.4 & 0\\
35 & 1000 & 1000 & 120 & 30 & 3 & 4 & 0.9 & 2\\
36 & 1000 & 1000 & 120 & 40 & 3 & 5 & 1.6 & 1\\
37 & 1000 & 1000 & 120 & 50 & 2 & 5 & 2.4 & -1\\
38 & 1000 & 1000 & 120 & 60 & 2 & 6 & 3.5 & -3\\
    \bottomrule
  \end{tabularx}
\end{table}
\begin{table}[p]
  \centering
  \begin{tabularx}{1.0\textwidth}{r@{\qquad}r@{\qquad}r@{\qquad}r@{\qquad}r@{\qquad}r@{\qquad}r@{\qquad}r@{\qquad}r}
    $N$ & $L_z^+$ & $\lambda^+$ & $T^+$ & $A^+$ & $\Delta c_d~[ \% ]$ & $\Delta c_f~[ \% ]$  & $\Delta A_\mathrm{surf}~[ \% ]$ & $\Delta P~[ \% ]$\\
    \midrule
39 & 1200 & 0 & 0 & 0 & 0 & 0 & 0.0 & 0\\
40 & 1200 & 600 & 30 & 44 & 2 & 7 & 5.1 & -74\\
41 & 1200 & 600 & 40 & 59 & -4 & 5 & 8.9 & -71\\
42 & 1200 & 600 & 50 & 36 & 9 & 12 & 3.5 & -6\\
43 & 1200 & 600 & 60 & 21 & 5 & 6 & 1.2 & 2\\
44 & 1200 & 600 & 70 & 29 & 6 & 8 & 2.3 & 2\\
45 & 1200 & 600 & 80 & 66 & -5 & 6 & 11.0 & -23\\
46 & 1200 & 600 & 90 & 51 & -1 & 6 & 6.8 & -10\\
47 & 1200 & 600 & 100 & 14 & 2 & 2 & 0.5 & 1\\
    \bottomrule
  \end{tabularx}
  \centering
  \begin{tabularx}{1.0\textwidth}{r@{\qquad}r@{\qquad}r@{\qquad}r@{\qquad}r@{\qquad}r@{\qquad}r@{\qquad}r@{\qquad}r}
    $N$ & $L_z^+$ & $\lambda^+$ & $T^+$ & $A^+$ & $\Delta c_d~[ \% ]$ & $\Delta c_f~[ \% ]$  & $\Delta A_\mathrm{surf}~[ \% ]$ & $\Delta P~[ \% ]$\\
    \midrule
48 & 1600 & 0 & 0 & 0 & 0 & 0 & 0.0 & 0\\
49 & 1600 & 1600 & 20 & 22 & 11 & 11 & 0.2 & -47\\
50 & 1600 & 1600 & 40 & 34 & 14 & 14 & 0.4 & -6\\
51 & 1600 & 1600 & 40 & 48 & 19 & 19 & 0.9 & -23\\
52 & 1600 & 1600 & 50 & 60 & 19 & 20 & 1.4 & -17\\
53 & 1600 & 1600 & 50 & 73 & 21 & 22 & 2.0 & -35\\
54 & 1600 & 1600 & 60 & 27 & 8 & 8 & 0.3 & 4\\
55 & 1600 & 1600 & 70 & 71 & 17 & 19 & 1.9 & -4\\
56 & 1600 & 1600 & 80 & 17 & 2 & 2 & 0.1 & 1\\
57 & 1600 & 1600 & 90 & 65 & 13 & 14 & 1.6 & 4\\
58 & 1600 & 1600 & 100 & 40 & 8 & 8 & 0.6 & 5\\
    \bottomrule
  \end{tabularx}

  \centering
  \begin{tabularx}{1.0\textwidth}{r@{\qquad}r@{\qquad}r@{\qquad}r@{\qquad}r@{\qquad}r@{\qquad}r@{\qquad}r@{\qquad}r}
    $N$ & $L_z^+$ & $\lambda^+$ & $T^+$ & $A^+$ & $\Delta c_d~[ \% ]$ & $\Delta c_f~[ \% ]$  & $\Delta A_\mathrm{surf}~[ \% ]$ & $\Delta P~[ \% ]$\\
    \midrule
59 & 1800 & 0 & 0 & 0 & 0 & 0 & 0.0 & 0\\
60 & 1800 & 900 & 30 & 49 & 10 & 12 & 2.9 & -84\\
61 & 1800 & 900 & 40 & 63 & 7 & 12 & 4.7 & -69\\
62 & 1800 & 900 & 50 & 22 & 7 & 7 & 0.6 & 2\\
63 & 1800 & 900 & 50 & 44 & 12 & 14 & 2.3 & -8\\
64 & 1800 & 900 & 70 & 28 & 7 & 8 & 0.9 & 3\\
65 & 1800 & 900 & 80 & 17 & 3 & 4 & 0.4 & 2\\
66 & 1800 & 900 & 80 & 60 & 6 & 9 & 4.3 & -8\\
67 & 1800 & 900 & 90 & 39 & 6 & 7 & 1.8 & 2\\
68 & 1800 & 1800 & 30 & 14 & 5 & 5 & 0.1 & -2\\
69 & 1800 & 1800 & 40 & 51 & 19 & 20 & 0.8 & -27\\
70 & 1800 & 1800 & 40 & 70 & 22 & 23 & 1.5 & -67\\
71 & 1800 & 1800 & 50 & 59 & 20 & 21 & 1.1 & -15\\
72 & 1800 & 1800 & 60 & 44 & 15 & 15 & 0.6 & 3\\
73 & 1800 & 1800 & 60 & 75 & 21 & 22 & 1.7 & -14\\
74 & 1800 & 1800 & 70 & 29 & 7 & 7 & 0.3 & 4\\
75 & 1800 & 1800 & 80 & 36 & 9 & 9 & 0.4 & 5\\
76 & 1800 & 1800 & 90 & 66 & 13 & 14 & 1.3 & 4\\
77 & 1800 & 1800 & 100 & 21 & 3 & 3 & 0.1 & 2\\
    \bottomrule
  \end{tabularx}

  \centering
  \begin{tabularx}{1.0\textwidth}{r@{\qquad}r@{\qquad}r@{\qquad}r@{\qquad}r@{\qquad}r@{\qquad}r@{\qquad}r@{\qquad}r}
    $N$ & $L_z^+$ & $\lambda^+$ & $T^+$ & $A^+$ & $\Delta c_d~[ \% ]$ & $\Delta c_f~[ \% ]$  & $\Delta A_\mathrm{surf}~[ \% ]$ & $\Delta P~[ \% ]$\\
    \midrule
78 & 3000 & 0 & 0 & 0 & 0 & 0 & 0.0 & 0\\
79 & 3000 & 3000 & 40 & 51 & 21 & 21 & 0.3 & -17\\
80 & 3000 & 3000 & 50 & 78 & 26 & 26 & 0.7 & -20\\
81 & 3000 & 3000 & 60 & 26 & 7 & 7 & 0.1 & 4\\
82 & 3000 & 3000 & 70 & 64 & 19 & 19 & 0.4 & 8\\
83 & 3000 & 3000 & 80 & 11 & 1 & 1 & 0.0 & 1\\
84 & 3000 & 3000 & 90 & 66 & 16 & 16 & 0.5 & 10\\
    \bottomrule
  \end{tabularx}
  \caption{Actuation parameters of the turbulent boundary layer simulations, where each setup is denoted by a case number $N$. The quantity $\lambda^+$ is the spanwise wavelength of the traveling wave, $T^+$ is the period, and $A^+$ is the amplitude, all given in inner units, i.e., non-dimensionalized with the kinematic viscosity $\nu$ and the friction velocity $u_\tau$. Each block includes setups with varying period and amplitude for a constant wavelength. The list includes the values of the averaged relative drag reduction $\Delta c_d$, the averaged relative skin friction reduction  $\Delta c_f$, the relative increase of the wetted surface $\Delta A_\mathrm{surf}$, and the net power saving $\Delta P$.}
\label{tab::simulations}
\end{table}
\FloatBarrier

\bibliographystyle{spmpsci}
\bibliography{marian_abbrv}

\end{document}